# Comparing Selected Criteria of Programming Languages Java, PHP, C++, Perl, Haskell, AspectJ, Ruby, COBOL, Bash Scripts and Scheme
## Revision 1.0

Sultan S. Al-Qahtani
Concordia University
Montreal, Quebec, Canada
s_alqaht@cse.concordia.ca

Luis F. Guzman
Concordia University
Montreal, Quebec, Canada
l_guzman@cse.concordia.ca

Rafik Arif
Concordia University
Montreal, Quebec, Canada
r_ar@cse.concordia.ca

Adrien Tevoedjre
Concordia University
Montreal, Quebec, Canada
a_tevoed@cse.concordia.ca

Pawel Pietrzynski
Concordia University
Montreal, Quebec, Canada
p_pietrz@cse.concordia.ca

## Abstract

Comparison of programming languages is a common topic of discussion among software engineers. Few languages ever become sufficiently popular that they are used by more than a few people or find their niche in research or education; but professional programmers can easily use dozens of different languages during their career. Multiple programming languages are designed, specified, and implemented every year in order to keep up with the changing programming paradigms, hardware evolution, etc. In this paper we present a comparative study between ten programming languages: Haskell, Java, Perl, C++, AspectJ, COBOL, Ruby, PHP, Bash Scripts, and Scheme; with respect of the following criteria: Secure programming practices, web applications development, web services design and composition, object oriented-based abstraction, reflection, aspect-orientation, functional programming, declarative programming, batch scripting, and user interface prototype design.

# 1. Introduction

The first high-level programming languages were designed during the 1950s. Ever since then, programming languages have been a fascinating and productive area of study [43]. Thousands of different programming languages have been created, mainly in the computer field, with many more being created every year; they are designed, specified, and implemented for the purpose of being up to date with the evolving programming paradigms (*e.g.* imperative, object-oriented, aspect-oriented, reflective, and functional to name a few).

While some programming languages enjoy world-wide popularity and are commonly used to develop large, enterprise-level applications (e.g. C, C++, C#, Java, PHP, or Perl); others are only used by a lower number of people or are more oriented to the academia for research and education (e.g. OCaml, Haskell, Scheme, OZ, or Scala).

Computer scientists aim to develop programming languages that combine expressive power with simplicity and efficiency [43], also they are aimed to support multiple programming paradigms; such is the case of Ruby, Python, or Delphi. The idea of a multi-paradigm language is to provide a framework in which programmers can work in a variety of styles, freely intermixing constructs from different paradigms. The design goal of such languages is to allow programmers to use the best tool for a job, admitting that no one paradigm solves all problems in the easiest or most efficient way [53].

With the continuing increase in the variety, functionality, and complexity of engineering software, with its more widespread use, and with its increasing importance, more attention must be paid to programming language suitability so that rational decisions regarding language selection may be made [43].

## 1.1 Related Work

Extensive work and research have been done in the field of comparative studies of programming languages. In the paper "Programming Languages: A Comparative Study" [43] the authors collate C++, Java, Lisp, and Perl under the parameters of reusability, portability, reliability, readability, efficiency, availability of compilers and tools, familiarity and expressiveness. Also a study of program length, programming effort, run time efficiency, memory consumption, and reliability is disclosed by implementing and running the same set of programs on said languages.

In the paper titled "A Comparative Study of Language Support for Generic Programming" [44] the authors report on a comprehensive comparison of generic programming (a style of computer programming in which algorithms are written in terms of *to-be-specified-later* types that are then *instantiated* when needed for specific types provided as parameters) in six programming languages: C++, Standard ML, Haskell, Eiffel, Java (with its proposed generics extension), and Generic C#. As languages increasingly support generics, the authors highlight the features necessary to provide powerful generics and that their absence causes serious difficulties for programmers.

In the work called "A Comparative Study between Computer Programming Languages for Developing Distributed Systems in Web Environment" [45] K. Aldrawiesh et al present a view of the capabilities of the programming languages Java, C++, ANSI C++ and C# for developing Web services and distributed systems. The criteria of comparison used by the authors are high integrity, distributed system, simplicity and usage, concurrency, platform, maintainability, and reliability.

In the article written by Steve Vinoski titled "The Language Divide" [46] it is proposed a comparison between middleware programming languages and dynamic scripting languages for Web development. It is stated that languages like Java, and now C#, might also provide a more direct answer to bridging the language divide between middleware and the Web. The author also states that much similarity exists between server-side invocations in object-oriented or service-oriented middleware and Web server invocations via CGI or similar approaches.

Other notable works worth mentioning in the field of programming languages comparison are [45, 46, and 47].

## 1.2 Overview

According to Hebert Meyer:

> "*No programming language is perfect. There is not even a single best language; there are only languages well suited or perhaps poorly suited for particular purposes.*"[187].

Moreover, he also stated that:

> "*A useful language needs arrays, pointers and a generic mechanism for building data structures*" [187].

The purpose of this document is to present an impartial comparative study between the programming languages Haskell, Java, Perl, C++, AspectJ, COBOL, Ruby, PHP, Bash Scripts, and Scheme. The criteria of comparison is focused on secure programming practices, web applications development, web services design and composition, object oriented-based abstraction, reflection, aspect-orientation, functional programming, declarative programming, batch scripting, and user interface prototype design.

## 1.3 Haskell

Haskell is purely functional language, non-strict, standard, and modern. Haskell provides polymorphic typing, higher-order functions, lazy evaluation and all the features that are typical of functional language. The reason that let Haskell being purely functional language is all computations are done via the evaluation of expression to yield values. Every value has an associated type. Haskell programs are type safe, because Haskell's static type system defines the formal relations between types and values. This means at compile time the compiler will deal with types, consequently there is no type checking during program execution. Finally, we can come up with, Haskell is strongly typed, and Haskell's type system is robust enough.

## 1.4 Java

Java was started as a project called "Oak" by James Gosling in July 1991. Java is portable OO-language, simple, and designed at Sun Microsystems labs by research staff and originally developed by James Gosling. Java syntax fairly similar to C++, also Java borrows ideas from Mesa, Objective-C and Modula-3. Some features of Java, Java has inheritance, exception handling, modularity, strong type checking, garbage collection, polymorphism and etc. The newest version of Java is Java platform 6, specially this version includes nested classes, reflection, and persistence as well as many additional standard libraries. The *class* in java is the fundamental structure component. Java standard library includes extensive I/O facilities, date/time support, cryptographic security classes, distributed computation support, GUI toolkit, and system interfaces. Additionally to normal application development by java, Java is used to develop embedded programs, called 'applets', for web browsers and other Java enabled platforms. This capability is an important part of Java, and the standard library packages include a security manager to restrict the capabilities of Java applets. These applet facilities were important to Java's widespread adoption and popularity. With Java, you can build complete applications featuring everything from accelerated 3D graphics and other multimedia features to strong cryptography and network connectivity. On the Web, Java can be used on the client side to create applets (these small programs can run right inside a Web page with the full power that Java has to offer), and on the server side to create dynamic Web pages using Servlets and JavaServer Pages.

## 1.5 Perl

Perl is a language that originated as a scripting tool meant to replace awkward grep, sed and awk UNIX tools written by Larry Wall. It went through several revisions ultimately reaching version 5, improving its functionality from a simple tool adding support for simple I/O operation, powerful regular expression matching, extensibility and style flexibility that resulted in coining its motto "There's more than one way to do it." [125]

Perl is a weakly typed language with dynamic data types (except for internal types) and most importantly it is an interpreted language that uses an interpreted usually called 'perl'. Its most notable features include procedural nature, memory management through reference counting, regular expressions, lists, associative arrays and simulated objects. It also easily integrates into Apache for web development, has good web frameworks such as Catalyst and performs fairly well in various programming styles with various degrees of difficulty and full support. It is ideal for processing text files making it a good web language as well as an extensive library of modules available to almost any need consolidated into one repository called Comprehensive Perl Archive Network (CPAN) making it easy to see if anyone has created an extension for a specific purpose. [126]

## 1.6 C++

As one of the most influential languages of all time, C++ started as an extension to C by Bjarne Stroustrup to integrate Simula structures into C to improve distributed computing in 1979 with first commercial implementation released in 1985. A subsequent release in 1989 added more advanced features dealing with extending classes, many function and variable types as well as error handling. In 1998 the language was finally standardized in ISO/IEC 14882:1998 where the structure was officially defined. [127]

The language is a statically typed language that requires a compiler to convert the source code to executable code specific to a CPU architecture and OS. The language is general-purpose with a mix of high and low level features giving the programmer choice of implementation, even if the programmer may choose incorrectly. Language ideology involves 0-overhead principle meaning not to include anything that is not used to avoid any performance penalties. Features of the language include full object-orientation support, templates, operator overloading and platform independence. [127]

## 1.7 AspectJ

AspectJ is an aspect programming extension of Java language created in order to improve modularity by separating the secondary functions from the main business logic [1]. AspectJ introduces new expressions (that does not exist in the Java syntax) in order to implement new modules called aspects. In more details, AspectJ can alter the behavior of a program by applying advice (additional behavior) at some points of the program, or it can alter the structure of an object by adding a property or a function dynamically to the object (intertype declaration) [2]. An advantage that AspectJ has is that all AspectJ programs are valid Java programs [1], so we can say that AspectJ takes advantage of all the capabilities of Java while offering more modularity. It is very important to mention that AspectJ does not supplant Java.

## 1.8 COBOL

COBOL is one of the oldest programming languages (created in 1959). It has dominated the software application development for decades and still one of the used programming languages nowadays in business software. COBOL started as a procedural programming language that aimed at making programming easier by using English words in its expressions. COBOL presents different advantages, like simplicity, portability and maintainability, also COBOL is considered to be very powerful in processing massive data for business applications. COBOL has undergone several improvements (revisions on 1968, 1974, 1985 and 2002) in order to catch up with the new programming trends. The latest incarnation of COBOL support object oriented features (no compliant compiler available yet) and also maintains backward compatibility. In this paper, the focus is on the different versions of COBOL according to the comparison criteria.

## 1.9 Ruby

Ruby is a dynamic, reflective, general purpose object-oriented programming language that combines syntax inspired by Perl with Smalltalk-like features. Ruby originated in Japan during the mid-1990s and was first developed and designed by Yukihiro "Matz" Matsumoto. Ruby is a pure object-oriented programming language with a super clean syntax that makes programming elegant and fun. Ruby successfully combines Smalltalk's conceptual elegance, Python's ease of use and learning, and Perl's pragmatism. Ruby has started to become popular worldwide in the past few years as more English language books and documentation have become available. Ruby supports multiple programming paradigms, including functional, object oriented, imperative, and reflective. It also has a dynamic type system and automatic memory management [33].

## 1.10 PHP

PHP: Hypertext Preprocessor is a widely used, general-purpose scripting language that was originally designed for web development to produce dynamic web pages. For this purpose, PHP code is embedded into the HTML source document and interpreted by a web server with a PHP processor module, which generates the web page document [34]. PHP is a general-purpose scripting language that is especially suited to server-side web development where PHP generally runs on a web server. Any PHP code in a requested file is executed by the PHP runtime, usually to create dynamic web page content. It can also be used for command-line scripting and client-side GUI applications. PHP can be deployed on most web servers, many operating systems and platforms, and can be used with many relational database management systems. It is available free of charge, under the PHP License [35]; and the PHP Group provides the complete source code for users to build, customize and extend for their own use. PHP supports the imperative, procedural, object oriented, and the reflective programming paradigms.

## 1.11 Bash Scripts

Anyone interested in Linux Administration should have a good understanding and a minimum of proficiency in Shell Scripting. The Shell is a command interpreter which does not belong to the Operating System (OS). This is why he is named as such, materializing its separation from the OS Kernel. Bash is a free software UNIX shell written for the GNU Project developed by Brian Fox in 1987. Also, Bash is an acronym which stands for Bourne-Again Shell [188]. Besides being an interpreter, the Shell is a genuine programming language which can be used to prototype complex applications in a quick-and-dirty manner [187]. It encompasses not only the notions of variables, arithmetic operators, control structures and functions available  but also specific operators such as : |

or ;.  Since the Shell and its derivatives are interpreter i.e. each line is analyzed, verified then and executed, few analysis rules are implemented to avoid hindering performances at execution. This implies a coding rigidity from the programmer(s). In addition, Bash scripts written with specific Bash command are not portable to other shell such as Bourne Shell unless Bash is installed. That said, based on the second quote of Herbert Meyer mentioned earlier, we can notice that Bash scripting is likely to not fit his definition of usefulness of a programming language given its natural features [187].

## 1.12  Scheme

Scheme is a programming language inspired by the functional language Lisp. It was created in MIT labs by Gerald Jay Sussman and Guy L. Steele and is mostly en vogue in educational circles.. It is considered by one of the two main dialects of Lisp but adopts a minimalist design philosophy specifying small standard core and powerful tools for language extension [215]. Therefore, the Scheme syntax is simple with very few keywords. Its prefix notation allows overcoming operators' precedence of operators. Moreover, thanks to its sophisticated macro system Scheme is highly flexible and can be adapted to many computational situations including adding multiple paradigms capabilities [29]. Its fundamental built-in features are: lexical scope, lambda calculus, blocks structure, proper tail recursion, first-class continuations, shared namespace for procedures and variables not to mention minimalism [215]. Its base types are booleans, numbers (complex, Integer, real), characters or symbols which are variables. In addition it supports many composited types such as vectors, hash tables, associative lists and many more. Hence, Herbert Meyer definition of useful programming language is likely to find a match through Scheme programming based on its natural and extensible features.

# 2. Analysis

This section pairs up languages for pair wise and individual analysis accoding to ten criteria. Each criteria is described in the context of the languages being compared and then appropriate observations are drawn.

## 2.1 Haskell vs. Java

We did comparative study between Haskell and Java in popularity, and we used [17, 18] web sites where are related and especially doing this comparison, the provided service for free and we used them as referenced in this paper. We got scientific and clear charts that describe the gap between these two languages.

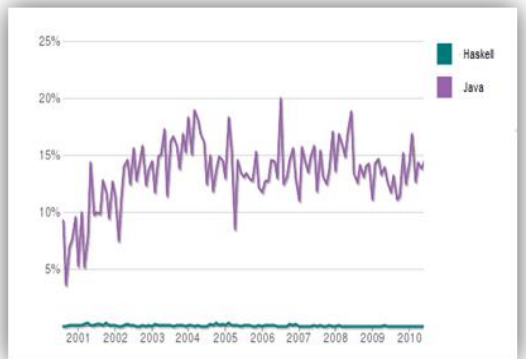

**Figure 1: Monthly Lines of Code Changed**

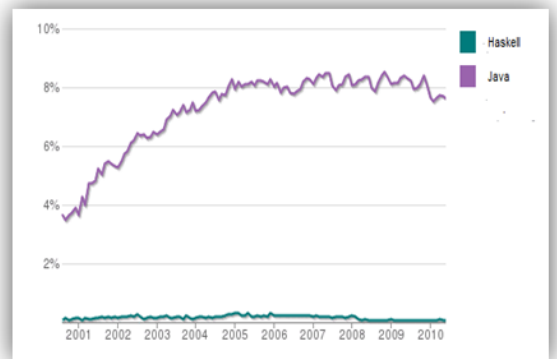

**Figure 2: Monthly Project**

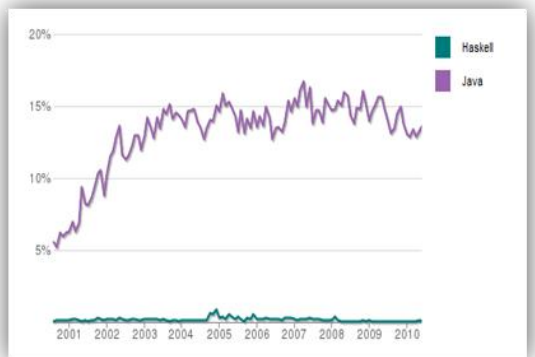

**Figure 3: Monthly Commits**

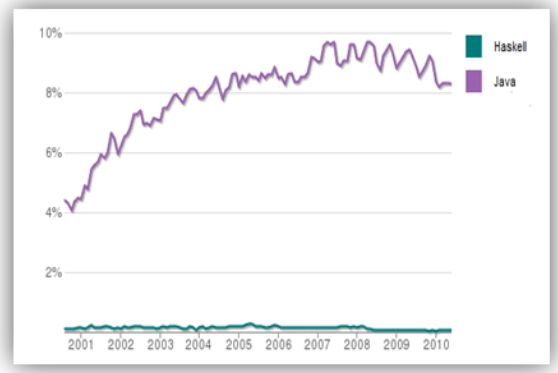

**Figure 4: Monthly Contributors**

The graphs are shown in figure 1, 2, 3, and 4 compare the languages that we picked (Java and Haskell). The height of each point on the graph is the sum of all commits, contributors, lines of code or projects in that month. These studies are taken from [19]. Languages are always charted over 20 years, and don't include recent month. The most recent month is excluded because Ohloh doesn't yet have complete information for it [19].

As you can see in figure 1, the lines show the sum of lines of code changed in a month, and Java has the lead over Haskell in this situation. For example, in year 2010 Java has 14% of lines changed comparing with Haskell that has low percentage in this manner. This is giving clear picture of the robustness that Java has, and this mean Java's developers are working to get the last updated version on this language. This study does not reflect weaknesses or vulnerabilities in each of Java or Haskell, the study based on the popularity usage only. For this reason, in the next sections you will see in details how Haskell compete Java in different aspects.

### 2.1.1  Default more secure programming practices

Security in programming languages generally has many aspects and many special fields. In Haskell language, there are some security and safety features. Basically as is found in any language, Haskell is able to provide some advanced group of proof security and safety features. In Haskell you never are able to run unsecure code why? Haskell's type system allows users or programmers to build security constraints in embedded domain-specific constructs that can be applied by the compiler. It is generally possible to use the power of type system to solve the problem without waiting for language support for something like tainted data when there is new kind of security issue has been arise [1].

In the most of functional languages and our example here is Haskell, have ability to provide strong module systems with flexible and excellent security. Based on some studies and researches the real functional languages have been more strongly typed than C#, C++ and Java or anything vein [3]. *do* in Haskell and similar languages has ability to provide support for modular security. The module is one of the fundamental constructs of the language and module-based security is better than object-based; instead of "public/private/protected", you can provide as many security modes as makes sense for your module.

High-Level languages give a great advantage to the programmers by freeing them from many explicit concerns about resource allocation and management. In Haskell memory allocation and de-allocation can be done in safe and implicit manner via garbage-collector heap. Concurrent threads supported by GHC Haskell dialect, which these threads implicitly scheduled and allocated by underlying run time system. Finally, lazy evaluation property in Haskell will take care of making sure that the data dependency order is respected so this gives programmers an opportunity to not worry about the order of the execution, even within single thread.

The great example of this property is Monad concept. Monad concept is a basic tool for *sequencing*: the Monad constructors allow the sequencing happen when the things in your program need to be done in sequence. Monad is not one function that does sequencing while there is basic concept of what a monad is. So, anything that fits the concept and you implement it can use all monad tools by language.  Listing some of Monad concept as following:

1. You can use the same constructs for writing sequence code of list processing
2. Writing numeric code that does update to multiple arrays,
3. Performing IO,
4. Combining non-deterministic computations and more of functionalities

The mentioned features have affects to support program safety, programmer productivity and arguments about programs. Unfortunately, they also have a disadvantages side: systems programming would become increasingly difficult. One of the possible requirements to systems programmers to achieve performance goals is to control resources explicitly or systems programmers need to access particular low level machine features. But the problem with Haskell doesn't provide graceful mechanisms to shift between implicit and explicit resource management. In fact, in the existence of lazy evaluation, programmers find it very difficult just to predict—let alone to control—space and time exhaustion [2].

Java as programming language has many of features that have an important security implication [4]. When Sun Microsystems released the first version of Java language, the language attracted the attention of most of programmers around the world. There are different reasons were attracted the developers to java for instance: cross-platform capabilities, ease of programming compared to OO languages like C# or C++, robustness and memory management, Java's security and the last tow is our target to analysis why? [5].

Programs written in Java must be reliable in a variety of ways. So Java puts a lot of conformations on early checking for possible problems, later dynamic checking, and eliminating situations that are possibly causing error. Java gives a programmer an opportunity to not worry about dealing with memory because they don't have to worry about it getting corrupted. So this is the reason of why there are no pointers in Java, also the programs can't accidentally overwrite the end of a memory buffer. Gain unauthorized access to memory is possible in some of OO languages such as C or C++ while it is not possible to happen in Java. The arithmetic operations are well defined and platform independent, as are the type conversions. The built-in **bytecode verifier** ensures that these checks are always in place.

Moreover, there are inclusive security mechanisms available in Java that has ability to control sockets, access to individual files, and other sensitive resources. Java Virtual Machine (JVM) must have security manager in place to give programmer advantage of the security mechanisms. This is an ordinary Java object of class **java.lang.SecurityManager** that can be put in place programmatically but is more usually specified via a command line parameter.

Some of securities aspects of java secure programming that we will consider are those which are built in java code itself. The first aspects of Java security that we will consider are those which are built into the Java language itself. In Java program there are access levels associated with each entity whether it is object reference or primitive data type. The levels that we are talking about are private, default, protected, and public. The meaning of each one is as the following [5]:

1. _Private_: it means that an entity can only be accessed by code from within the class in which it is defined.
2. _Default_: it means access is only possible from within the package of classes which includes the class in which it is defined.
3. _Protected_: it is similar to default but includes the possibility of access from sub-classes of the defining class.
4. _Public_: it means access is possible from within any class. These are the levels referred to in the first of the Java rules given below.

Java's language rules enforced by the compiler and some rules cannot be enforced at until runtime like "_Array bounds must be checked on all array accesses_" and "_Objects cannot arbitrarily be cast into other objects_". The rules that we mentioned listed below [5]:

- "_Access levels are strictly adhered to in Java;_"
- "_Programs cannot access arbitrary memory locations: Java does not use pointers_"
- "_Final variables are immutable. A subclass cannot override a final method;_"
- "_Variables may not be used before they are initialised;_"
- "_Array bounds must be checked on all array accesses;_"
- "_Objects cannot arbitrarily be cast into other objects._"

A discussion of Java's security model often lies behind the idea of a sandbox model. You need to restrict the program environment with certain bounds when you want a program to run on your computer. Sometimes you may allow the running program to access some specific resources, but in general you want to restrict it with its sandbox. Java sandbox has responsibility to protect a number of resources and it does so at a number of levels [5]. The default sandbox has three related components: *byte code verifier, class loader* and *security manager.*

A general example used by java, Generating Public and Private Keys for cryptography subject. The following example generates a key pair for various public/private key algorithms. Some of the more popular algorithms are: DSA, Diffie Hellman, and RSA. See appendix A *GeneratingPublicPrivateKeys.*

Finally, Java programs security and safety have disadvantages that can be compromised by some ways and java as programming language has areas of potential vulnerabilities such as type safety, public fields, inner classes, serialization, reflection, the JVM tool interface, debugging, and monitoring and management. *Java and Java Virtual Machine Security* [6] and *Securing Java* [4] explain and describe some Java vulnerabilities that have resulted from implementation bugs. However, java facilities and features could be used by an unwary user which might not realize could breach the safety.

### 2.1.2  Web applications development

Programming in typed functional languages for instance Haskell have a lot of advantages in web application development. After long of searching through related websites and investigating about main web applications that have been developed by Haskell, I found that Haskell is ready for serious web applications. There are many tutorials aims to get you started with developing web application in Haskell [7]. Those tutorials relatively describing light-weight approaches to Haskell's web applications which use CGI [7] libraries and an XHTML [7] combinator library.

HAppS [8] is one of the common projects that have been written in Haskell. This web server provides all the major server components you need to build high quality web applications. In this application you can use all the provided components individually or composing them together into single package and this package can handle all aspects of developed internet application including databases. The components provided by this web server are as following:

- HAppS-HTTP    - High performance dynamic web serving framework.
- HAppS-State    - Global in memory Haskell state with ACID guarantees.
- HAppS-IxSet    - Efficient relational queries on Haskell sets.
- HAppS-Data    - Transparent conversion to/from XML and form-data.
- HAppS-Plugins  - Sessions, FlashMsgs, HelpReqs, and maybe soon Mail, IRCBot, and UserAccounts. Common components of many standard web apps.
- HAppS-SMTP    - Send/receive/relay SMTP mail envelopes.
- HAppS-DNS    - Pure Haskell DNS resolver. No need to link to clibs and suffer builds complexity.

Happstack [9] is another common example and the new name for HAppS. It is a Haskell web framework. The developers in this web framework can prototype quickly, scale massively, change easily, and operate reliably. It has ability to support OS X, Windows, FreeBSD, and GNU/Linux environments.  This frame work add some advantages to the previous one such as: RAM cloud Database Architecture, Integrated HTTP Server, Flexible Routing and Request Processing DSL, Flexible Templating Options, and Type Safty.

Based on the latest information that I got from Haskell Communities and Activities Report [7], the web development tools that have been done by using Haskell as choice of functional language are as the following:

1. **Holumbus Search Engine Framework [10]:** This framework (Holumbus) has a collection of tools and modules which produce flexible, fast, and highly customizable search engine with Haskell. There are two main parts of this framework. The first part is taking responsibility of indexing to extract the information or data of given type of documents, e.g., web site's documents and stores it in an appropriate index. The second part is the search engine for querying the index.
2. **HCluster:** HCluster (temporary name), it is a small tool that developed to be using in remote clustering middleware. This tool initially aimed to clustering at verifying offline and online computations in distributed electoral process.

Also some mentioned tools in that report [7] as *JavaScript Monadic Writer* and *Haskell Dom Bindings.*

Haskell is still in evolvement, enhancement, improvement and development way, so there are some of web servers and applications which purely built from scratch using Haskell as programming language or as built-in tool supporting current web application as I mentioned before. I used Haskell.org as main reference to get the current web application and services that have been developed by Haskell. http://hpaste.org, http://www.parallelnetz.de are categorized as Haskell web applications. As web servers, there are examples like Haskell Web Server (HWS) written by Simon Marlow, Modular Haskell Web Server (MHWS) and this web server with module system and support for CGI and it is based on HWS, Wash Server Pages (WSP) is an extended version of HWS that runs WASH modules as servlets.

Finally, the latest and the newest release for web development related to Haskell is SNAP framework. What is SNAP [11]? Simply it is web development frame work for environments of UNIX systems, and it is written by Haskell as programming language. As I read from snapframework.com SNAP has test suite with a high level of code coverage and it is well-documented, but still-evolving interfaces. The first version of SNAP currently divided into three components: *snap-core, snap-server* and *heist (XHTML libraries).*

As I mentioned in the introduction about popularity between Haskell and Java, Haskell still weak in view of support and documentations. For that reason Haskell needs some time of technical supports in this field of programming. And I think it will be in the correct way to be one of the important languages in developing and design web applications.

The fundamental requirements for web applications to be effectives, these web technologies have to be useable, secure, accessible, portable, interactive, and flexible. Going back to the dark ages, the Common Gateway Interface (CGI) technology was published and defined to give web servers opportunity to process and serve dynamic content. At that time Perl so far was the common language using to develop CGI. CGI can be developed in any script or any programming language. But the big problems with CGI were scalability and performance, and security is another big concern. Some solutions came out at that time from web server vendors such as defining APIs to solve these problems, like ISAPI from Microsoft and INSAPI from Netscape. The problems with these solutions are, the application written to these APIs is restricted to one particular server vendor. As I mentioned, portability is one of the fundamental requirements, if you need to transfer the application from one server to another you have to start from scratch. Another problem is reliability, these API support typically C/C++ execution and when the application crashed it brings the server down with it. So, after this brief introduction we have enough pictures about web app developments and its history. Servlets come to be the rescue! For these weaknesses, Servlet API leverages the advantages of Java platform to correct and solve the issues of APIs and CGI.

Java web application development is one of the fundamental technologies to hit the internet. Java popularity based on achievements of getting very high scores of that mentioned features. Java has strong presence between client side and server side. Java web applications can be run anywhere and all you need is a Java Virtual

Machine (JVM) running on client's side. And fortunately, this feature is coming automatically with most browsers, i.e. you don't need to change existing program or installing new software on your computer to adapting with Java.  No matter of whether you are using Unix systems or Microsoft Windows Server, you don't need to worry about portability. Java has ability to run the same web application on different platforms and machines, and this is very important feature for web application because you cannot as developer to predict what systems your clients are using and which technologies could run on their systems. From different aspect of the importance of using java is object oriented language. Developers easily have ability to build modules that can be used many times under any circumstance, and this part will be show in details in the next section. Java web application development is popular with developers because its portability, ease of use and robustness. Java became the programming language of choice for internet solution because of its additional security features [12].

The problem with Servlet is bit complicated and you have to write, compile and deploy life cycle, from this reason Java Server Pages (JSP) considered to be the third generation solution. With JSP you can create dynamic content and make it faster and easier to build web based application. And this web application has ability to work with different technologies such as web browsers, application servers, web servers and other development tools. JSP technology allows predefined requests and java code to be inserted into static web pages content. For each request from these pages the code is compiled at the runtime.  At the server side, java code is controlled by virtual machine (VM). This VM is pre-installed software and integrates with host OS.  In this situation and the common VM used to run JSP is Java Virtual Machine (JVM) as I mentioned before.

As an example, JSP pages are basically web pages with transitional HTML and java code. When the server receive request from client, at the first time server will know the received request is JSP page by checking page extension ".jsp" rather than ".html" and this tells the server that request needs or requires special handling as in figure . Here is a simple example date.jsp [13]:

```
<HTML>
<HEAD> <TITLE>JSP Example</TITLE> </HEAD>
<BODY BGCOLOR="ffffcc">
<CENTER>
<H2>Date and Time</H2>
<%
java.util.Date today = new java.util.Date();
out.println("Today's date is: "+today);
%>
</CENTER>
</BODY>
</HTML>
```

When date.jsp submitted to the server, it will be compiled by JSP engine into servlet. Serlvlet as we know has its one engine, so the produced servlet from JSP engine will be handled by servlet engine. After that servlet engine execute the servlet class that have been loaded from sevlet engine to create dynamic HTML to be sent back to the client's browser, as shown in figure 2.

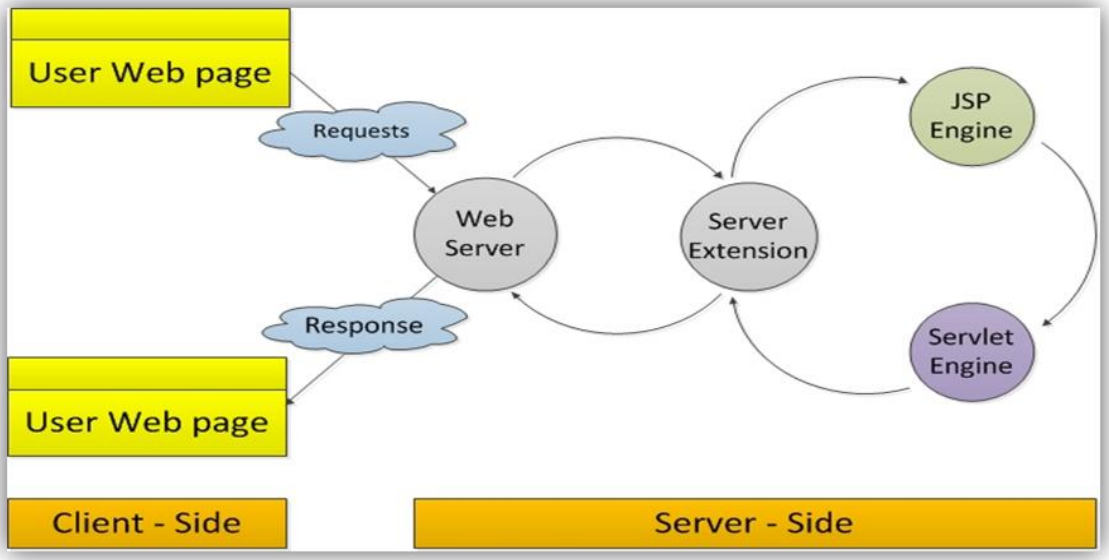

**Figure 5: Request/Response Flow when Calling a JSP**

The drawbacks of JSP will be same drawbacks of Java. The main problem will be disk space, because JSP pages will be translating into class files and the server has to store resultant class files with JSP. Initial compilation produces delay at the first time when accessing the JSP page. [13]

### 2.1.3 Web services design and composition

During web process composition the data flow has to be established between different participating services of the process. One of the issues in data validation is type checking and it is considered during security evaluation of web servers/application servers. The usefulness of type checking is helping the developers to build proper input validation in web application servers. Some of designed techniques of application servers are using validation as a means to examine data type and range of valid values for inputs. Simple Object Access Protocol (SOAP) router has ability to perform data validation functionality for web services. It is highly recommended for the security issues in web services scenario. The SOAP router should provide type checking for requests submitted by client's web services. The logically manner is the unencrypted field of sent data need to be validated otherwise the application servers will reject invalid data in web services input massages. So if Haskell as functional programming language is used by client and server sides, then the process of type checking may be made easy. So we can achieve type checking by having intermediate layer between the service and the client, this layer can be constructed by Haskell as functional language to convert the data type provided and supported by the client to common functional language representation and performs type checking.

The problem with constructing web services is, there is no unified infrastructure support to provide static checking. In SOAP reporting type error is not returned until XML message is validated on the server at runtime [14]. Web Service Description Language (WSDL) has been suggested to be used to provide a mapping to programming language to offer static checking. Then the best choice to implement the systems that uses WSDL types is Haskell based on the previous mentioned features. In this idea, the client provided the data types to convert to a common reference language and perform type checking.

In web service process composition Haskell usage in valid XML generation has been discussed in [15, 16]. HaXML is a domain specific language for parsing, filtering, transforming, and generating XML documents. HaXML utilities treat a Document Type Definition (DTD) as a series of type declarations in Haskell, thus conflating validation with type checking. With the use of HaXML the type checking layer can be added to the process generators. In addition to the process generation stage, the features of functional languages can be used during process execution also. After comprehensive search about complete written program using this HaXML tool, I come up with the following code developed by Michael Neumann, this code is used to transforms a XML documents:

```
module Main where
import XmlLib
import XmlCombinators

main = processXMLwith
    ( mkElem "HTML"
      [ mkElem "BODY"
         [ message `o` (tag "messages" /> tag "message") ]
      ]
    )
...... - See the Appendix B form complete program
```

No doubt with the popularity between Java and Haskell in web services design domain, Java has the lead over Haskell in this field. Java Web Services at a Glance [21] is big example on the robustness and usability of java through web applications and services. In general, web service is web-based application that use transport protocol to exchange data with the clients. Developing web services that are using Java technology APIs and the facilities of using tools that are provided based on an integrated web services stack called Metro [21]. Metro stack project consist of Web Services Interoperability Technologies (WIST), Java Architecture for XML Binding (JAXB) and Java API for XML-Based Web Services (JAX-WS), which enable you to freely create and deploy reliable, interoperable, transactional, and secure Web services and clients. The Metro stack is part of Project Metro and as part of GlassFish, Java Platform, Enterprise Edition (Java EE), and partially in Java PlatForm, Standard Edition (Java SE). GlassFish and Java EE also support the legacy JAX-RPC APIs [21]. Web services are divided into groups based on the technologies that are used, as following:

1. Core Web Services: in this group the technologies used to develop web services are fundamental technologies, for example: the solutions for processing XML, development SOAP based and RESTful web services, and data binding. The services available in this group are JAX-WS and JAXB.
2. Enhanced Web Services: this group included the technologies that are used to develop enhanced web services by using features such as atomic transaction, and secure and reliable messaging. The service included in this part is WIST.
3. Secure Web Services: this section of groups has the technologies that have been used to secure web services. The services used to develop this group are WIST, and XML and Web Services Security (XWS-Security).
4. Legacy Web Services: Especially this group has been replaced by group (Core Web Services), but this group still supports older web services. The services used in this group are Java API for XML-Based RPC (JAX-RPC).

I can conclude in this section with; java has huge of valuable, informative, well-documented, and an experts' supports references in this domain of my comparative study. That's why Java as programming language has ability to be adopted with web services design and composition.

### 2.1.4  OO-based abstraction

Developing systems that contain things always depends on Object-Oriented design. The "things" are objects, and these objects have defined behaviour and state. The behaviour means, the objects respond in particular manner to the received message, and the state means, the objects contain data attributes which at any given time contain a specific set of values.  The exact behaviour of an object depends on its state at the time when it receives a message. The object's internal data (known as *instance variables* or *fields*) can typically only be manipulated through the functions which the object exposes (called *methods*).

Functional languages are not object-oriented languages because functional programming is considered to be an alternative to object-oriented. As I mentioned before, the real functional languages have been very strongly typed. The main things that object-orientation gives you as a developer are modules, state encapsulation, and polymorphism. Haskell as functional language provide strong module systems with excellent and flexible security. Haskell is designed to avoid the use of state, encapsulates it via monads when necessary; and Haskell provides a mechanism called type classes which provides a great mechanism for polymorphism [3].

Type classes in Haskell have a set of functions that can be implemented into different implementations depending the type of given data. Type classes may look like the objects-oriented programming such as C++ and Java, but they are truly different in their syntax and how to use them [20]. While Haskell has characteristics of using objects, that means it also supports a notion of class extension (inheritance and multiple inheritance). For example, you may define class *Cars* which inherits all of the operations in class *Mobile*. Also Haskell has polymorphism property in different forms such as ad-hoc polymorphism or overloading, parametric polymorphism, and inclusion polymorphism. For example:

```
folder :: forall ab. (a->b->b) -> b -> [a] -> b
```

his example means folder is a parametric polymorphism function; when actually used, it may take on any a verity of types (Strings, char, double …). Haskell has some significant differences that distinguish it from other OO-languages as follow [7]:

- Haskell separates the definition of a type from the definition of the methods associated with that type.
- The class methods defined by a Haskell class correspond to virtual functions in some OO-languages such as C++ class.
- Haskell classes are nearly similar to a Java interface.
- Haskell does not support overloading style, like some OO-languages such as C++, in which functions with different types share a common name.
- The type of a Haskell object cannot be implicitly coerced; there is no universal base class such as Object which values can be projected into or out of.
- Java attaches identifying information (such as a VTable) to the runtime representation of an object. In Haskell, such information is attached logically instead of physically to values, through the type system.
- There is no access control (such as public or private class constituents) built into the Haskell class system. Instead, the module system must be used to hide or reveal components of a class.

Java is an Object-Oriented programming language. Object-oriented programming has been discovered to let the programmer to view programs differently from the older procedural programming paradigms and it has been promoted as a more productive and natural way to view solutions to problems. Objects in Java offer several benefits such as modularity, hiding information, reusability, and debugging made simple.

From the introduction in this section, while an *object* represents one specific thing, the *class* represents a category of things. The big three concepts with OO are encapsulation, polymorphism, and inheritance. Additionally, information/implementation hiding, state retention, object identity, message passing, and classes. With java you can use inheritance, which is a fundamental thing in OOP, but Java doesn't have multiple inheritances to avoid ambiguity problem that happened in C++. The *interface* is another property, which is physical representation for objects, which objects are abstract representations of real things. These characteristics gives java an opportunity to be used as frontend for the backend databases applications like Oracle database management system. Additionally, makes java platform independent which means it can run under most operating systems and platforms. General example on OOP using Java as follow:

```
class ObjectOrietedExmaple  {
  public ObjectOrietedExmaple (){
    String text = new String("I'm a simple Program ");
    String text2 = text.concat("that uses classes and objects");
  }

  public static void main(String[] args){
    ObjectOrietedExmaple testObject = new ObjectOrietedExmaple();
    System.out.println("Welcom to Object Oriented Programming");
    System.out.println(testObject.text2);
  }
}
```

### 2.1.5  Reflection

Reflection is a functional extension to OOP paradigm. Smalltalk and scripting languages are representing high-level virtual machine programming languages that are commonly using reflection. Unfortunately, the statically typed or manifestly typed programming languages are less commonly used this property. Many languages of OOP applying this style attempt to do self-examination, self-modification, and self-replication. Haskell is under type of declarative programming paradigm, which does not have property of reflection as primitive goal of its design. Generally, Haskell doesn't support reflection in the same way as many other languages do, but there are idiomatic ways to accomplish the same things when the developer would like to use reflection for [7]. Some published articles belong to how to binding Haskell to object oriented component system via reflection. Another approach, is how to interfacing Haskell with object-oriented languages. These papers possibly implemented to improve Haskell reflection implementation. Unfortunately, cause the lack of time I couldn't go deeply in this property to get enough information.

One the famous features used in Java programming language is "Reflection". This property let java programs are able to inspect or examine upon themselves, and manipulates internal properties of the program. Based on this feature, java reflection quite powerful and can be useful for instance, with Java reflection the class can obtain the name of all its members and display them. With Java reflection, it is also possible to instantiate objects, get/set filed values and invokes methods. Another example of tangible use of Java reflection is in JavaBeans where the software components can be manipulated visually via a builder tool. To take a look on how the reflection works in Java, consider the following simple example [21]:

```
import java.lang.reflect.*;
class Reflect {
    public static Constructor [] constructors;
    public static Method [] methods;
    public static Field [] fields;
    public static Class thisClass;
    static{
```

```
        try{
            // dissect and see this class itself
            thisClass = Class.forName("Reflect");
        }
        catch(ClassNotFoundException cnfe){
            System.out.println("class doesn't exist"+cnfe);
            System.exit(1);
        }
        constructors = thisClass.getConstructors();
        methods = thisClass.getMethods();
        fields = thisClass.getFields();
    }
    public static void main(String [] args){

        ...... // See the Appendix A for complete Code
    }
}
```

The output will be as following:

```
This class is class Reflect
Method 0 public static void Reflect.main(java.lang.String[])
Method 1 public native int java.lang.Object.hasCode ()
Method 2 public final native java.lang.Class
.
.
.
Method 9 public final native viod java.lang.Object.notifyAll()
Field 0 public static java.lang.reflect.Constructor []
                Reflect.constructors
Field 1 public static java.lang.reflect.Method []
                Reflect.methods
Field 2 public static java.lang.reflect.Field []
                Reflect.fields
Field 3 public static java.lang.Class.Reflect.thisClass
```

The advantages of using Java reflection are signature-based polymorphism where Java reflection familiar with polymorphism based on interface inheritance, inspecting and manipulating classes where the direct code is depending on the class details such as interface, methods or fields and these details known at compile time. The next advantage is creating adaptable and flexible solution because it is dynamic nature. Finally, writing tool that require implementation details.

Exposes implementation details may considered to be one of disadvantages of using Java reflection. Another problem of using Java reflection is performance, the code has reflective is slower than the same functionality code that without reflective where reflection can be order of times 10 to 30 times slower than the direct code. The code complexity is considered one of the problems of using reflection. The reflective code is harder and more complex to understand than the native code. The last one of problems is security. There are some security concerns about using reflection especially in internet programming.

### 2.1.6  Aspect-Orientation

Aspect-Oriented Programming (AOP) is a paradigm invented at Xerox PARC in the 1990s [22] and addresses many problems that traditional OOP doesn't solve completely or directly. This paradigm gives developers ability to better separate tasks that shouldn't be inextricably tangled such as mathematical operations and exception handling. AOP improves performance and less time rewriting the same code for the programmers. In general AOP gives better code encapsulation for distinct procedures and reuse, AOP aims at modularising cross-cutting concerns that show up in software.

Haskell as pure functional language doesn't support AOP from its design phase principled. Does AOP has a meaning for functional languages specially Haskell? Based on a briefly survey in this topic, I found functional programming community seems to be resistant to the AOP as part of the language. Some studies have been done in this field like Aspectual Caml [23]; this paper makes a good attempt by concerning its design around a classic problem, namely the expression problem. This developed evaluation provides open extension of new expressions and new operations for the language at the same time. The tow-dimensional extensibility is hard to be achieved either in object-oriented languages or functional languages. AspectML [24] is another proposed work by Washburn and Weirich. This paper shows functional type directed programming and proved that the extensibility of aspect-oriented programming is important for extensible generic programming [25]. In another work, AspectFun [31, 32] is one of the proposals in area of functional aspect-oriented programming. This paper shows the advice invocations and static resolution of types are emphasised; and this work demonstrated the lack of interest in function AOP and showed one of the reasons where the perceived lack of apps domains is one of them. And basically the most of developments of AOP are based on OO environments. One the reason for this lack of interest is the great scepticism that many researchers have towards AOP.

Java as object oriented programming has become the main-stream over the last few years and having almost completely replaced the procedural approach. The software system can be seen as a built of collection of discrete classes and this is one of the biggest advantages of OO. Each class has well defined task and its responsibility is clearly defined. At the end, all those classes collaborate to achieve the goal of the development purpose of the application. But, some parts of the system can't be viewed as only one class responsible for one task, they cross-cut the complete system and have effective on other classes. Examples might be securing in exception handling, distributed application or logging method calls. Of course, coding this concept can handle these parts by adding the code to each class separately and this will violate the principle that each class has well defined responsibility. So this is where AOP comes into play. In Java, AOP has new program construct, called *aspect*. This *aspect* used to capture cross-cutting aspects of application in separate program entities. The application classes keep their well-defined responsibilities and each aspect captures cross-cutting behaviour.

In java there are open source aspect-oriented frameworks such as AspectWokz, JAC(Java Aspect Components), dynop, DynmicAspects, Nanning, JBossAOP, CASER, EAOP, Colt, CALI, PROSE, and Azuki Framework.

### 2.1.7  Functional programming

Functional programming is sub-kind of declarative programming paradigm and different from imperative (or procedural) programming and neither is it like OO programming. Functional programming paradigm is something different but not radically because the concepts that has been explored is familiar programming concept where expressed in different way. Sometimes functional programming called *expression oriented programming* since FP is all about *expressions* and everything reduced into expression. FP's expressions are a collection of operations and variables that results in a single value. For examples, *x==7* is a Boolean expression, *5+x* is an arithmetic expression, "*Hello World*" is string expression and so on. Therefore functions are very important in FP and for that reason you may guess that from the name!

FP's functions are used as objects; Functions in FP are often passed within a program in the same way as other variables. General example, about how to pass functions within program as parameters is that GUI programs where the programmer assigns the name of the function to the *command* attribute of button control. Then the programmer treated the event handler function as object and assigned a reference to the button. This is the key idea to FP. Also, FP tends to be heavily list oriented. Finally, the problem solving goal of FP is focusing on *what* rather than *how*. That means, FP describes the problem from point of to be solved rather than focusing on how the mechanism way of solution. FP does not have side effect such as Haskell. Haskell gives the programmer ability to take high level of what is to be computed rather than how. Another interesting feature of Haskell is its lack of any *loop* construct. There is no **for** and no **while**. There is no **GOTO** or **branch** or **jmp** or **break**.

Haskell is the meant language for the previous characteristics. Haskell is categorized from family of pure functional language paradigm. The pure functional language is structured by defining an expression which captures the goal of the developed program. Each term of the expression is in turn a statement of a characteristic of the problem and probably encapsulated as another expression and the evaluation of each of these terms eventually yields a solution.

Java as object-oriented language can be classified as an imperative programming language, and this is one aspect of the Java language that is often overlooked [28]. The modularity is the key to productivity and successful programming on any platform. Java developers are facing problem of modularity, when the modular programming needs more than just dividing a problem into parts; it involves being able to combine small parts solutions together into effective complete part. For that reason, this type of paradigm (FP) concept is developed to be seems natural to use FP techniques when developing modular code in Java platform.

Java as an open source, popular language and mainly OO, the developers tried to embed some libraries developed in Java and using FP patterns. So converting applications into FP software is not necessarily easy or possible. These libraries handles both OO programming features and using FP patterns where appropriates. One example of the developed libraries in this manner (FP) is **FunctionalJ** [29] where a general functional programming API for Java developed by Frederic Daoud. This library makes it easy to use functional programming constructs in java code and this library provides many features such as easily representing functions as objects, uses parameter binding, replace procedural code with functional code, and no need to deal with exception if you don't need. The reason of trying join FP parading into OO paradigm is the modularity. Some FP concepts have been developed to be used in Java platform such as closures and higher order functions.

## 2.1.8 Declarative programming

As we described in section 2.1.7, Haskell is declarative programming based on the brief details in that section.

In Java, Enterprise JavaBean (EJB) has been designed by declarative model, and this makes EJB components special which you can specify services such as transactions, persistence, security etc ..., to that the container should provide. The process of making Java as part of declarative programming paradigm is EJB only implements the business logic and at the runtime, the services have been provided by deployment descriptor which the container uses the metadata specified in it. This deployment descriptor is an XML file and not a part of Java code that makes up the EJB. From this point the ideas of looking for standard way to annotate the Java code which makes up EJB are coming after answering the following questions [30]:

1. *"Is there a standard way to annotate the Java classes that make up the EJBs so that a developer can look at the class definition, together with annotations, and know everything about that class?"*

2. *"It would be even better if the remote, home interfaces and the deployment descriptor could be automatically generated by a tool using the annotations."*
3. *"Better yet, can we provide the same kind of declarative services for a simple Java object? If so, how? This article examines how [30]."*

Java Tiger release (JDK 1.5) adds a new language construct called *annotation* (proposed by JSR-175 [31]). Annotation is a generic mechanism for associating metadata and combining the information (declarative information) with program elements such as classes, methods, fields, parameters, local variables, and packages. The compiler can store the metadata in the class files. Later, the VM or other programs can look for the metadata to determine how to interact with the program elements or change their behaviour.

The following example [30] illustrates the implementation procedure for the developed library that used for this manner.

Declaring an Annotation:

```
package njunit.annotation;

import java.lang.annotation.*;

@Retention(RetentionPolicy.RUNTIME)
@Target({ElementType.METHOD})
public @interface UnitTest {
        String value();
}
```

Using an Annotation:

```
import njunit.annotation.*;

public class Example {

    @UnitTest(value="Test 1. This test will pass.")
    public void pass() {
        assert 10 > 5;
    }

    @UnitTest("Test 2. This test will fail.")
    public void fail() {
        assert 10 < 5;
    }}
```

Accessing Annotations at Runtime:

```
package njunit;

import java.lang.reflect.*;
import njunit.annotation.*;
```

```
public class TestRunner {
    static void executeUnitTests(String className) {
        try {
            Object testObject =
                Class.forName(className).newInstance();
            Method [] methods =
                testObject.getClass().getDeclaredMethods();
            for(Method amethod : methods) {
            UnitTest utAnnotation =
                amethod.getAnnotation(UnitTest.class);
            if(utAnnotation!=null) {
                System.out.print(utAnnotation.value() +
                                    " : " );
                String result =
                    invoke(amethod, testObject);
                System.out.println(result);
                }
            }
        }catch(Exception x) {
            x.printStackTrace();
        }
    }
        ...... // See the Appendix A for complete Code
```

### 2.1.9 Batch scripting

Comparing between Haskell and Java using "batch scripting" criteria is an easy manner. Since, the meaning of this criterion lies under this question: Are Haskell and Java support batch scripting? There are some high-level languages classified to be scripting languages. The scripting language is usually interpreted rather than compiled. The batch scripting concept is interpreted one command at a time.  Some advantages of writing scripts are often to facilitate enhanced features of websites. Other characteristics of using batch scripting such as running external command and automations.

In Haskell, based on what I found, is supporting batch scripting techniques by some ways. Library called System (System :: String -> IO ExitCode) used in Haskell to perform and executing external commands such as external Linux command 'ls'. For example, System ("ls") is a way of how using batch scripting in Haskell. Computation system cmd [7] returns the exit code produced when the operating system processes the command cmd.  On Windows, system is implemented using Windows's native system call, which ignores the SHELL environment variable, and always passes the command to the Windows command interpreter (CMD.EXE or COMMAND.COM); hence UNIX shell tricks will not work [7].

Finally, I can come up with that Haskell support this property of comparison in easy, fashion, and effective manner.

Java provides some libraries that satisfied the mentioned characteristics of batch scripting such as executing external commands and automation. In java, implementing external commands is relatively simple. This property involves the use of two java classes Runtime class and Process class. In some details of how using these classes as following:

- To run the external command as separate process, you use Java exec method of Runtime class.

- Invoking Runtime exec method returns a Process object for managing the sub-processing.
- Finally, you use the getInputStream() and getErrorStream() of the Process object methods just to read the normal output of the command, and the errors output of the command.

I can conclude in this section with, most of high-level programming languages supports batch scripting concept. This concept allows the programming languages to perform some systems commands to execute some certain functions.

## 2.1.10  UI prototype design

Haskell has many benefits tools and it gives a good combination of *robustness* and *simplicity* for the graphical user interface development model. Haskell has many of toolkits for programming graphical user interface (UGI) such as *wxHaskell*, *Gtk2Hs [32]*, *hoc*, and *qtHaskell*.

**Simplicity**: GUI on Haskell is very easy to program and can getting a window with very useful widget inside it with few lines of code.

**Robustness:** as known the main advantages of Haskell over other traditional programming languages are summarized in (type inference, laziness, purely functional, etc). These characteristics give great combination to keep developed program bug-free and easy to fix/change.  So, it is amazing to see how the developer can design and develop GUI in Haskell without mutable variables or global variables. The other definitive feature that makes possibility of robustness is the type system and static typing.

The following example illustrates how to use "Hello world" GUI with button event:

```
module Main where
import Graphics.UI.WX
main :: IO ()
main = start gui
gui :: IO ()
gui = do
  f <- frame [text := "Hello World!"]
  st <- staticText f [text := "Hello StaticText!"]
  b <- button f [text := "Hello Button!"]
  set f [layout :=
          row 0 [widget st, widget b]
        ]
  return ()
```

No doubt with Java GUI started in development area before Haskell based on the history between them. The most famous and known GUI library in java is Swing. Swing is the advanced graphical user interface for the java SE platform. In Java it is easy to build GUI but the problem is hard to get something easy to build and maintain, and look fancy.  In Java GUI there are two separation concerns "BAD" and "GOOD". First, BAD is resulted by mixing logic and interface, and this style for small student program. The button's action listener is the responsible to perform the button's code, and this doesn't scale well as program get bigger. Second, GOOD is separating GUI from logic. When the program grows larger it is suppose to separate GUI processing from logic. For that reason, Java developers invited some GUI programming alternatives such as Matisse in Net-beans, JFormDesinger, BuoyBuilder, Visual Editor for Eclipse, SwiXML( XML representation of GUI), and JAXX. One example on Java GUI as following:

```java
import java.awt.*;
import java.awt.event.*;
import javax.swing.*;
public class Frame2 extends JFrame
{
  JPanel pane = new JPanel();
  JButton pressme = new JButton("Press Me");
  Frame2()          // the frame constructor
  {
     super("JPrompt Demo"); setBounds(100,100,300,200);
     setDefaultCloseOperation(JFrame.EXIT_ON_CLOSE);
     Container con = this.getContentPane(); // inherit main frame
     con.add(pane);     // JPanel containers default to FlowLayout
     pressme.setMnemonic('P'); // associate hotkey to button
     pane.add(pressme); pressme.requestFocus();
     setVisible(true); // make frame visible
  }
  public static void main(String args[]) {new Frame2();}
}
```

## 2.2  C++ vs. Perl

### 2.2.1  Default secure programming practices

Securty is a big issue in software applications today with the emergence and rapid spread of the Internet and availability and connectivity of applications over the network computers are given increasingly more exposure to the outside world.  Security flaws in applications today are the most costly and critical bugs because they may not only compromise the application but give an unauthorized person access to the machine the application is running on.  A language can contribute to the risk of having vulnerabilities by several specific factors from type systems, internal mechanisms and semantics to specific library function implementations. There are several areas in which applications can fall short of security expectations in which the programming languages will be assessed in, being: memory safety, input validation, race conditions, privilege confusion and user interface failures. [103]

#### *2.2.1.1  C++*

**Memory Safety.** The area of memory safety in C++ is the lowest point in C++ security evaluation.  Most C++ security flaws originate from buffer overflow attacks and unchecked memory manipulation due to the fact that C++ allows memory manipulation through the use of memory pointers.  While pointers are powerful tools misused or improperly used pointers can lead to some devastating effects adding extra complexity when writing applications using them.  There are two types of buffer overflows and a double free vulnerability that leads to a buffer overflow when correctly setup. One thing that is being addressed in C++ is that standard functions like 'memcpy' or 'strcpy' or 'strcat' did not terminate upon a given count but rather on a NULL character which allowed for more flexibility when designing buffer overflow attacks. Microsoft recommends usage of the safe versions of these and other functions included in their newer distribution of Visual Studio. [125]

The first of the two are stack-based buffer overflows that occur when an attacker is able to write beyond the end of a buffer allocated on the stack.  This way he can write malicious code in memory on the stack and overwrite values that are further on the stack such as return addresses which make the instruction register jump to malicious code when affected function is exited.

The second type of buffer overflow that happens in C++ is the heap-based buffer overflow.  There are two ways of exploiting it depending on the operating system memory protection strategy, one being overwriting the stack return address by overflowing the heap and the second overwriting a virtual table function pointer to point to malicious code. The first method depends if the operating system keeps track of the stack and heap sizes separately and maintains the boundary.  If it does not then by writing downwards from the heap the user is able to overwrite a part of the stack containing the return address of the current function and instruct the application to execute injected code. In the second approach if code execution on the heap is allowed or if the code is stored somewhere else that is executable it will be possible to launch an exploit. The idea is to reach an object that is dynamically allocated and has virtual functions, which contains the virtual function pointer table, and overwrite the addresses of those functions with the address of the malicious code.  This way when the object methods are reached they instead will execute the attacker payload.

The last type of memory exploit in C++ has to do with a double free bug and is a little more difficult to craft. Inside a list of free chunks of memory if a pointer is freed twice it will overwrite the forward and backward references in a linked list chunk of free memory. By doubly unlinking it the chunk ends up pointing to itself. Then with other exploits the forward reference is overwritten to 12 bytes before the return address and the backward reference with the malicious code address. Once the program tries to allocate a chunk of the same size it will believe it has and by trying to unlink the chunk it will go to the return address and execute the malicious payload.

To prevent these types of exploits there are several methods to go by. Firstly static checks such as Control-C are ran to ensure array indices are within bounds and flag the possible overflows to the programmers to double check their code is safe. Dynamic checks then are implemented by the compilers that control the size of dynamically allocated buffers and signal a runtime error if any of the boundaries is violated. Further sandboxing and memory marking is used to restrict access to memory segments by the application and enable non-executable heap and stack to prevent the attacker storing malicious code there for execution. Memory and instruction set randomization and encryption can further complicate the plans of an attacker having do descramble and decrypt the memory before being able to launch an exploit. When dealing with dangling pointers garbage collection will also help get rid of the possibility of double free bugs. [126]

**Input Validation.** In the area of user input the C++ language inherits a flaw from the C Runtime Library that is present in the sprintf string formatting function. The flaw allows the attacker to write to an arbitrary memory address through a well crafted unchecked input and therefore modify the return address to point to stored attack code.

This vulnerability is hard to check due to the flexibility of input but utilities such as FormatGuard try to assess if a string is safe to process and count the number of arguments required in the string comparing to what was submitted in the call. Other methods of prevention include safety checks by static analysers. [126]

Other flaws in input validation depend on the technologies used in the application, so for example if a relation database using ODBC is used then SQL queries have to be properly prepared to be safe from SQL injections. There are no other inherit problems to C++ except problems with libraries used for specific causes such as HTTP communication, cross-site scripting or directory traversal.

**Race Conditions.** Like most languages that deal with file operations C++ is susceptible to the time-of-check time-of-use bug where after a variable is checked the condition it was checked for could be invalid by the time it is used. One well-known example of that is the symlink race where a symbolic link is created instead of a target file and while the target file is safe to operate on if the check passes and synlink switches the meaning of the file handle the operation on the file will operate on the protected file instead.

A method to prevent race conditions at least in part is to use critical sections which outline code as atomic meaning no other code can execute between the statements making switches impossible. With today's multiprocessing computers this may be harder to guard against but with proper synchronization it's achievable. [127]

**Privilege Confusion and Escalation.** Privilege faults are not inherit to C++ and they are rather caused by the coder not using properly file manipulation checks and not atomizing file operations making them susceptible to race conditions. These could also be a result of the memory manipulation if system memory is not protected.

**Interface Failures.** User interface in C++ comes in forms of many libraries and each one has their own characteristics. Like before C++ does not inheritably come with any security flaws from the user interface domain although its libraries such as MFC or GTK go though iterations that improve and harden their components.

**Language-specific issues.** The underlying issues that cause C++ vulnerabilities have much to do with the type system of C++ and lack of array bounds checking. There are several errors that can happen when manipulating numbers in C++ that are not signalled such as overflow, when during addition a variable reaches its maximum value and instead of signalling an error the value is then added to the sign bit effectively subtracting the variable range from the result. The opposite is also possible and known as underflow. Another error is the fact that the same memory can be used as a signed or unsigned version of a variable, so by using a signed variable as unsigned half of the range is added effectively. The opposite conversion also happens when a program accidentally uses an unsigned variable as a signed variable. Yet another problem which plagues C++ is truncation which happens when a type with larger size gets assigned to a type with a smaller size and part of the value gets truncated to fit in the new memory size. These manipulation errors that are not reported or checked combined with the fact that C++ arrays of any sort are simply pointers to an arbitrary length of memory to which access is unchecked by the compiler or the operating system allowing for modification of arbitrary memory locations. Many operating systems today because of this knowledge restrict the access of program's memory operations to prevent compromising the system due to insecure or intentionally malicious code. [126]

Some of the tools developed to solve these problems along with other root C++ problems such as memory allocation and deallocation are static and dynamic code analysis tools that scan for unsafe patterns in the code of known vulnerabilities. Most of these tools are commercial tools, such as PC-Lint and other packages for static and dynamic analysis, but there are tools like Yasca or RATS that are open source but do only static code analysis.

### 2.2.1.2 Perl

**Memory Safety.** Because Perl dynamically allocates memory when needed it is not susceptible to programmer errors that may cause buffer overflows or arbitrary memory access. Perl also cleans up the memory by reference counting avoiding any programmer intervention during freeing of memory. This does not however make Perl immune to memory attacks as periodically there are security flaws discovered in the distribution of Perl such as CVE-2009-1927 that with a well-crafted regular expression using UTF8 characters is able to crash a Perl application. [128] Another more serious flaw CA-1997-17 was identified and fixed quickly where the 'suidperl' function allowed for a buffer overflow due to a kernel race condition. [129]

**Input Validation.** Because of Perl's premise as the most flexible scripting language, in which there is more than one way to do something, this also comes at a price of increased risks in user input validation. This is the area where most Perl security issues exist because of the variety of ways that input is handled and used. Since Perl is largely used as an intermediary that launches programs, formats their output and based on it launches other programs using the formatted output as input, there are also a lot of opportunities for any script to be launched with incorrect input or even launching incorrect scripts.

The first type of input validation failure happens when passing arguments that are later used to launch command line scripts using functions such as 'system', 'exec' or backticks. This happens when a programmer calls the function without validating input as such:

```
my $userID = <STDIN>;
system("touch /home/$userID/lastUpdate");
```

In the above case whether the command is launched in backticks, by 'exec' or 'system' the $userID variable is a vulnerability because it is not escaped and if the program is ran by root and value equals something like "`;chmod a+rw /etc/passwd; echo `" it will end up in privilege escalation. As a fix that will prevent these types of input security flaws 'system' should be called in the multi-argument mode.

```
system("touch", "/home/$userID/lastUpdate");
```

This way the argument is escaped and it will not launch another process. This does not prevent however any flaws that are a result of the called program having a vulnerability to the input arguments. The same solution applies to 'exec' and since backticks do not support this solution they should be avoided.

A second type of security flaw that is similar to the above is the 'open' call that opens a file or a pipe. That is possible if inside the open call there is a '|' symbol that indicates the output of before command should be piped to the command after, so if the open call is used with an unchecked parameter it can result in running an arbitrary script.

```
open(A_FILE, "/home/$userID/userfile");
```

If $userID is set to "`| malicious_command_here `" then a malicious command is executed. There are several other ways to exploit this even if explicit '>', '<' or '>>' prefixes and splitting execution into pipe input and execution separately with '`open(INPUT, "-|") or exec("cat", "/home/$userID/userfile");`'.

Another case of insecurity is putting user input in 'eval' statements or /e regex modifier. These statements evaluate the argument into Perl code and execute it, so such evaluations have to be done carefully and user input should never be passed there unless completely trustworthy.

Yet another case, because of the reliance on system commands, are insecure environment variables. For example if a program calls a system tool without the absolute path and the PATH variable has been corrupted, a different program will be executed leading to an exploit. To avoid the the script should redefine the used environment variables before using them. They are not explicitly considered user input, but are under the control of the user.

A solution to these problems related to user input lies also in filtering user input that may contain unwanted characters or white-listing the user input allowed in sensitive cases. Also ideally programmers should avoid the shell altogether using Perl library functions instead if available.  This approach should also be used when dealing with external entities such as databases to avoid SQL injections and help filter possible cross-site scripting. [130]

There are two tools that also aid the identification of those faults.  The first of those tools is 'taint mode' that is built into the perl interpreter.  When running the script with '-T' option the script does not run unless a set of predefined rules is not met, such as variables in the execution of the functions above as well as many other system functions.  Variables are traced from input and marked as insecure through assignments and such and if they end up in a certain system or Perl library call execution is forbidden.  Taint mode is not perfect and does not catch all user input, such as from CGI.pm when using `param('name')` or never flags 'sysopen', which is simply a slightly more secure wrapped version of 'open'.  [131]  The second tool is the most widely used tool not written by the authors of Perl and is called Perl::Critic.  Written by Elliot Shank it is a source code analyzer that checks the source code for uses unconventional uses of Perl constructs or functions, difficult to read statements and error-prone or ambiguous code. It follows a set of predefined rules allowing the user to customize the rules and fine tune the levels of severity of bugs to check for different types of rules. [132]

**Race Conditions.** As a file manipulation language Perl deals quite a bit with file operations, which is where a lot of race conditions can come into play.  As for any program the "time-of-check time-of-use" dilemma is very difficult to overcome when dealing with file operations.  Some solutions to this is when writing to a new file use 'sysopen' instead of open so that contents are appended in the worst case if an attacker manages to sneak in a symbolic link to a sensitive file between the check and usage.  Another solution is instead of writing to the target file, write to a temporary file with a random generated name instead and then move that file in place of the sensitive file.  These solutions do not eliminate the problem but make it more difficult to exploit. [130]

**Privilege Confusion and Escalation.** As for any language that does lots of scripting privilege faults can appear and are largely based on setuid scripts which are used because the scripts manipulate files not belonging or accessible to the user that is calling them.  Just because of the nature of usage of Perl setuid is used quite often so any security flaws such as race conditions, user input validation can lead to privilege escalation.  A good countermeasure is in case is to set the effective UID and GID of the script to the real UID and GID so that the access checks are executed correctly.  Perl also automatically switches to taint mode when running setuid scripts avoiding putting any extra trust in the kernel. [130]

**Interface Failures.** Perl does not have any user interface on its own but uses external libraries written in other languages such as wxPerl, gtk2-perl or Perl-Tk so any failures from the user interfaces are inherit from bugs in those libraries.  One place where you can consider it as having a user interface is when used as a CGI application, or any web service.  In this case the faults are specific to each applications and because applications can interact with databases and files, to which multithreaded access may not be controlled it may result in unpredictable behaviour.

**Language-specific issues.** Because Perl is built largely in C it inherits some of its flaws such as the 'rand' function which is not truly random.  A proper seed should be set every time or alternatively substituted with other libraries or calls to /dev/random to get a more random number that will be very difficult to reproduce.

### 2.2.2 Web applications development

Web applications started to evolve since the widespread use of the internet. As the web browsers \started to support various dynamic components such as JavaScript by Netscape in 1995 or Flash by Macromedia in 1996 web content slowly started to come alive. On the other side the first attempts at developing dynamic content started with Common Gateway Interface (CGI) where a request was passed with arguments to an external program from the web server service and at that point frameworks for easier handling of HTTP requests emerged to support web development. [133] Majority of the popular web application frameworks are written for the Java language, but some exist for C++ and Perl as well.

The current trend to offload the server component towards a more involved and interactive client component is causing the web applications to become web services responding SOAP requests from AJAX-based web clients, which will be covered in the next criteria.

#### *2.2.2.1 C++*

**CGI.** The first available technology from the web services was to execute a process from the Apache mod_cgi module which executed each process individually. This means that each HTTP request would need its own process handling it which created some parallel execution concerns when dealing with shared resources such as files and databases that did not allow concurrent access. Because of efficiency and optimization of C++ which is largely based on C it is still one of the fastest interfaces for web applications ahead of Java and others. [134] Even though simple HTML is easy enough to write, complicated output gets messy writing all HTML to the client directly from the code. One advantage of simple CGI over later technologies was that if a program written in C++ crashed it would not bring down the web service in any way since it was its own process. There are lots of different technologies implementing the C++ CGI interface to ease the development. A snippet of sample code is attached in the appendix.

**Microsoft ISAPI.** Microsoft having its own web service on MS Windows, the IIS, decided to create its own interface to handle web applications in a better way than a CGI handler integrating it into the web service. ISAPI works by compiling the application code into a DLL which is executed by the IIS web server as an extension directly. Up to version 5.1 the ISAPI was integrated into IIS making it vulnerable to DOS attack if the ISAPI caused a crash. Since IIS version 6.0 ISAPI is run as a separate process so that if an extension caused a crash it would not bring down the web server. Running web applications on ISAPI is significantly faster than CGI due to avoidance of spawning new processes and inter-process communication, making it the fastest way to run web services today on IIS in windows. Designing for the ISAPI has changed over the years but the basics of implementing under CHttpServerContext in VC++6.0 or HttpExtensionProc in version 9.0 as a single point of entry with some supporting functions or parameter maps is still fairly easy and worth the advantage over CGI. A proper response object still needs to be returned. [135]

**FastCGI.** FastCGI is a protocol that interfaces between web servers and applications allowing for improved performance and smaller memory footprint than pure CGI, just like ISAPI. FastCGI implementation ran on Apache as well as many other servers and because of its packet oriented multiplexing nature it was able to process many requests simultaneously and even connect multiple times to the same target application. One major advantage of FastCGI over ISAPI was that not only it contained bindings to a lot of languages it also ran on many operating systems including Unix and Linux. [136] The ease of design was based on the implementation of FastCGI and there were many different libraries, one of which is fastcgi++. Distributed under the GPL it is a pretty good implementation that pretty easy to program for. [137] A snippet is included in the appendix A.

**Wt.** Pronounced 'witty' Wt is an advanced framework for writing interactive web applications using C++ on the server side and AJAX on the client side. It is a widget library and compares itself pretty well to existing widget libraries and because it hides from the developer the underlying technology chosen to render the webpage, it is able to decide dynamically based on what the browser supports, taking that responsibility away from application developers. It is a pretty impressive approach that generates the client code based on the operations coded for in the backend through the WApplication class and it does not depend on JavaScript or AJAX to work, but does benefit from when they are present. [138] Some other features include protection against user input due to built-in widgets, implementing all optimization tricks to improve responsiveness, portable graphics using browser capabilities such as VHL, SVG or HTML5 canvas and easy deployment using either built in httpd or a FastCGI integration. [139] There is another framework that is new and still inferior to Wt called CppCMS and together that is all the web applications frameoworks available in C++.

### 2.2.2.2 *Perl*

**CGI.** Same as C++ the Common Gateway Interface allowed people to run Perl scripts as a result HTTP queries. Even though the complexity of such calls is pretty low, the performance is also not very impressive either. For ease of development it wasn't too friendly having to parse the different requests in different fashions at first but fortunately Lincoln D. Stein came with help in form of a CGI.pm module that abstracted a lot of the reading into simple functions that were either procedural or object oriented. [140] Sample programs are provided in Appendix A.

**FastCGI.** As an improvement to CGI the Perl bindings for FastCGI allowed it to run faster because of the improved communication between a persistent process and the web server. The simplest implementation is the FCGI Perl module available from CPAN which can be used with the Perl CGI::Fast module. The FastCGI especially benefits large long-running processes that take a long time to initialize. [141] There exist many different implementations of FastCGI today for different operating systems and different web servers. An example snippet is in the Appendix A.

**mod_perl.** A step forward from FastCGI with a more managed framework was the mod_perl apache extension where the Perl interpreter is practically integrated into Apache making it extremely fast to execute scripts. By doing so not only does the mod_perl extension replace the heavier CGI framework but also it has access to the inner workings of the web server making it possible to handle requests any way the developer likes. The requests also do not create new processes, but instead are handled by static processes inside Apache making the extension faster and ligter. [142]

**Catalyst.** In 2005 the Catalyst project was started with inspiration from the Maypole framework as well as Ruby on Rails and Java's Spring. From its predecessor it took the database support and added other new features. The new framework supports both mod_perl and FastCGI on Apache, lighttpd or Ngnix as well as its own web server, abstracts any database usage with an object model and also integrates with many large UI frameworks like AJAX jQuery or YUI. [143]

There exist other frameworks but they are not as widely used as Catalyst.

## 2.2.3  Web services design and composition

Web services differ from web applications because instead of providing an entire application they provide just the backend to a web application without caring what the front end will be like.  The technologies to provide the services from an application to another computer had went through many standards, each suitable for its own time and purpose, starting at the basic RPC mechanisms, domain specific uses such as ODBC for databases, distributed architectures such as CORBA, Simple Object Access Protocol also known as SOAP and eventually Web Service Definition/Description Language or WSDL in short.

**Domain Specific and Lightly Used.** In the meantime there were several domain specific and proprietary web service standards such as ODBC as an RPC for database applications, NetDDE developed by Microsoft, Qworum, Hessian, BEEP and other protocols and technologies, but we will not discuss those due to their low relevance or usage.

**SOAP, WSDL and REST.** As XML-RPC became more popular and more functionality was being added to it evolved into Simple Object Access Protocol that provided extended support for user data types.  Developers could specify the services, define messages, recipients, senders and the message path and because the protocol works over HTTP it avoids proxies and firewalls through tunnelling.  WSDL is a definition of SOAP services by a server where the client is allowed to discover the service and invoke it by the specified parameters found in the indicated XML schema. REST is a new architecture of a web service where directories in a URI are treated as collections of resources and each element inside as an individual resource.  Bothe collections and elements react to four requests, namely GET, PUT, POST and DELETE.  [144]

### 2.2.3.1  C++

**Sockets.** Remote procedure calling was not built in by default into C++.  Functions could be called from dynamically linked libraries but those libraries had to exist and be launched on the current server.  Any inter-process communication over the network had to be done via network sockets and required to be built from the ground up. Some early initiatives at C++ wrappers existed but eventually applications used one of many available RPC libraries with their own formats that abstracted those sockets into something more manageable. [145]

**CORBA and RPC.** Started in 1990 by OMG it existed before C++ and was used to enable computers and devices to communicate with each other. By using the IDL mappings released soon after C++ in 1996 it became available to developers and its widespread use began. The last release of CORBA was in 2002 at which point other technologies such as XML-RPC started to creep in. [146]

Some attempts to create a standard distributed computing framework were taken first by the DCE/RPC (Distributed Computing Environment / Remote Procedure Calls) project which was proprietary at first but became open source in 2005.  Based on this model Microsoft had released its own DCOM (Distributed Component Object Model) that used MIDL (Microsoft Interface Definition Language) to be applied to the COM objects. In comparison CORBA provides better abstraction and encapsulation while DCOM focuses on providing a more flexible programming environment. While DCOM has the edge on flexibility CORBA still has better data representation, inheritance, platform support and exception handling. [147] Soon after the initial release interworking between the two was implemented.

**Standardized RPC.** The popularity rose as the Internet grew and so did demand for distributed application over HTTP. The first attempt to create a standard communication protocol for an application to be used as a service in C++ became the XML-RPC created in 1998. With many C++ implementations such as mainly XmlRpc++ or xmlrpc-c and others including new frameworks like gSOAP and libmaia for Qt XML-RPC became a widely accepted standard in C++ web applications for remote procedure calls. While XML-RPC is very simple because of efficiency of C++ web applications could be executed quickly and efficiently over the network to serve their purpose. [144] Another competing technology that emerged in 2005 was JSON-RPC (JavaScript Object Notation - Remote Procedure Call) that was based on JavaScript as the name implies with the main benefit of smaller notation. Instead of using the full tags JSON uses brackets and JavaScript-like syntax to represent its objects, which in cases of huge objects is a big saving having a bracket instead of an xml tag. There currently though only exists one C++ library from JSON being JsonRpc-Cpp and it is still in early stages. [148]

**SOAP, WSDL and REST.** There are several frameworks that allow for web service development in C++ being gSOAP that translates the WSDL and SOAP definitions defined for gSOAP into C++ source files that then go as handlers of the requests that de/serialize objects as they come and go and dispatch them to appropriate subroutines. [149]

Another toolkit that is popular in C++ web service development is Apache Axis and Axis2 which implements a long list of features for a web service framework with quite a few extensions. Because it builds itself into Apache it is very quick, modular and designed with extensibility in mind. [150] Apache Axis2/C is a newer and more extensive project although it only runs in C and not C++. The next instalment in the series is a WSO2 Web Services Framework which extends Axis2 making it available in C++. Since Axis2 comes with a more extensive set of features it is a great improvement for C++ web service development supporting WS-Addressing, WS-Policy, WS-Security, WS-SecurityPolicy and through extensions WS-Eventing and WS-Reliable Messaging specifications. [151]

Yet another toolkit built for web services as another attempt by Microsoft to have a foothold by introducing its technology is the Windows Web Services (WWS) framework built not on .Net Framework but rather be optimized for speed and low memory footprint. [152]

### 2.2.3.2 Perl

**Sockets.** As the basic type of inter-process communication Perl started with sockets. They were present in Perl since the beginning as the IO::Socket package and allowed to do all the basic socket operations abstracted into a socket object. The socket support has grown and allowed customized protocols to emerge, although because of its simplicity and lack of support for data types and structure it wasn't particularly well suited for widespread web services. [153]

**CORBA and RPC.** CORBA was emerging around the same time as Perl with version 2.0 and didn't take long before a CORBA library appeared in Perl from a group of computer scientist at the University of Frankfurt. The library only relied on Unix API and free libraries, which made it largely successful to implement CORBA applications in Perl where it is still widely used today and its features have expanded to support latest technologies starting at IPv6, extensive security policies, X11 application support and plenty of other features. [154]

Aside from CORBA there are many types of simpler RPC communications in Perl that came about. One such library was the RPC::Simple library by Dominique Dumont where simple RPC calls could be made easily between the client and server system, seamlessly but the protocol did not guarantee any security, type checking or error checking although it simplified distributed networking. [155]  Another more advanced library that came about was DCE-RPC in support of the protocol by the same name by OpenGroup which had some minor success in distributed and multiplatform networking as base for DCOM. It was more complicated to use and included some authentication via external modules.  [156]

**Standardized RPC.** When the Web came about to a standardized way of executing remote calls XML-RPC and JSON-RPC standards and libraries came about.  When XML-RPC first took off Randy J. Ray implemented the first library that was used for this protocol under the name RPC-XML and has improved its functions with adapting encodings, increasing robustness, speed and better integration with Apache. [157]  There exists one other library that supports XML-RPC for Perl that is newer and bares the same name as the protocol.  The other competing technology for standardized RPC using the JSON format also found its implementer, Makamaka Hannyaharamitu, who developed a library with the same name in 2005 that was finalized into version 1.00 in 2007. The library went through many enhancements from that point, but mostly extending support to different encodings and bug fixes. [158]

**SOAP, WSDL and REST.** Just like in the preceding protocols there is a limited number of implementing libraries or projects that allow for Perl to function as a web service.  One such Perl framework is the SOAP-Lite framework that is pretty easy to use and defined thanks to Perl's concise language semantics.  It supports all the web application technologies, starting with CGI to mod_perl and mod_soap HTTP daemon.  The code stays neat and concise in any solution and supports SOAP and WSDL, autodispatching, authentication, error handling, proxies, access control and complex data structures. Even though this all sounds promising as the main Perl SOAP interface it does not adhere to a lot of the rules and still has some type guessing flaws that make it hard to predict at best sometimes. [159]  The other package that attempts to make SOAP and WSDL available to Perl developers is WSO2's WSF/Perl framework which is the WSF/C framework wrapped in Perl code to allow Perl to consume SOAP requests and provides the specification of web services in WSF/C like SOAP MTOM, WS-Addressing, WS-Security, WS-SecurityPolicy and WS-ReliableMessaging.  [460]

## 2.2.4  OO-based abstraction

Object-orientation is a concept in programming languages that allows a programmer to abstract the data, logic and interactions into a set of objects talking to each other. There are many areas of object-orientation that a programming language can implement as well as the extent to which it implements it. These include class, instance, method, message passing, inheritance, abstraction, encapsulation, polymorphism and decoupling. [161]

### 2.2.4.1  C++

While C++ was first known as C with Classes is known as one of the first object oriented languages, it is not a purely object oriented language because it has some procedural language elements. Classes combine the operations and data of an abstract data type in C++ into a single entity. Instances of those classes can be created either in dynamic or static memory and they will be initialized through the use of constructors. Similarly once an object is no longer needed and is deleted or goes out of scope a destructor is called so the developer can free any reserved resources. Methods in C++ come in many flavours starting at member functions, virtual overwritable functions, overloaded functions, operators, constant functions and creation and destruction ones mentioned before. Methods can also be public, private or protected, ranging in visibility as desired. Message passing occurs by objects calling each other's methods with arguments as message data. Inheritance in C++ is very elaborate and expanded mechanism. Starting with simple inheritance a class can call the construction or destructor of the parent from within itself, can inherit from an abstract class that cannot be instantiated and can inherit from multiple classes, albeit shouldn't because of risky consequences unless sure of resulting behaviour. Abstraction is nicely done in C++ by hiding away data as 'private' and providing public methods to retrieve the data. If subclassed those method called from a pointer of a parent class will still call the right method for the real instance and return the proper value. Encapsulation is also well done in C++ by marking methods and member variables as either 'public', 'protected' and accessible by inheriting children or 'private' and accessible only from within the class itself. For the cases where variables should be hidden away from everyone except a class that is closely tied and needs private access, there is a 'friend' class operator that allows for inter-class member access. Polymorphism is also well implemented in C++ via the use of vtables, which are virtual function tables that launch the proper method for the object regardless of the pointer type from which the method is being called. Of course that pointer has to be of one of the types of parent classes of target instance for C++ to allow for the method call. Coupling and modularity is another one of C++'s strengths, not only detaching the class from the rest of the project if not used but also detaching the header from the implementation, where implementation can be hidden away or swapped while the header remains for initial compilation. [162]

Some of C++ shortcomings why it cannot be considered a purely OO language are that the built in types are not objects and that is requires a procedural part, namely the main function, to be executed and crate objects.

### 2.2.4.2  Perl

Perl does a fair job at implementing object-orientation into its semantics. I have to say to start off Perl is not an object oriented language, although it can instantiate and manipulate objects. The notion of a class in Perl is equivalent to a package, which is sort of a namespace. This class has to possess the 'new' subroutine to be able to be instantiated and therefore an instance of that class created. An instance of a class is essentially a hash reference that is blessed, where blessing means given references to subroutines , fields, variables and constants in that package. Some people mistakenly refer to classes as modules, but modules are the physical files containing the packages, which are logical scopes for the subroutines, fields, constants and variables and there can be multiple packages in one module. Methods are simply references to package subroutines inside the hash and message passing occurs when one instance of a class calls the method reference in another instance of a class. Inheritance is present in Perl in the form of @ISA directive which is better used as the 'base' pragma, specifying the base class. When building a constructor a call to the higher class has to be made sometimes to initialize its variables as well and this is made easier with the SUPER class that looks for the first higher class with an implementation of 'new'. Abstraction does not exist explicitly in Perl because Perl treats all instances as hash references so no variable really has a type and any method could be called and will simply result in a runtime error if the method is not defined.  This is both, a blessing and a curse (no pun intended).  It is possible to check if a method is defined via the 'can' method, but it has to be done programmatically by the developer.  This type of abstraction can be called 'interface inheritance'.  Encapsulation as such also does not exist in Perl because of its hash reference nature all data is visible to the outside.  Fields provide some safety because if defined in a package the compiler will not allow execution if it realizes that the programmer is referencing a field that is not specified.  The only way around this constraint is using inside-out objects.  Inside-out objects rely on the fact that local variables in a module are not visible outside the module, so by creating maps of object Ids to variables in one hash per variable the values are in fact hidden from outside the module and the only thing visible is the ID. This approach still has its downfalls to be addressed such as maintenance, overhead of hash lookup and object destruction.  Polymorphism does not exist explicitly and relates closely with abstraction and for the exact same reasons.  Perl supports decoupling very well allowing reuse of its code in modules by simply including them in the library path to be loaded. [163]

There are also a few other remarks about OO compliance of Perl.  Perl does not have a main object, instead it executes like regular procedural code.  That code itself can create new objects using new, manipulate them using their subroutines, and access their data members using fields and constants.  Also while variables are non-typed and internal types are not objects.  This all in all makes Perl a fair object oriented language with some of the benefits, like modularity, and some that totally miss the boat such as abstraction and encapsulation.

## 2.2.5  Reflection

The ability to dynamically create and use objects depending on the context in which an application is ran is a great tool because it supports more organized and flexible styles of programming than having to specify all the possibilities explicitly.  It is the ability to discover object types at runtime, convert the type name of the object into the symbolic name of the object and then create and invoke its methods or evaluate a string as if it were source code and execute it. [164]

### 2.2.5.1  C++

Reflection is a feature of a language that requires some metadata to be stored about the types in order to be able to discover them and use them dynamically.  It is not present in C++ and there are a few reasons why C++ is unsuitable to have them implemented.  First of all we have to remember that  C++ is derived from C, a highly optimized and procedural language.  C++ is a language where the benefit of storing metadata is outweighed by the performance gain from the compiler potentially optimizing those classes away and inlining all their code, just like in C. Also meta-programming such as templates is not suitable for reflection because each type of a template is generating new code for its own purpose, so creating templated reflection would have to generate all possible class types at complie-time and store them for later execution.  Templates are sort-of the C++ alternative to reflection where types may not be dynamic at runtime but are flexible enough to solve most problems. [165]

Now that we discussed why reflection in C++ may not be the appropriate, given the language structure, we can look at the work that has been done to add the mechanic to the language. The first step towards reflection in C++ was RTTI (Run-time type information), which allows for an object to be cast as another object with 'dynamic_cast'. It works only if the object contains virtual functions and needs a virtual destructor. Since this was not enough there were many approaches to add reflection to C++ that were considered: parse the debugging information where program should not be changed although debugging information may vary between compilers; building a separate processor that would parse the code and build descriptors, which would not change the code but require a lot of effort and be compiler-specific; create a specialized compiler supporting reflection which also is a lot of work and would force users away from their known tools; lastly let the programmer enter the metadata himself which is simplest and most portable, but puts more work on the programmer. [166]

The next libraries that brought reflection closer to what it is supposed to be were SEAL, XCppRefl and a few others. SEAL is a more advanced reflection mechanism that separates the build into two parts, the user library part and dictionary that can be used to lookup and instantiate objects. This is best described by the diagram from the published document by the creators:

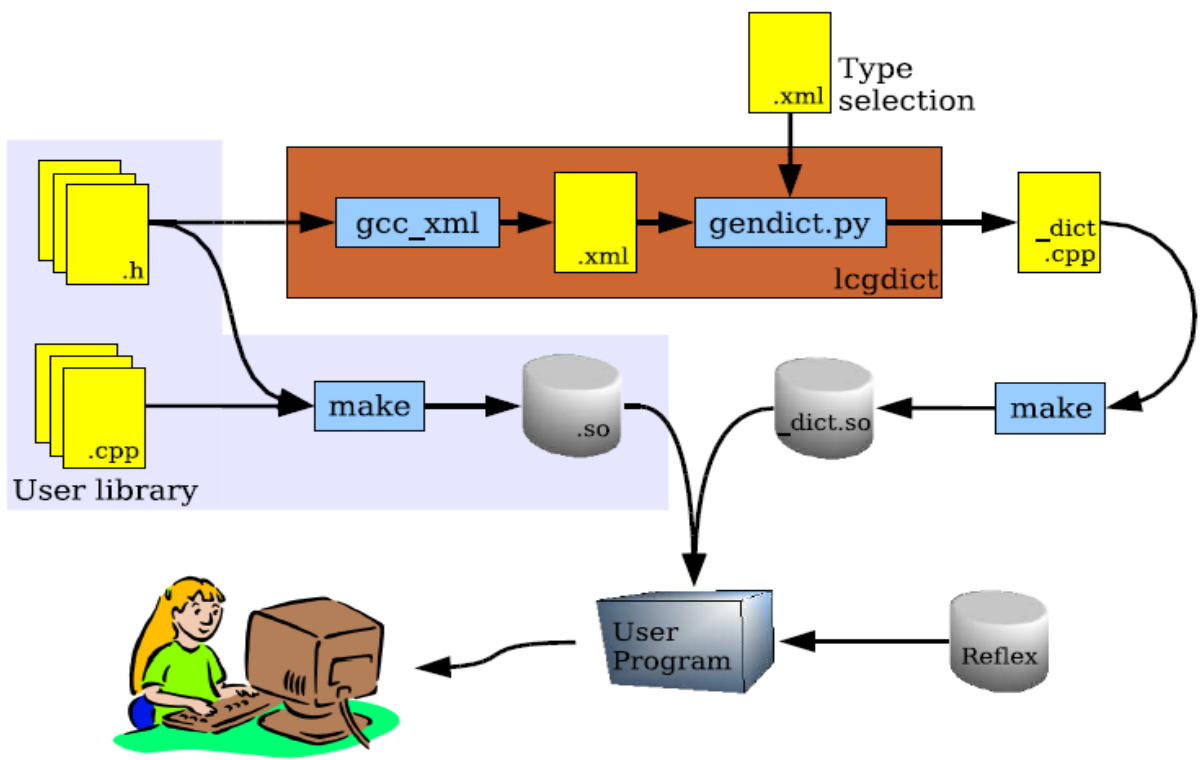

Figure 6: Producing reflection information in SEAL [167]

As you can see the lcgdict parses the header files that are included for the dictionary through gcc_xml compiler and generates the dictionary library. The second part is a Reflex library with an API for using reflection. By combining the two the user is able to instantiate objects by name, call their functions, retrieve information about attributes and check them and destroy objects after they are no longer needed. [167]

Another library that is well implemented is XCppRefl, which is based on the paper "C++ Reflection for High Performance Problem Solving Environments". This approach is also non-intrusive and therefore pretty easy to integrate. It is fully compliant with the C++ specification and works by including the classes in the generated code when parsing syntax trees of included classes that is referenced when the library looks up, instantiates or calls upon a class. It is fairly easy and intuitive while providing the basic functionality of reflection. [168]

There are also other works that are more intrusive such as MOP using metaclasses, reflection through template programming and XTP by Stroustrump which has not been completed to date.

### 2.2.5.2  Perl

Perl is a very reflective language as most weakly typed languages. With the introduction of object-orientation into Perl also some functionality was added to add reflection support to the modular object structure.

The mechanics are really simple when working with objects. To create an object all that is needed is to call the constructor on the class name as a string. To invoke a method or fetch the value of the field the instance has to call a function using a variable or a name with the "from reference" '->" operator as such:

```
my $class = "Desired::Package";
my $object = $class->new();
$object->{$fieldName};
$object->$methodName();
```

This way the method will be called and any method is available. The 'exists' and 'can' functions can be used to safeguard against calling unidentified methods. A snippet in the appendix illustrates how small and effective is object reflection in Perl.

Another thing that Perl does is using the 'eval' function it can evaluate an arbitrary string as if it was source code. This way arbitrary code could be added and launched at runtime, which is a technique Perl uses for loading new modules at runtime. [164]

## 2.2.6  Aspect-orientation

Aspect oriented programming is the programming style that abstracts roles and responsibilities into aspects of the program. It works by creating advice, or code that performs a function, which are then attached at join points making pointcuts. The style is home to aspect-oriented languages such as Aspect-C or Aspect-J, but other languages also can support the programming style.

### 2.2.6.1  C++

There are two approaches to Aspect-oriented programming in C++, one without modification to the compilation process and another that pre-processes the code for greater AOP compliance.

The first method is to attempt aspect oriented programming using templates. This is a neat approach that requires some modifications to the affected class, such as insertion on join points, and leaves the rest to templates which use the class as a parameter. It requires that a template is created for each aspect for each object and advice code can be factored out to a separate class that can be reused in other aspects. The limitations of the approach are quite a few. Scalability is the first factor where the wrapper code can easily outgrow the aspect code and explicitly defining order of the aspects is cumbersome and error prone. Also debugging with lots of templates and namespaces can get very difficult. For join types there is no distinction between method call and execution, no generic interface to join-point context and lack of advice for private methods. The whole concept ends up in excessive template meta-programming making is suitable only for very small applications. [169]

The second approach is closer to AspectC++ but stays within the C++ language and is called Dynamic Aspect Oriented C++. This approach uses two classes, Aspect which is subclassed to create user generated aspects and AOP_Engine that weaves and unweaves aspects dynamically into the application. Once the code is written the pre-processor inserts join-points at function entrance and possible exits as well as creates records for each class and method. The neat thing about this solution is that it enables the user to add and remove aspects at runtime replacing the advice if necessary. The performance of the application is affected by 7% without any aspects due to the hooks and by 37% with aspects compared to manual static weaving.

While AspectC++ is an aspect-oriented extension to C++, the language in which it is written does not conform to the C++ specification so we cannot count it as a way to use AOP in C++. It does though do a very nice job at aspect-orientation and translate its source into C++, but that source is not for editing but for compilation.

### 2.2.6.2 *Perl*

Perl provides aspect-oriented style of programming through the Aspect library by Adam Kennedy. Because of its loose typing rules it integrates neatly into Perl and allows for pretty good cross-cutting concerns encapsulation and focuses on subroutine matching and wrapping where an aspect could be attached to a subroutine in many ways. It also provides a set of aspects already built ready to be applied and reused. The limitations include low support for inheritance and some performance issues if code is not correctly written. Here is an example of installing aspects and some advice:

```
# create some aspects
aspect Singleton => 'Company::CEO';
aspect Profiler => call qr/^Company::getReport/;
# create a pointcut and assign it to the subroutine
my $pointcut = call qr/^Company::listEmployees/;
before { print LOG "listing by ".getlogin()."\n"; } $pointcut;
```

Overall the integration is pretty seamless and allows for the basic AOP functionalities in Perl. [170]

## 2.2.7 Functional Programming

Functional programming is programming in a functional style where operations are performed in the form of functions that work on first-order functions or higher-order functions. They often use recursion for their looping needs because there are no loops in functions themselves. Another thing they usually use is lazy evaluation when processing lists and such. It is possible for a natively non-functional language to adapt the style of functional programming depending on the features of the language. [171]

### 2.2.7.1  C++

C++ has had various libraries implementing functional programming style but the most popular and widely used is the FC++ that implements some key features of functional programming.  FC++ offers a lot of features like operations on polymorphic and higher order functions.  It is possible to pass them as arguments and supports functions like 'compose' that do operations and return a possibly polymorphic function as well.  It includes a large part of the Haskell Standard Prelude implementing a lot of function-oriented concepts.   Thanks to its conversion methods to STL it integrates easily into regular object oriented C++ programs and also as a must in good functional programming it implements lazy lists to optimize speed when evaluation is unnecessary until the values become queried for. The implementation relies heavily on templates that allow for any list type to be passed and manipulated.  The library performs slower than Haskell but better than g++.  It uses some optimizers of its own as well to speed up with reuse, reference counting, tail recursion optimization and using global data.  FC++ implements a bit of its own memory management with its own allocator and reference counted pointers for memory management. FC++ makes C++ good for usage in functional programming style. [172]

### 2.2.7.2  Perl

Perl supports functional programming in more than one way.  By itself Perl is a very well adaptable functional programming language because it can easily return subroutines making polymorphic functions easy to compose.  Directives such as 'grep' or 'map' have their own terminating conditions making them ideal for traversing lists and abide by the functional programming rules. [173]  There also exists a Language-Functional library for Perl that expands on the manipulation of lists with 'foldl', 'foldr' and the usage of infinite lists that are needed by true functional programming.  The library does not support currying or types.  Currying can be done by modifying Perl subs with a bit of explanation from manuals such as Higner-Order Perl [174] that is freely available online.  The library implements a large portion of Haskell's functionalities to allow writing functional code in Perl that is closer to real functional code.  Unfortunately because of the scripting nature of Perl it does not perform as fast as Haskell. [175]

## 2.2.8  Declarative Programming

Declarative programming is a style of programming that describes what the program should do without describing how it should do it.  It does not specify the algorithm like imperative programming does. Such languages lack side effects and have a clear correspondence with mathematical logic.  Functional programming is a subtype of declarative programming, so because both languages have demonstrated the ability to write functional code they both should satisfy this criterion. Under the umbrella of declarative programming there are functional, logic, constraint and domain-specific languages.

### 2.2.8.1  C++

C++ as shown before is good at functional programming thank to the FC++ library. As an addition the 'bitmap boost' library to allow for bidirectional maps support compliant with STL.  As for logic programming a library called LC++ was built on top of FC++ to provide pure logic programming support that is almost equivalent to Prolog.  Because of usage of lazy lists it is able to perform better than other logic libraries such as M PC++ in terms that it gives the user control over the queries.  [176] Another alternative is the Castor library which allows for a smaller set of operations to be done with logic programming, but is still a considerable candidate. [177] In the field of constraint programming a C++ library that enables that style in C++ is a commercial library called CHIP V5 by COSYTEC. It claims a high level of abstraction and improvement of performance with its main feature as the Global Constraints. It also supports open architecture that can be used on many platforms. [178]

### 2.2.8.2 Perl

Perl support for functional programming was described earlier as fairly competent and suited for functional programming. Declarative programming on the whole is therefore well suited for Perl because of the way that it could be written without loops but instead as a set of statements. Also thanks to hashes that could easily abstract equations and subroutines their evaluation many applications are possible without the need of specifying the exact algorithms when writing the code. [170] Perl handles logic programming by delegating it to Prolog. There are several modules that bring Prolog to Perl, namely SWI, Yaswi, Interpetter and AI:Prolog modules. The advantages to using Prolog are that it is a standard in logic programming, but unfortunately its knowledge is required when using it through Perl. [179] Using Perl itself it is possible to build constraint networks of higher complexity and evaluate them in bidirectional function maps but requires some prior knowledge of declarative programming in Perl. [174]

## 2.2.9 Batch Scripting

Batch scripting is the ability to write a script that is easily modifiable and is able to easily process file operation either on local machine or over the network. It can to a great extent simplify the life of a system administrator writing rudimentary tasks in form of scripts that are ran from a scheduler. Batch scripting is an important quality of a language that would be used in flexible systems management without the use of heavy compilers or development environments.

### 2.2.9.1 C++

C++ is not a scripting language, although there exists an interpreter for C and a subset of C++ statements called Ch. It supports variables, arrays, functions, classes and several libraries such as POSIX, X11/Motif, Win32, GTK+, OpenGL and ODBC in its development kit. Not only small scripts but large applications with thousands of lines of code as well can be ran using Ch. The intention of this shell was to execute programs live for prototyping, interactive presentations or in learning environment. [180]

### 2.2.9.2 Perl

Perl is by its nature a batch scripting text processing language and grew to become one of the most used general purpose languages. Perl has quite a few features in the text processing department, easily opening reading and writing files, matching, replacing, filtering, sorting and manipulating text in various ways as well as many other features. Its regular expression engine is one of the most powerful and simple to use. In the system interaction department it offers quite a few ways to launch processes and capture their output, with forks, threading, file and user permission operations, management and checks, easy access to pipes and other ways of interactions with the system. From the networking side it is able to use plenty of remote connection modules for file transfers like 'ftp' and 'sftp' and with libraries such as expect that are able to launch commands on a remote host they can even tunnel through many hosts if needed. Perl also contains an excellent array of standard libraries with CPAN repository for most possible uses and technologies, and with flexible semantics it is very well suited towards many styles of scripting living up to its motto "There's more than one way to do it". [181]

## 2.2.10  UI Prototype Design

User interface design is its own field and there are many tools that aid us today in developing user interfaces. User interfaces themselves are divided into many types, as in command line user interfaces, graphical user interfaces and web-based user interfaces. There are also other user interfaces but the three types listed are the primary types that will be used in vast majority of cases.

### 2.2.10.1  C++

**Command Line Interface.**  Natively C++ does not have a graphical user interface and can perform simple I/O to the command line with its standard libraries. There is no standard for command line user interfaces in C++ although there are a few command line parsers available. There does not seem to be any CLI menu framework available through which a user can navigate, it's up to the developer to write it if he so desires. Writing input and output in C++ is fairly simple so the work required simply depends on the scale of the project.

**Graphical User Interface.**  Microsoft was one of the first to introduce user interface programming to C++ and they did it with dreadful Win32 programming that involved lots of unnecessary micromanagement, cryptic naming and overly complex API. Thankfully within some time Microsoft has decided that it will make people's life easier with MFC – Microsoft Foundation Classes. While that still went through iterations, it improved greatly the usability of the API with a standard structure for windows, input, panels, events, etc. for interfacing with Windows by developers. Today Visual Studio provides a neat interface to draw and create windows, dialogs and applications, link events to the widget actions and generate code stubs and classes for the user interface actions and components. [182]

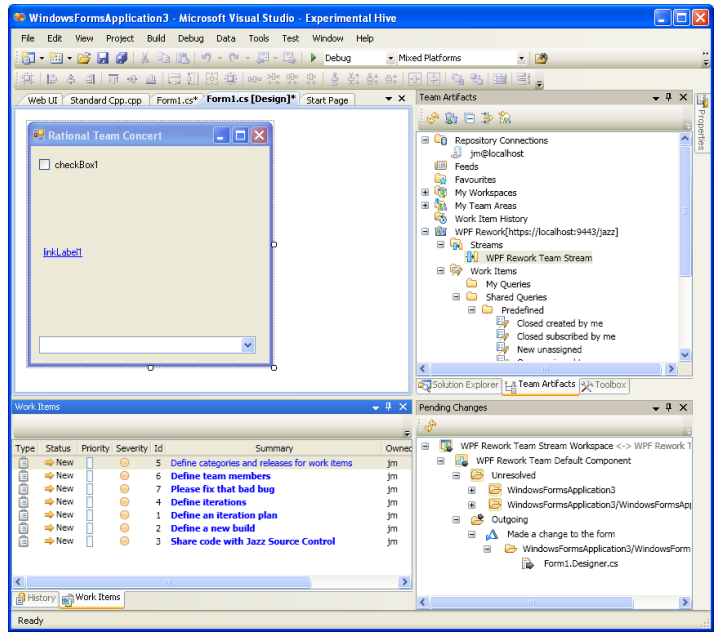

**Figure 7. Visual Studio User Interface – Design Mode**

An originally proprietary framework that slowly started adopting the free licences called Qt is known for its success across many platforms with successful applications such as Skype, Virtual Box, Google Earth and even the KDE desktop.  Qt is written mostly in C with bindings for many languages. Apart from the intuitive Qt Creator in C++ it also includes libraries for events, filters, a level of reflection and memory protection and a few other features.  [183]

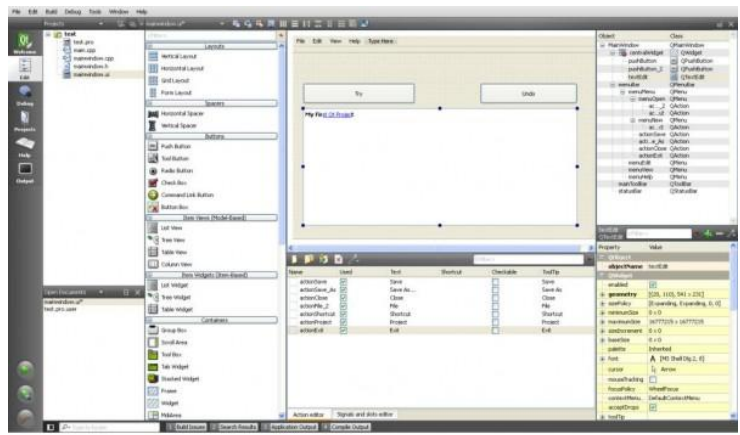

**Figure 8. Qt Crator**

Yet another library available today for user prototyping is the GTK+ library.  This library is slightly more complicated to use for the sole reason of not having a designer application. Aside from that it is complete and runs on all X11 desktops as well as Microsoft Windows.  Applications that use GTK+ include  Wireshark, VMWare Player and most notable GIMP (the reason for GTK in the first place) and Chromium, Google's browser.  [184]

Another toolkit available on in C++ natively is WxWidgets. This toolkit is another major toolkit distributed open source and it comes with a wxWidgets Form Designer application for creating user interfaces visually. The toolkit offers a very thin layer of abstraction between the API and the widgets, but contains other non-GUI libraries such as IPC, networking, etc. Some of the applications designed using wxWidgets include FileZilla, TortoiseCVS and Audacity. [185]

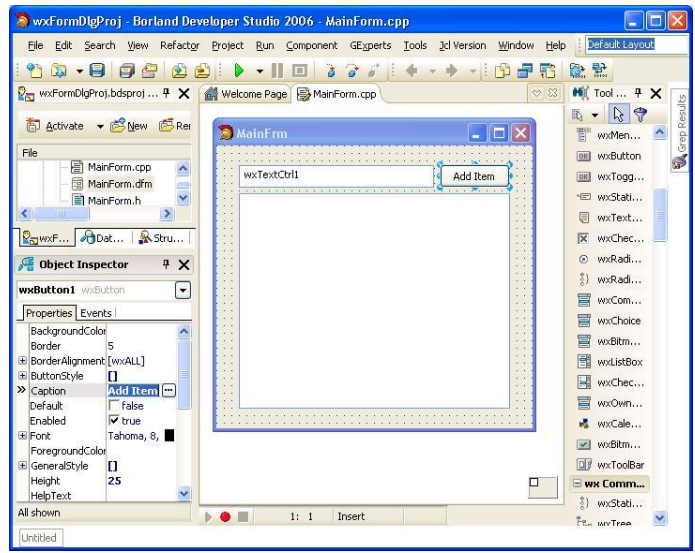

**Figure 9. wxWidgets Form Designer**

**Web-based User Interface.** There are plenty of user interfaces that could be built as part of web applications described in Criteria 2, the best one today is Wt and also the most modern. Given the modular capabilities and widget-inspired design it abstracts the user interface away from the developer allowing them to concentrate on the widgets instead of worrying where and how to organize the user interface elements. [139]

### 2.2.10.2 *Perl*

**Command Line Interface.** Because of Perl's adeptness at text parsing it has a rather easy time dealing with parsing user input and sorting arguments into hashes. Quality to directing them through a menu system only depends on the ambitions of the developer, though it is easier to make a nice menu system under Perl thanks to its easy and forgiving semantics.

**Graphical User Interface.** This is where Perl does not shine so much. The best two toolkits available are not natively on Perl, but rather have Perl bindings, being Qt, probably the best for prototyping because of the Qt Creator, gtk2-perl, a Perl binding to the GTK+ library, and wxPerl, an extension module using WxWidgets through XS.

There is no toolkit built specifically for Perl specifically and the closest one is Perl-Tk, an adaptation of the Tk toolkit for Perl. The advantages of the toolkit is claimed to be rapid development, easy to learn, multi-platform compatible, elegant networking and others. Unfortunately the feel of it seems to be outdated comparing to other toolkits available. [186]

**Web-based User Interface.** Among the web user interface frameworks the best one Catalyst for Perl is the Catalyst framework as described in Criteria 2. It provides the most novel features to create a UI at a higher level with a lot of functionality.

## 2.3 AspectJ vs. COBOL

### 2.3.1 Default secure programming practices:

Programming languages should discourage and prohibit poor practices and support and promote good practices. In this section I am going to focus on common secure programming practices such as: the type system, memory management, exceptions and errors handling, modularity and general readability and understandability.

AspectJ is a strong typed language and this provides for it a safety feature. For example the compiler "ajc" could detect meaningless or invalid code. Also the AspectJ program could resist format string attack. Moreover, AspectJ benefits from the Java compiler security features (Strong typing, Class, interface and variable checking) in order to map the AspectJ join points and advices to regions of bytecode. COBOL, on the other hand, is not strongly typed; it has only three data types: numeric, alphanumeric and alphabetic. This makes COBOL non-suitable for writing system and scientific applications.

According to Toshio Suganuma et al [118] COBOL could be considered as a non typed language "but rather it allows programmers to provide the system with an example (a picture) of how the storage of each data item should be displayed, and hence the amount of storage to be reserved." The following example shows several data items declared in COBOL:

```
01 Name PIC A(20) VALUE SPACES.
01 AccountInfo PIC X(50) VALUE SPACES.
01 Balance PIC 9(10)V99 VALUE ZEROS.
```

This introduce additional work when we need to assign values to data items, because there are several rules that must be followed, such as checking the compatibility rules between the sending and receiving items and truncation rules when the sending item is longer. This problem with COBOL may lead to loss of precision during type conversion.

Nonetheless, the two languages have a common feature. The absence of explicit pointers (COBOL does not have pointers and Java uses references) that are considered to be dangerous practices in other programming languages that could lead to memory corruption and vulnerabilities.

In memory management, AspectJ uses an automatic memory management (allocation and deallocation) through the automatic garbage collection which is an essential security feature of the Java language.

Earlier versions of COBOL have no heap allocation. Although this drawback, especially in implementing modern data structure like recursive trees, COBOL has performed well in its domain without that.

In COBOL, All the data and variables are first allocated in memory when the program is loaded, and they stay in the same address space until the program terminates. There are none of the dynamic allocation features that modern languages like AspectJ offer. COBOL does not support parallel execution (single thread) of control, therefore the programmer does not risk to have problem like memory corruption problems which are common to the storage management approaches [118].

The latest version of COBOL (the COBOL 2002 standards, not yet fully implemented by any compiler) has heap allocation through ALLOCATE and FREE. Also it has a garbage collection mechanism [105].

In exception handling and errors, AspectJ has classes that inherit the Java RuntimeException like NoAspectBoundException and SoftException in order to treat exceptional cases that may occur. COBOL is not able to end the program properly if an exception occurs. Generally to handle an error, COBOL's subroutine calls are often used combined with the "harmful" [116] GOTO statements for dealing with errors. According to [105], the COBOL 2002 standards allow to handle common exceptions.

The two languages are different from programming paradigm viewpoint. AspectJ is aspect oriented; it inherits many object oriented programming advantages like simplicity and modularity, whereas COBOL with its rigid format and the PERFORM-THRU clause make the source code a spaghetti code, i.e. hard to understand the problem and more error prone.

In this criterion and based on the given information that we already gathered, we can say that both languages have support for some secure practices, but we can say that COBOL does not support all the practices that AspectJ support.

## 2.3.2 Web applications development

Web-applications are simple applications that run on the web. They are used through Web browsers and they are platform independent [106]. Java has its technologies and standards in order to develop web applications. The most common Java technologies used for the web are servlets. AspectJ could be integrated to Java servlets and the servlets servers run the AspectJ aspects fine [107]. Also aspects could be executed by JSP runners. AspectJ can be used in distributed web applications in order to implement the distribution and persistence which are common corsscuting concerns among distributed web applications [108].

We can see in [109], how AspectJ could be used in order to implement the monitoring of an application performance and failures by putting the performance-critical code with calls to record usage, timing, and errors logging in aspects.

COBOL was created before the appearance of Internet, so the earlier versions of COBOL did not have the ability to communicate through networks. But recently, with the creation of COBOL web servers like the "ICOBOL Web Server", it is possible to create CGI scripts using the ANSI COBOL syntax and to build web applications that handle HTTP requests [110].

COBOL generally runs on mainframes to provide back end functionality. It is possible to make COBOL talks the language of the web. However, this needs special middleware (adapters) and application servers (generally programmed in other languages) in order to make it accessible through browser, and this cannot be considered as web development using COBOL [111], moreover it is considered to be non secure.

## 2.3.3 Web services design and composition

A Web service is a unit of managed code that can be remotely called through HTTP request. It allows us to expose the functionalities of our existing code over the network; therefore other remote applications can use the functionalities of your program. It is noteworthy that web services rely on common standard which is XML in order to ensure the communication [112].

Java is a very good language for creating web services and there is a variety of tools that supports Java to do that (XML based Web services like the apache WSIF). As already mentioned, AspectJ could be integrated to any Java program, even those intended for web applications, and this is the same for Java program for web services (Java programs with aspects that are supposed to expose their functionalities over networks) [113]. AspectJ could be used in order to add profiling to web services without any impact on the web services. Moreover, AspectJ could add internal tasks (intended to be used internally only) to web services (e.g track invocation and handle encountered exceptions) while keeping the Java code modular. In [244] under "Demonstration: Using an Aspect Library" title, we can see how AspectJ is used for failures handling, statistics gathering and monitoring of a web service.

As mentioned above in the web applications development section, there are set of tools available in the market to make its COBOL functionalities available online allowing applications to interoperate with *COBOL* programs using XML interactions. COBOL85, the mostly used today, views XML as data (simply tagged characters) note that XML parsing and generating is a very important for web services. Nonetheless, even if this major drawback can be overcame with few extensions to the language, the problem that remains is that there are no native COBOL toolkits or libraries for creating and accessing Web Services. Third-party software is needed.

### 2.3.4 OO-based abstraction

In this section, we are going to talk about the most important concepts of object oriented paradigm.

Just like a class, an aspect could contain definitions of constructors, fields, methods, pointcuts and advices. Moreover an aspect could contain inner classes, aspects and interfaces. Aspects are similar to classes in their declaration. Hence we can say that the encapsulation of variables and methods in Java classes is done with AspectJ by the encapsulation of pointcuts and advices. Aspects hide the information by controlling visibility of the members. Here is a general form of a pointcut:

```
[visibility modifier] pointcut name(ParameterList): PointcutExpression;
```

 The programmer can define his own constructor, but the defined constructor should be no-argument constructor. Aspects could extend other aspects or implement interfaces, just like classes in Java do but with some restrictions.

- Extending an Aspect: inheritance in aspects is the same as classes, but the restriction is that only abstract classes could be extended, in other words concrete classes could not be extended. When an aspect extends another it inherits the fields, the methods, the pointcuts and the advices, but the child aspect is not allowed to extend or override inherited advices. The following figure shows the possibilities of inheritance :

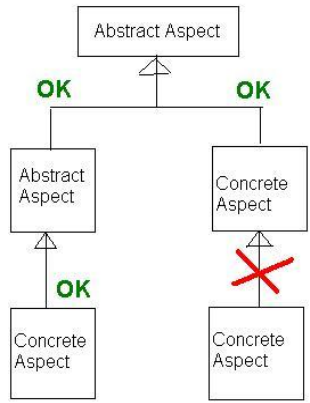

**Figure 10: Inheritence in AspectJ [121]**

- Implementing an interface: Interface implementation works exactly the same for aspects as for classes, but there are two restrictions which are the Serializable and Cloneable interfaces could not be implemented by an aspect.
- Inner aspects: an aspect could be defined inside a class or interface (it should be static). Generally inner aspects are used when we have concerns that crosscut a single class.

In the following listing we can see an abstract aspect of the multiple readers, single writer and another concrete aspect that extend the first, this example show the utility of inheritance in AspectJ and how the advices from the parent aspect apply to the joinpoints defined by the child aspect in addition to visibility control and encapsulation.

```
public abstract MultipleReaderSingleWriter perthis (readMethod() || writeMethod())
{
//declaring the lock
private ReadWriteLock rw = new WriterPreferenceReadWriteLock();
//two abstract named pointcuts that the extending aspect must redefine
protected abstract pointcut readMethod();
protected abstract pointcut writeMethod();

//advice before the execution of the readMethod that check if it is possible to
acquire the lock then it is possible to read
before() : readMethod(){
        try {rw.readLock().acquire();}
        catch (InterruptedException intEx)
        {
                throw new RuntimeException("Unable to obtain R Lock");
        }
}
// advice after reading
after() : readMethod() {
        rw.readLock().release();
}
// before writing the writer should acquire the write lock
before() : writeMethod() {
        try{
        rw.writeLock().acquire();
        }
        catch (InterruptedException intEx){
        throw new RuntimeException("Unable to obtain W lock");
        }
}
```

```
after() : writeMethod(){
      rw.writeLock.release();
}
declare warning: readMethod() && writeMethod() : "A method cannot be  read and write
method";
}
```
Parent aspect

```
public aspect MyResourceAccessProtection extends MultipleReadersSingleWriter
{
Protected pointcut readMethod() : execution(* MyResource.*(..)) && !writeMethod();
Protected pointcut writeMethod() : execution(* MyResource.update*(..));)
}
```
Child aspect

**Listing 1 Inheritance in AspectJ [121]**

In COBOL85, which is a procedural language, the concept of encapsulation does not apply. In the DATA DIVISION- WORKING STORAGE SECTION, COBOL defines the names that the program will use to refer to the data that it manipulates, the declared data is global and it is accessible to all the procedures of the program. In the new COBOL, there is now a "LOCAL STORAGE SECTION" so that each function, method or program can have its own data rather than using global variables throughout a program [119].

For object oriented COBOL the abstraction is well respected. The procedures could be named and also indicate the required parameters for their execution and what is the type of the returned value. Here is general form of a procedure PROCEDURE DIVISION [USING dataname1] [RETURNING dataname2].

Also, object oriented COBOL introduced the visibility modifiers; the following is an example about that: class-id. B data is private.

We can see in the Hello World example of Artur Reimann [120] how inheritance is performed using Fujitsu COBOL2000

In listing 1, we see a program written with object oriented COBOL from which we can see how object oriented abstraction is done in COBOL through the use of the keywords using and returning and method naming. Also the listing shows encapsulation in OO COBOL.

```
Identification Division.
*class name DemoCoo
Class-Id. DemoCOO.
Environment Division.
Configuration Section.
Repository.
Class BaseClasses.
Identification Division.
*definition of the class data
Object.
Data Division.
Working-Storage Section.
*string msg
01 msg pic x(30).
Procedure Division.
Identification Division.
*first method which simply display a message
```

```
Method-Id. MethodOne.
Procedure Division.
Display msg.
End Method MethodOne.
*end of definition of the first method
Identification Division.
*second method, the setter method
Method-Id. setMsg.
Data Division.
Linkage Section.
*string in-msg, this string is local, only the second method can use it
01 in-msg pic x(30).
*the method signature
Procedure Division using in-msg.
Move in-msg to msg.
End Method setMsg.
*end of definition of second method
End Object.
End Class DemoCOO.
***###Main Program###
Identification Division.
        Program-Id. Client.
Environment Division.
Configuration Section.
Repository.
Class DemoCOO.
Data Division.
Working-Storage Section.
01 DC usage object reference DemoCOO.
Procedure Division.
Invoke DemoCOO "new" returning DC
Invoke DC "setMsg" using by content "This is a message when MethodOne of DC object
call"
Invoke DC "MethodOne"
        Exit Program.
End Program Client.
```

Listing 2: Example of a simple program written with Object Oriented Cobol [120]

The listing 1 is a composed of two classes, first class with a setter method setMessage method and a display method called methodOne, the second class is the main which declares an object DemoCOO "01 DC usage object reference DemoCOO." and initialize it "Invoke DemoCOO "new" returning DC" then call the setMsg method of the object DC with a string as a parameter and finally call MethodOne of the DC object. We can notice here, OO COBOL is very complicated to understand if compared by the other imperative version even if it use additional English word in its expressions.

### 2.3.5  Aspect Orientation

AOP is a programming paradigm that aims to improve modularity of objects through encapsulating concerns that are spanned across multiple units into aspects. It is important to note that aspect programming paradigm while it builds on top of other programming paradigms (object oriented, procedural and functional), it does not supplant them [245]. AOP is based on three main concepts: aspect, pointcut and advice.

AspectJ is a powerful aspect oriented language for Java, it introduces the AOP concepts through new keywords. An aspect is like class in Java, it encapsulates pointcuts and the advices. The quantification of a join point is implemented using the keyword pointcut. A point cut in AspecJ is expressed as follow:

```
pointcut    pointcutName([parameters]): body of the pointcut
```

The pointcut body may contain other keywords, like:

- `call/execution(<Method signature>):` to intercept the call of a method/constructor or their execution
- `set/get(<Field signature>):` to intercept the access or update of a field
- `cflow(<Pointcut>):` to pick out each join point in the control flow
- `args(Type or Id, ...)` :to pick out each join point where the arguments are instances of the appropriate type

The advice concept is implemented using three keywords:

`before:` before proceeding at join point

`after:` returning from join point

`around:` on arrival at a join point, it gets explicit control over the executing method.

Other keywords could be associated with the advice keywords in order to make them more specific to certain situations, like `after returning or after throwing`.

The listing 3 shows an aspect implemented with AspectJ in which there is a use of three types of advice, and in which advice a pointcut is embedded.

The listing is composed of an interface `FigureElement` that two classes `Line` and `Point` implement.

There are three aspects: `PointBoundsPreCondition`, `PointBoundsPostCondition` and `PointBoundEnforcement`.

```
aspect PointBoundsEnforcement {

      private static final int MIN_X = 0;
      private static final int MAX_X = 15;
      private static final int MIN_Y = 0;
      private static final int MAX_Y = 15;
      //the around is executed whenever the method setX is called and the new X
parameter is passed to the clip method which verify if X is between 0 and 15, if it is
not so it is set to 0 or 15
         void around(int newX):
            call(void Point.setX(int)) && args(newX) {
            proceed( clip(newX, MIN_X, MAX_X) );
            System.out.println("Around Advice: point X checked, and it is valid");
         }
      //the around is executed whenever the method setY is called and the new Y
parameter is passed to the clip method which verify if Y is between 0 and 15, if it is
not so it is set to 0 or 15

         void around(int newY):
            call(void Point.setY(int)) && args(newY) {
          proceed( clip(newY, MIN_Y, MAX_Y) );
```

```
              System.out.println("Around Advice: point Y checked, and it is valid");
         }
// we can see in this example how the aspect could contain methods used by the advices
        private int clip(int val, int min, int max) {
            return Math.max(min, Math.min(max, val));
        }
    }
// the rest of the listing is available in appendix
```

**Listing 3. Example of using aspect to control the creation of point [246]**

In the previous listing, we can see how the verifications of the value entered by the user and the enforcement of value could be put in aspect instead of having them in the classes.

Crosscutting concerns also existed in old software systems written in classic COBOL. In the PROCEDURE DIVISION of a COBOL program there is a section called DECLARATIVES in which pointcuts and advices could be implemented. The following code shows a pointcut and an advice written in COBOL to handle potential error regarding a file access

```
DECLARATIVES.
HANDLE-F0815-ASPECT SECTION.
USE AFTER ERROR ON FILE-F1.
HANDLE-F0815-ADVICE.
MOVE "F0815" TO PANIC-RESOURCE.
MOVE "FILE ERROR" TO PANIC-CATEGORY.
MOVE FILE-STATUS TO PANIC-CODE.
GO TO PANIC-STOP.
END DECLARATIVES.
```

**Listing 4. A pointcut and an advice in COBOL85 [114]**

So, can say that COBOL is one of the oldest languages that support aspect orientation, but that was accidental [114]. That is why the aspect orientation of COBOL is very limited, especially pointcuts capabilities. For example in COBOL we cannot advise subprogram calls or field access. AspectCobol is a prototypical implementation of the full aspect oriented paradigm for classic COBOL in order to enforce its support for AOP [114].

## 2.3.6 Reflection:

In programming languages, reflection is the capabilities of a program to examine and modify itself during run time. For example in object oriented languages, reflection allow the programmer to find out information about a class (access modifiers, superclass, fields, constructors and methods), also it gives the possibility to modify source code at runtime. Reflection is useful for programs that process programs. AspectJ has a support for reflection, for example, the thisJoinPoint instruction return reflective information about the current joint point, so the advice know exactly what kind of join point it is executing under, therefore the advice can modify its behavior according to the reflective information provided from the thisJointPoint statement.

Listing 5 shows how an AspectJ aspect could return reflective information about a JAVA program.

```
package com.demo;

public class Demo {

    static Demo d;

    void run(){
        d = new Demo();
        d.method1(1,"whatever string ");
        System.out.println(d.method2(new Integer(3)));
    }

    void method1(int i, String o){
        System.out.println("Demo.method1(" + i + ", " + o + ")\n");
    }

    String method2 (Integer j){
        System.out.println("Demo.method2(" + j + ")\n");
        return "Demo.method2(" + j  + ")";
    }

    public static void main(String[] args){
        new Demo().run();
    }
}

package com.demo;

import org.aspectj.lang.JoinPoint;
import org.aspectj.lang.reflect.CodeSignature;

aspect GetInfo {

        static final void println(String s){ System.out.println(s); }
// this is a to intercept any execution within the Demo object of whatever object
        pointcut demoExecs(): within(Demo) && execution(* *(..));
// this is the advice during execution of a method
        Object around(): demoExecs() {
      // the case if the executing method is the main  method
        if (!thisJoinPointStaticPart.getSignature().getName().equalsIgnoreCase("main"))
                {
// this line return the name of the executing class
            println("Intercepted message: " +
                thisJoinPointStaticPart.getSignature().getName());
            println("in class: " +

thisJoinPointStaticPart.getSignature().getDeclaringType().getName());
// this call to the printPramas method which print the information about the executing
// method. Information are the name of the method the value of the parameters and their value.
            printParameters(thisJoinPoint);
            println("original method in running: \n" );
                }
            return proceed();
          }

        static private void printParameters(JoinPoint jp) {
            println("Arguments: " );
            Object[] args = jp.getArgs();
            String[] names = ((CodeSignature)jp.getSignature()).getParameterNames();
             Class[] types = ((CodeSignature)jp.getSignature()).getParameterTypes();
            for (int i = 0; i < args.length; i++) {
```

```
                    println("   " + i + ". " + names[i] +
                        " : " +              types[i].getName() +
                        " = " +              args[i]);
                }
            }
        }
```

Moreover, the AspectJ API contains the package org.aspectj.lang.reflect which could be used to take advantage of the reflective aspect of AspectJ.

The classic version of COBOL does not have any support for reflection.

## 2.3.7  Declarative and Functional Programming:

Annotations are like meta-tags that you can add to your code and apply them to package declarations, type declarations, constructors, methods, fields, parameters, and variables. Annotation based development leads to declarative programming style [123]. As we know JDK5support annotation in order to make development easier. Also AspectJ has the capabilities to support annotation and it has a package implemented for that purpose, which is org.aspectj.lang.annotation in which there a set of classes for annotations. For instance, AspectJ could annotate a class using the @Aspect annotation [121]. Here is an example:

```
@Aspect
Public class world{}
```

Is equivalent to :

```
Public aspect world {}
```

There are also pointcut annotation which is @Pointcut and others for advices and intertype declaration. Here is a good example for using @Pointcut annotation:

```
import project.pack1.Class1;

public aspect Foo {
        // any call of Class1 methods
        pointcut Class1Operation() : call(* class1.*(..));
        // any call of methods in classes under the pack1 package
        pointcut anyPack1Call() : call(* project.pack1..*(..));
}
```

Using the annotation style this is equivalent to :

```
@Aspect
public class Foo {
        @Pointcut("call(* project.pack1.Class1.*(..))")
        void Class1Operation() {}

        @Pointcut("call(* project.pack1..*(..))")
        void anyPack1Call() {}
}
```

So we can see that AspectJ has some support for declarative programming since it supports annotations.

Based on the searching and investigation, COBOL is an imperative language and it has no capabilities to perform declarative or functional programming.

Even if AspectJ is a reflective language, we can not find in the littreture any proof that this feature is used in functional programming with AspectJ. We can say that AspectJ has some support for declarative programming through annotations while COBOL does not.

## 2.3.8  Batch scripting:

First, scripting is programming within a program in order to automate certain functionalities. Scripting languages does not need compilation. Generally, batch programs are written to automate a set of tasks within an operating system, or within software like auto filling forms in the browser using JavaScript. The opposite of a batch job is the interactive processing, in which users enter individual commands to be processed immediately [115].

Java provides an API for scripting which integrate virtually any scripting language with the Java code. Crosscutting concerns may exist in a batch scripting program, like the exception handling in the following scripting with Java:

```
import Java.io.*;
import Javax.script.*;

public class App
{
    public static void main(String[] args)
    {
        try
        {
            ScriptEngine engine =
                new ScriptEngineManager().getEngineByName("Javascript");
            for (String arg : args)
            {
                Bindings bindings = new SimpleBindings();
                bindings.put("author", new Person("Ted", "Neward", 39));
                bindings.put("title", "5 Things You Didn't Know");

                FileReader fr = new FileReader(arg);
                engine.eval(fr, bindings);
            }
        }
        catch(IOException ioEx)
        {
            ioEx.printStackTrace();
        }
        catch(ScriptException scrEx)
        {
            scrEx.printStackTrace();
        }
    }
}
```

**Listing 5. Example of use of the scripting library of Java [122]**

So it is obvious that AspectJ could be used to improve the modularity of this code, for example by removing the exception handling code and putting them in an aspect. This is an indirect way to use AspectJ in batch scripting, but I think that when we try to automate work there is no need to think about modularity of the script or even think in term of objects.

COBOL could be a powerful batch scripting language if the intended task of the script is to manipulate filesystems or accessing VSAM datasets, because COBOL is good at file-oriented application.

### 2.3.9  User interface prototyping and design:

User interface prototyping is a technique used to produce high fidelity prototype (a prototype which is a potential foundation from which to continue developing the system, generally with good quality code) or low fidelity prototype (to be thrown away later) in order to show the future user interfaces to the stakeholders and also to validate the user-interface design [247]. Generally, to produce prototypical user interface specialized tools are used (especially for low fidelity prototypes) for example visual basic is a good tool for prototyping.

AspectJ could be used in order to deal with the crosscutting concerns in Java graphical user interface applications, but I think that Java Swing is complicated and it is not a good idea to implement prototypical user interface using Java Swing, unless a visual editor is used.

There are some commercial tools to use with COBOL in order to design GUI. For instance the COBOL sp2 is a graphical user interface tool for COBOL applications. The tools could be used for prototyping (just like visual basic). The next two figures show the COBOL sp2 tool and Visual Basic editor. They look similar:

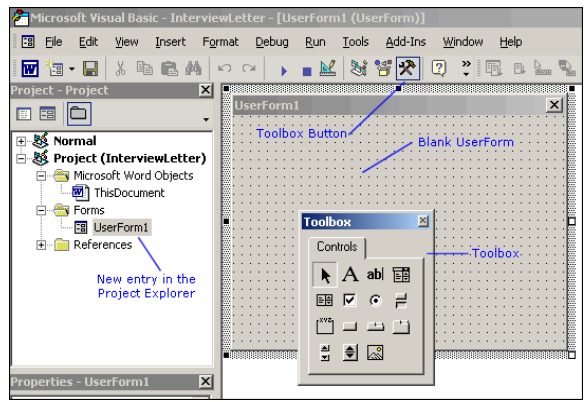

**Figure 11. Visual basic IDE.**

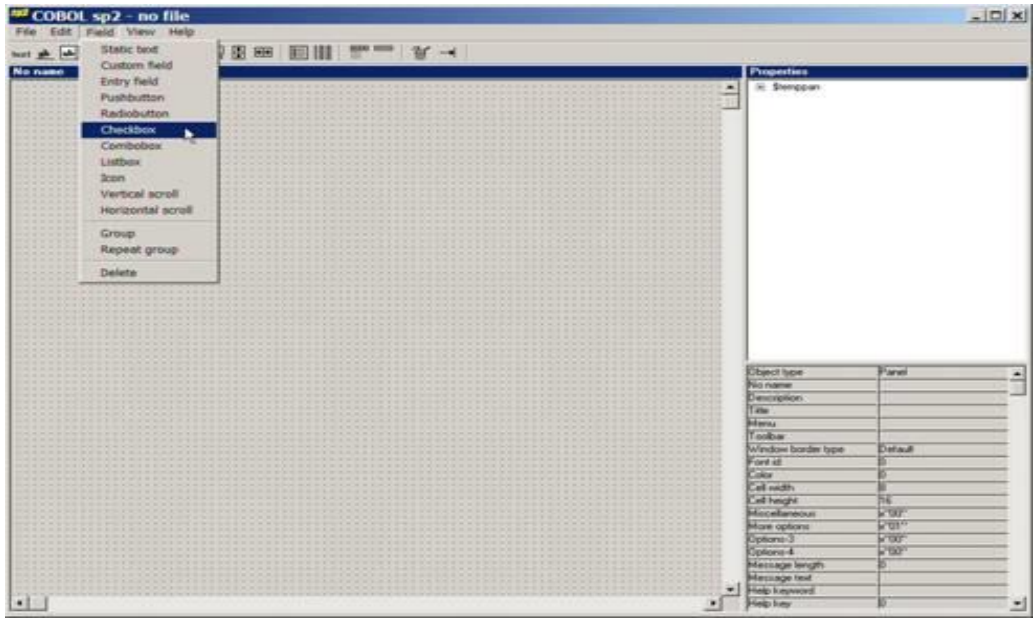

**Figuare 12. COBOL sp2 IDE.**

This tool and similar others could be useful for prototyping, but the shortcoming of these tools is portability. We can see that the vendors list the number of operating systems that support the tool, so we can conclude that some operating systems could not be supported.

But The COBOL language does not have a native library that could produce graphical user interface and the products available in the market are platform dependent. Generally COBOL is not supposed to be used to implement user interface prototypes (COBOL programs are generally run through the command line).

## 2.4  Ruby vs. PHP

Here we compare and contrast the specified criteria for Ruby and PHP.

### 2.4.1  Secure Programming Practices

The National Vulnerability Database (NVD) stores all vulnerabilities found in computer software. The overall proportion of PHP-related vulnerabilities on the database amounted to: 20% in 2004, 28% in 2005, 43% in 2006, 36% in 2007, 35% in 2008, and 30% in 2009 [83]. Most of these PHP-related vulnerabilities can be exploited remotely: they allow crackers to steal or destroy data from data sources linked to the Web server (such as an SQL database), send spam or contribute to Denial of Service attacks using malware, which itself can be installed on the vulnerable servers.

PHP application developers do not make a good job at securing their code: they do not care, or they do not know how, or they try but fail. PHP is often a self-taught first and only programming language for non professional developers, whose codes will stay online forever. Also, as PHP is easy and popular, more bugs are written with it [83].

Recognizing that programmers cannot be trusted, some languages include taint checking to detect automatically the lack of input validation which induces many issues. Such a feature is being developed for PHP, but its inclusion in a release has been rejected several times in the past [84].

Although the vulnerabilities of PHP and PHP based applications are numerous and of important concern, several articles and books have been published that expose techniques and practices to improve security in PHP programming: [85, 86, 87, and 88] among many others.

Between all the practices exposed on those articles the most noteworthy are:

- Validating user input.
- Preventing SQL injection.
- Preventing cross-site scripting.
- Preventing remote execution (attempts to inject PHP commands).
- Enforcing security for temporary files.
- Preventing session hijacking.

Going into detail on each of these practices is beyond the scope of this document; for more information, please refer to the book "Pro PHP Security" written by Chris Snyder and Michael Southwell [88]. Other topics in secure programming practices are maintaining a secure online environment in which to develop and run applications, and practicing secure operations.

Although taint checking[1] is supported by the Ruby programming language, and other facilities for validation, authentication, authorization, data protection, etc. [90]; it is also not without its own set of vulnerabilities [G58]. But comparing it to PHP, Ruby has much less reported vulnerabilities across all versions according to the NVD (as can be appreciated in Appendix F).

---

[1] Feature designed to increase security by preventing malicious users from executing commands on a host computer. Taint checks highlight specific security risks primarily associated with web sites which are attacked using techniques such as SQL injection or buffer overflow attack approaches.

All in all, there are also several articles for reinforcing security in Ruby as well as the Rails framework [90, 91, and 82]. Ruby on Rails has some clever helper methods, for example against SQL injection, but Rails also has vulnerabilities in some areas such as file permission, sessions, input validation, and escape output.

Among the recommendations given by the previous sated references we can find:

- Write more secure software.
- Data encryption.
- Authenticate users.
- Cryptography.
- Protect against session hijacking.
- Set countermeasures for Cross-Site Request Forgery (CSRF)[2]
- Redirection and file uploads with executable malicious code.
- Prevent against brute-forcing accounts.

As per the [Appendix F] it can be concluded that as of 2010 PHP has more reported vulnerabilities (1009) and Ruby only has (6). As a whole, PHP is a more insecure programming language than Ruby, but ultimately, the responsibility of writing secure applications falls upon the programmers. They have to be aware of the common (and not so common) threats and defend the applications against them. Is does not matter if, for example, Rails provides excellent facilities for accepting, rejecting, or sanitizing system input and output if the programmer is not conscious of them, and when they should be used.

### 2.4.2 Web Applications Development

As mentioned before, PHP is a general-purpose scripting language, but is especially suited for server-side web development. PHP generally runs on a web server and can be embedded into HTML; any PHP code in a requested file is executed by the PHP runtime, usually to create dynamic web page content. PHP can be deployed on most web servers, many operating systems and platforms (Windows, Linux, UNIX, Mac OS X, etc.), and can be used with many relational database management systems (MySQL, PostgreSQL, IBM DB2, SQLite, Oracle, etc.). Also, it is compatible with many Web servers like Apache, Microsoft Internet Information Server, Personal Web Server, Netscape and iPlanet servers, and many others.

According to the PHP Manual [66], the main goal of the language is to allow web developers to write dynamically generated web pages quickly, although much more can be do with it. So it is safe to say that Web development is possible (if not one of the most popular choices) in PHP.

As previously stated, many popular web applications are implemented using PHP. From CMS, Wiki engines, eCommerce solutions, shopping carts, and blogging platforms to name a few. So its popularity cannot be denied.

Another powerful feature offered to PHP programmers are the vast number of *Web application frameworks*; a software framework that is designed to support the development of dynamic websites, Web applications and Web services. The framework aims to alleviate the overhead associated with common activities performed in Web development. For example, many frameworks provide libraries for database access, templating frameworks and session management, and they often promote code reuse [67].

---

[2] *This attack method works by including malicious code or a link in a page that accesses a web application that the user is believed to have authenticated. If the session for that web application has not timed out, an attacker may execute unauthorized commands.*

Usually these frameworks support the MVC architecture and offer features such as security, database access and mapping, URL mapping, Web template system, Ajax, Web services among others. For PHP there are several of these frameworks that streamline the Web development; some of the most popular are: Yii, CakePHP, Zend, Symfony, and CodeIgniter.

In the example exposed in Listing 6 it can be appreciated a simple PHP script embedded in HTML.

```
<!DOCTYPE HTML PUBLIC "-//W3C//DTD HTML 4.01 Transitional//EN"
    "http://www.w3.org/TR/html4/loose.dtd">
<html>
    <head>
        <title>Example</title>
    </head>
    <body>
        <?php
            echo "Hi, I'm a PHP script!";
        ?>
    </body>
</html>
```

Listing 6: An introductory PHP example

Instead of lots of commands to output HTML (as seen in C or Perl), PHP pages contain HTML with embedded code that does "something" (in this case, output "Hi, I'm a PHP script!"). The PHP code is enclosed in special start and end processing instructions **<?php** and **?>** that allows to jump into and out of "PHP mode."

What distinguishes PHP from something like client-side JavaScript is that the code is executed on the server, generating HTML which is then sent to the client. The client would receive the results of running that script, but would not know what the underlying code was. The Web server can even be configured to process all the HTML files in a Web application with PHP [61].

PHP is mainly focused on server-side scripting, so it can do anything any other CGI program can do, such as collect form data, generate dynamic page content, or send and receive cookies.

In the [Appendix A] can be found an example of HTML form processing using PHP. Although is very simplistic and basic it helps to illustrate the power of PHP for Web development which is only the tip of the iceberg.

Web development in Ruby is done mostly (if not exclusively) using the Web application framework *Ruby on Rails*. Rails is an open source Ruby framework for developing database-backed web applications. Rails supporters ensure that developing Web applications in the framework is at least ten times faster than using the typical Java framework, and all this without compromising the quality of the application [68].

Part of the reason is the Ruby programming language. Many things that are very simple to do in Ruby are not even possible in most other languages. Rails takes full advantage of this. The rest of the answer is in two of Rail's guiding principles: *less software* and *convention over configuration*.

*Less software* means the programmer write fewer lines of code to implement the application. Keeping the code small means faster development and fewer bugs, which makes the code easier to understand, maintain, and enhance.

*Convention over configuration* means an end to verbose XML configuration files--there aren't any in Rails! Instead of configuration files, a Rails application uses a few simple programming conventions that allow it to figure out everything through *reflection* and discovery. The application code and the running database already contain everything that Rails needs to know.

Ruby on Rails is separated into various packages, namely *Action Controller* (processes incoming requests to a Rails application, extracts parameters, and dispatches them to the intended action), *Action View* (manages rendering templates, including nested and partial templates, and includes built-in AJAX support), *Active Record* (an object-relational mapping system for database access), *Action Mailer* (a framework for sending emails based on flexible templates, or to receive and process incoming email) *Active Resource* (for managing the connection between business objects an RESTful Web services), Railties (is the core Rails code that builds new Rails applications and glues the various frameworks together in any Rails application), and Active Support (is an extensive collection of utility classes and standard Ruby library extensions that are used in the Rails, both by the core code and by the applications) [40].

Another powerful tool provided by Rails is the scaffolding. Rails *scaffolding* is a quick way to generate some of the major pieces of an application. If there is need to create the models, views, and controllers for a new resource in a single operation, scaffolding is the tool for the job. Scaffold allows the programmer to generate code that permits a user to create, read, update, and delete (CRUD) data in the database [40]. In Figure 1 it can be appreciated how easy it is to create a Ticket system using Rails scaffolding.

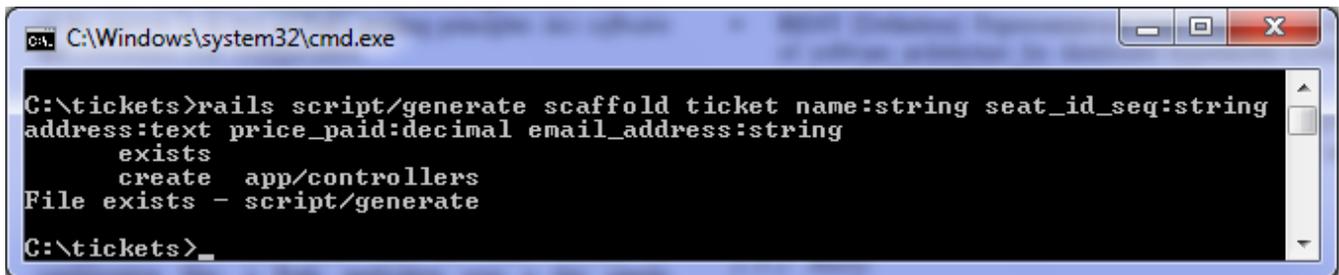

**Figure 13: Creating a ticket system with Rails scaffolding**

With the command from above, Rails creates the Model, the Views and the Controllers for the application, and all this without having to type a single line of Ruby code.

Although not as popular as PHP, Rails offers a powerful alternative to Web development, and its popularity is increasing every day. For example, CMS have been developed by using the framework, like Radiant[3] and Mephisto[4].

In the [Appendix B] can be found some files examples for a basic online advertisement system built using Rails scaffolding for the purpose of illustrate how the code looks like for Ruby Web development.

I would say PHP has the advantage here, although Ruby with the Rails framework is a strong contender and is gaining popularity, PHP being focused to Web development and having been on the market for a longer period of time gives it the edge on this area. The high number of frameworks, online communities, tutorials, books, companies programming in PHP, software systems, etc. is a testament that PHP is one of the most popular and widespread choices for developing Web applications.

### 2.4.3  Web Services Design and Composition

PHP sports various extensions for Web services functionalities, among them are OAuth (Open Authorization), SCA (Service Component Architecture), SOAP, XML-RPC, and REST.

---

- **OAuth:** is an authorization protocol built on top of HTTP which allows applications to securely access data without having to store usernames and passwords. In other words, it allows users to share their private resources (*e.g.* photos, videos, contact lists) stored on one site with another site without having to hand out their username and password.
- **SCA:** is an extension for PHP that makes it possible for a PHP programmer to write reusable components, which can be called in a variety of ways, with an identical interface and with a minimum of fuss. At present components can call each other either locally or via Web services. The SCA for PHP programming model can be extended to support a number of service types, such as REST and Atompub. However, Web services (more accurately, WSDL defined, SOAP/HTTP services), are the only type currently specified. An example SCA component for PHP is depicted in Listing 7.

```php
<?php

include "SCA/SCA.php";

/**
 * Calculate a stock price for a given ticker symbol in a given currency.
 *
 * @service
 * @binding.soap
 */
class ConvertedStockQuote {

    /**
     * The currency exchange rate service to use.
     *
     * @reference
     * @binding.php ../ExchangeRate/ExchangeRate.php
     */
    public $exchange_rate;

    /**
     * The stock quote service to use.
     *
     * @reference
     * @binding.soap ../StockQuote/StockQuote.wsdl
     */
    public $stock_quote;

    /**
     * Get a stock quote for a given ticker symbol in a given currency.
     *
     * @param string $ticker The ticker symbol.
     * @param string $currency What currency to convert the value to.
     * @return float The stock value is the target currency.
     */
    function getQuote($ticker, $currency)
    {
        $quote  = $this->stock_quote->getQuote($ticker);
        $rate   = $this->exchange_rate->getRate($currency);
        return  $rate * $quote;
    }
}
?>
```

Listing 7: SCA example for PHP

PHP has implementations of SOAP (*e.g.* PEAR::SOAP[5], PHP-SOAP[6], NuSOAP[7]). In the [Appendix A] it is exposed an extensive example of a Web service for PHP using SOAP client for the purposes of illustration on Web services design and composition. It is noteworthy to mention that SOAP is thought as slow to implement and hard to understand, and it is losing terrain to the REST architecture, the REST based API is easy to understand with simple HTTP request/response messages in XML format.

As it was mentioned above, Web application frameworks usually (if not always) offer features to support Web services facilities. Symfony, Zend Framework, CakePHP, CodeIgniter, and Sapphire offer implementations for REST Web services.

In conclusion we can say that the PHP library offers many alternatives for implementing Web services in an application, which is expected being a server-side scripting language. On the other hand, if a PHP framework is used for Web development; additional alternatives are available to the programmers.

Web services in Ruby on Rails are implemented using the REST architecture, specifically in the Active Resource component of Rails.

Rails makes it easy to offer multiple different content types, or representations, of a particular resource through a single controller action. Being able offer multiple representations of a resource through a single controller action makes adding support for non browser based HTML clients as simple as adding a few lines of code that returns the resource in the representation the client requested. The controller layout and logic of the application is preserved with the added benefit of interoperability with new clients.

Active Resource provides the tools to quickly and easily consume REST based web services conforming to the Rails RESTful URI structure and protocol conventions. Active Resource automatically maps the response from any conforming service to rich Ruby objects. Active Resource also provides all the lifecycle methods needed to easily find, create, update, and delete resources without having to write any code [73].

In the [Appendix B] a simple example of Web services for Ruby on Rails. As a Web application framework, it is expected of Rails to provide features to create and consume Web services, this are implemented using the REST architecture based on Roy Fielding's doctoral thesis, "Architectural Styles and the Design of Network-based Software Architectures" [74].

Then again, PHP takes the lead is this criteria. Simply because it offers more choices when deciding which Web service protocol should one use in developing a Web application. Given that Ruby on Rails offers RESTful Web services, PHP has implementations of SOAP, XML-RCP and others.

### 2.4.4 Object Oriented Based Abstraction

Starting with PHP 5, the object model was rewritten to allow for better performance and more features. This was a major change from PHP 4. PHP 5 has a full object model. Among the features in PHP 5 are the inclusions of visibility, abstract and final classes and methods, additional magic methods, interfaces, cloning and typehinting.

---

PHP treats objects in the same way as references or handles, meaning that each variable contains an object reference rather than a copy of the entire object [55].

PHP 5 introduces abstract classes and methods. It is not allowed to create an instance of a class that has been defined as abstract. Any class that contains at least one abstract method must also be abstract. Methods defined as abstract simply declare the method's signature they cannot define the implementation.

When inheriting from an abstract class, all methods marked abstract in the parent's class declaration must be defined by the child; additionally, these methods must be defined with the same (or a less restricted) visibility. For example, if the abstract method is defined as protected, the function implementation must be defined as either protected or public, but not private [55].

Listing 8 shows a basic example of the usage of abstract classes in PHP.

```php
<?php
abstract class AbstractClass
{
    // Force Extending class to define this method
    abstract protected function getValue();
    abstract protected function prefixValue($prefix);

    // Common method
    public function printOut() {
        print $this->getValue() . "\n";
    }
}

class ConcreteClass1 extends AbstractClass
{
    protected function getValue() {
        return "ConcreteClass1";
    }

    public function prefixValue($prefix) {
        return "{$prefix}ConcreteClass1";
    }
}

class ConcreteClass2 extends AbstractClass
{
    public function getValue() {
        return "ConcreteClass2";
    }

    public function prefixValue($prefix) {
        return "{$prefix}ConcreteClass2";
    }
}

$class1 = new ConcreteClass1;
$class1->printOut();
echo $class1->prefixValue('FOO_') ."\n";

$class2 = new ConcreteClass2;
$class2->printOut();
echo $class2->prefixValue('FOO_') ."\n";
?>
```

Listing 8: Abstract class example in PHP [55]

Inheritance is a well-established programming principle, and PHP makes use of this principle in its object model. This principle will affect the way many classes and objects relate to one another. For example, when you extend a class, the subclass inherits all of the public and protected methods from the parent class. Unless a class overrides those methods, they will retain their original functionality. This is useful for defining and *abstracting functionality*, and permits the implementation of additional functionality in similar objects without the need to re-implement all of the shared functionality [55]. Listing 9 exposes an example.

```php
<?php
class foo
{
    public function printItem($string)
    {
        echo 'Foo: ' . $string . PHP_EOL;
    }

    public function printPHP()
    {
        echo 'PHP is great.' . PHP_EOL;
    }
}

class bar extends foo
{
    public function printItem($string)
    {
        echo 'Bar: ' . $string . PHP_EOL;
    }
}

$foo = new foo();
$bar = new bar();
$foo->printItem('baz'); // Output: 'Foo: baz'
$foo->printPHP();        // Output: 'PHP is great'
$bar->printItem('baz'); // Output: 'Bar: baz'
$bar->printPHP();        // Output: 'PHP is great'
?>
```

Listing 9: Inheritance example in PHP [55]

Composition is an important concept in PHP. It occurs when an object creates another object(s); that is, the first object completely possesses the second or more objects. [Appendix G] shows a rather basic example that illustrates how one object composes another one.

Ruby is a fully Object-oriented programming language, and thus, it provides extensive support for developing Object-oriented applications along with the features to cover the concepts of OOP. Ruby is considered a *pure* Object-Oriented language, because everything appears, to Ruby, as an object. All Ruby data consists of objects that are instances of some class. Even a class itself is an *object* that is an instance of the *Class* class [94]. Listing 10, 11, 12, and 13 show examples of implementations of different Object-oriented concepts that support abstraction in Ruby (inheritance, polymorphism, and composition).

```ruby
class BankAccount
    def interest_rate
        @@interest_rate = 0.2
```

```ruby
        end

    def accountNumber
            @accountNumber
    end

    def accountNumber=( value )
            @accountNumber = value
    end

    def accountName
            @accountName
    end

    def accountName=( value )
            @accountName = value
    end
end
```

**Listing 10: Class example with *class variable* for Ruby**

```ruby
class NewBankAccount < BankAccount

    def customerPhone
            @customerPhone
    end

    def customerPhone =( value )
            @customerPhone = value
    end

end
```

**Listing 11: Inheritance example in Ruby**

```ruby
class Engine
    attr_reader :horsepower, :litres
end

class Gearbox
    attr_reader :manufacturer, :model_no
end

class Car
    def initialize(engine, gearbox)
        raise "Invalid Engine Object" if !engine.kind_of(Engine)
        raise "Invalid Gearbox Object" if !gearbox.kind_of(Gearbox)
        @engine = engine
        @gearbox = gearbox
    end
end

car = Car.new(Engine.new, Gearbox.new)
```

**Listing 12: Composition example in Ruby**

```ruby
class Point
    # Define accessor methods
    attr_reader :x, :y

    def initialize(x,y)
        @x,@y = x, y
    end
```

```
    # Define + to do vector addition
    def +(other)
        Point.new(@x + other.x, @y + other.y)
    end

    # Define unary minus to negate x and y
    def -@
        Point.new(-@x, -@y)
    end

    # To perform scalar multiplication
    def *(scalar)
        Point.new(@x*scalar, @y*scalar)
    end
end
```

**Listing 13: Operator overloading example in Ruby**

PHP and Ruby stand on equal ground in this criterion. Both programming languages have implementations of an Object-oriented model from the design level. This means that concepts like classes, methods, objects, encapsulation, polymorphism, inheritance, abstraction, and message passing are implemented into the language structure. The only difference between PHP and Ruby is that PHP does not follow the *pure* Object-Oriented approach like Ruby does. But for the purposes of OO Abstraction both languages offer facilities of equally ease of use.

### 2.4.5 Reflection

PHP 5 comes with a complete reflection API that adds the ability to *reverse-engineer* classes, interfaces, functions, methods and extensions. Additionally, the reflection API offers ways to retrieve doc comments for functions, classes and methods [69].

PHP is one of the most popular programming languages for web applications and provides a comprehensive set of meta programming and reflection facilities. The default configuration of PHP includes a set of functions and an object-oriented reflection API. Additional features are provided by extensions to the virtual machine of PHP.

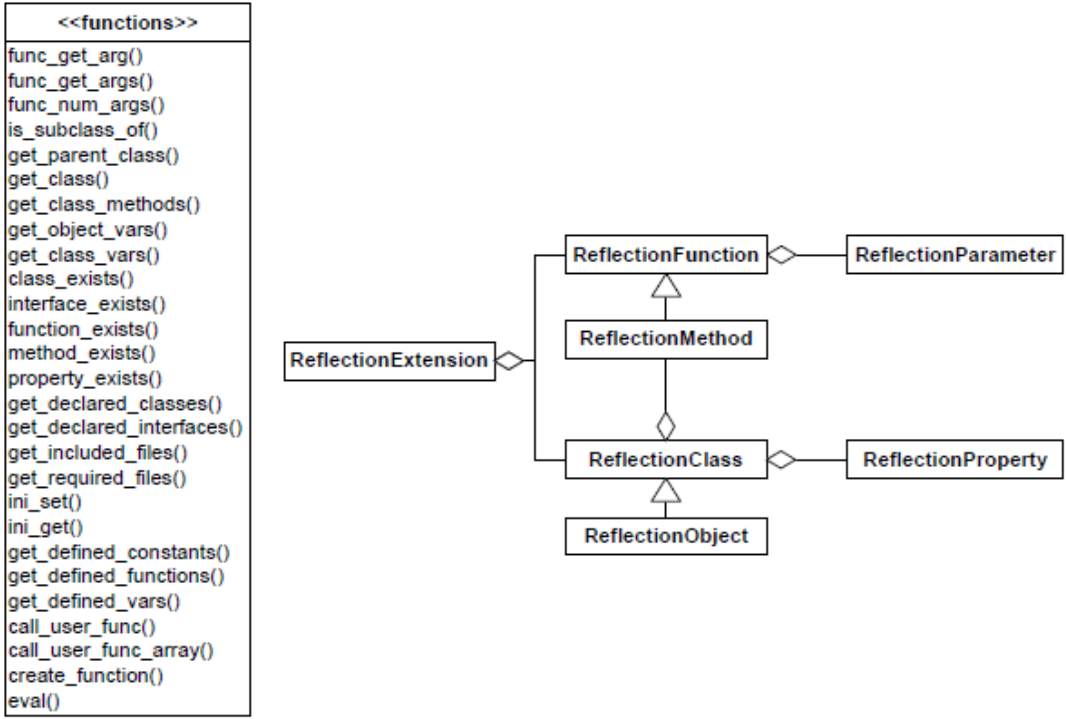

**Figure 14: Overview of PHP5 Reflection API**

As of PHP5, a reflection API has been included in the standard PHP distribution. It provides an API to obtain information about classes, including all of their methods and properties, without the need to parse any source code. Simple functions, runtime objects and PHP extensions can be inspected as well. Introspection is not the only feature of the reflection API. Furthermore, it is also possible to modify properties of objects, invoke arbitrary methods or functions, and instantiate objects, i.e. doing some basic intercession [70]. An overview of the features available is shown in Figure 2.

The procedural part has grown since PHP4 and includes basic functions like **class_exists()** and **get_declared_classes()** to check which classes are available in the current process instance. Also, there are functions to check which script files have been already loaded. **ini_get()** and **ini_set()** can be used to adapt configuration parameters for specific subsystems of the PHP execution environment at runtime [70].

As already mentioned, PHP provides different mechanisms to invoke arbitrary functions at runtime, e.g., to implement callbacks. The language directly supports a concept similar to function pointers using a variable with a function name as its value. The notation looks like **$functionName($a, $b)**, where the value of **$functionName** defines the actual function to be called. More flexible functions are **call_user_func()** and **call_user_func_array()** which enable the user to use object methods as callbacks and arbitrary number of parameters.

Callbacks are not the only way to execute user code. As in many scripting languages, PHP has a function called **eval()** which takes a string and tries to interpret and execute this string as PHP code. There are almost no restrictions on the code to be evaluated. Thus, it is possible to generate code, including function and class definitions at runtime, and execute it using **eval()**.

In Listing 14 there is an example of the usage of the ReflectionFunction class in PHP, which lets reverse-engineer functions.

```php
<?php
class ReflectionFunction extends ReflectionFunctionAbstract implements Reflector
{
    final private __clone()
    public void __construct(string name)
    public string __toString()
    public static string export(string name, bool return)
    public string getName()
    public bool isInternal()
    public bool isDisabled()
    public mixed getClosure() /* As of PHP 5.3.0 */
    public bool isUserDefined()
    public string getFileName()
    public int getStartLine()
    public int getEndLine()
    public string getDocComment()
    public array getStaticVariables()
    public mixed invoke([mixed args [, ...]])
    public mixed invokeArgs(array args)
    public bool returnsReference()
    public ReflectionParameter[] getParameters()
    public int getNumberOfParameters()
    public int getNumberOfRequiredParameters()
}
?>

<?php
/**
 * A simple counter
 *
 * @return    int
 */
function counter()
{
    static $c = 0;
    return $c++;
}

// Create an instance of the ReflectionFunction class
$func = new ReflectionFunction('counter');

// Print out basic information
printf(
    "===> The %s function '%s'\n".
    "     declared in %s\n".
    "     lines %d to %d\n",
    $func->isInternal() ? 'internal' : 'user-defined',
    $func->getName(),
    $func->getFileName(),
    $func->getStartLine(),
    $func->getEndline()
);

// Print documentation comment
printf("---> Documentation:\n %s\n", var_export($func->getDocComment(), 1));
```

```
// Print static variables if existant
if ($statics = $func->getStaticVariables())
{
    printf("---> Static variables: %s\n", var_export($statics, 1));
}

// Invoke the function
printf("---> Invokation results in: ");
var_dump($func->invoke());

// you may prefer to use the export() method
echo "\nReflectionFunction::export() results:\n";
echo ReflectionFunction::export('counter');
?>
```

**Listing 14: Using the ReflectionFunction class in PHP [71]**

As it can be appreciated, PHP offers a comprehensive reflection API, which is useful for the implementation of development tools, as well as advanced programming frameworks, analysis and visualization of software tools, compilers, IDEs, etc. The reflection facilities for PHP are extensive, and covering them all in this paper gets out of the scope.

Likewise, Ruby also comes with an extensive reflection API. One of the many advantages of dynamic scripting languages such as Ruby is the ability to *introspect*---to examine aspects of the program from within the program itself. Reflection in Ruby can be used to discover about a program what object it contains, the current class hierarchy, the contents and behavior of objects, information on methods among others [72]. Reflection is used to examine parts of our programs that aren't normally visible from where we stand, usually this means at runtime. Other uses for reflection in Ruby are tracing a program's execution, system hooks (a *hook* is a technique that lets you trap some Ruby event, such as object creation), etc.

For example, the method **ObjectSpace::each_object** lets traverse all the living objects in the program. In addition, unlike static languages, where a variable's type determines its class, and hence the methods it supports, Ruby supports liberated objects. You really cannot tell exactly what an object can do until you look under its hood. One example for such is exposed in Listing 15.

**For instance, we can get a list of all the methods to which an object will respond.**
```
r = 1..10 # Create a Range object
list = r.methods list.length » 60
list[0..3] » ["size", "end", "length", "exclude_end?"]
```

**Or, we can check to see if an object supports a particular method.**
```
r.respond_to?("frozen?")     » true
r.respond_to?("hasKey")      » false
"me".respond_to?("==")       » true
```

**We can determine our object's class and its unique object id, and test its relationship to other classes.**
```
num = 1
num.id        » 3
num.class     » Fixnum
```

```
num.kind_of? Fixnum        » true
num.kind_of? Numeric       » true
num.instance_of? Fixnum    » true
num.instance_of? Numeric   » false
```

**Listing 15: Examples of reflection over Ruby objects [72]**

Looking and examining objects is but the first part of reflection. To look at classes (the methods and attributes they contain) can be achieved with methods like `Class#superclass` which returns the parent of any particular class, or `Module#ancestors` which lists both superclasses and mixed-in modules (see Listing 16).

```
klass = Fixnum
begin
  print klass
  klass = klass.superclass
  print " < " if klass
end while klass
puts
p Fixnum.ancestors

Which produces:
Fixnum < Integer < Numeric < Object
[Fixnum, Integer, Precision, Numeric, Comparable, Object, Kernel]
```

**Listing 16: Examples of reflection over Ruby classes [72]**

To create a complete class hierarchy these two methods can be combined with `ObjectSpace::each_object` to iterate over all the classes in the program.

Reflection in Ruby also allows calling methods dynamically. For example, in Ruby a programmer can Stick all the command functions into a class, create an instance of that class, and ask that object to execute a method called the same name as the command string: `commands.send(commandString)`; and it does so dynamically, which means that it will find new methods added at runtime just as easy, there is no necessity to write especial command classes for `send` as it works with any object.

Another way of invoking methods dynamically uses `Method` objects. A `Method` object is like a `Proc` object: it represents a chunk of code and a context in which it executes. In this case, the code is the body of the method, and the context is the object that created the method. Once we have our `Method` object, we can execute it sometime later by sending it the message `call` (see Listing 17).

```
trane = "John Coltrane".method(:length)
miles = "Miles Davis".method("sub")
trane.call                » 13
miles.call(/iles/, '.')   » "M. Davis"
```

**Listing 17: Calling methods dynamically [72]**

As well as with PHP, Ruby also sports an `eval` function. The `eval` method (and its variations such as `class_eval`, `module_eval`, and `instance_eval`) will parse and execute an arbitrary string of legal Ruby source code (see Listing 18).

```
trane = %q{"John Coltrane".length}
miles = %q{"Miles Davis".sub(/iles/, '.')}

eval trane                » 13

eval miles                » "M. Davis"
```

I would say this is a tie, PHP and Ruby offer metaprogramming capabilities. Programs written in both languages are modifiable at runtime and an incremental compilation is also available. Also PHP and Ruby come with a full implementation of an API for reflection.

### 2.4.6  Aspect-Orientation

There exist various PHP implementations for Aspect Oriented Programming: PHPaspect, Aspect-Oriented PHP (aoPHP), Seasar.PHP, PHP-AOP, and FLOW3 to name a few. Also, in the field of academia research, some papers have been published in the topic of AOP in PHP [75, 76, and 77].

aoPHP is an extension of PHP that supports AOP in the Web development context. Currently, functions are the first class of joint points supported in aoPHP, so the pointcut descriptors are defined based on function calls. Before, after and around advice are supported and must be declared inside aspect files with the **.aophp** extension. These aspect files must be located at the same folder where the **.php** files are. The PHP files that are going to have their behavior affected by aspects must define which **.aophp** files are going to affect them. This is done by using the **<?aophp ?>** tag instead of the **<?php ?>** one.

To illustrate AOP in PHP, Listing 19 shows a simple example of a log in PHP file (**login.php**) that is affected by a logging aspect PHP file (**login_a.aoPHP**). The **a_login.aophp** aspect module contains three pieces of advice. The first before advice acts on calls to **checkPW()** and logs the time and IP address of the user which is attempting to log in. This is done through the **exec(checkPW($p))** pointcut descriptor which matches the calls to **checkPW($p)**. Note that the advice can access contextual information at the joint point, in this case the password (through the **$p** parameter). The other pieces of advice are responsible for logging the successful and unsuccessful logging in of a user, respectively. For simplicity, the log info is printed in the browser but in reality it could be registered in a database or XML file, for instance [76].

```
<!-- login.php --> <?aoPHP filename="login_a.aoPHP">

function invalidLogin($p){
    echo "<b>Can Not Login, $p Invalid Password</b><br>";
}

function redirectMember($p){
    echo "<a href=\"member.php\"> Click to Enter</a><br>";
}

function checkPW($p){
    if($p != "test"){
        invalidLogin($p);
    } else {
        redirectMember($p);
    }
}

$f = $_REQUEST['f']; switch($f){
    case "check":
        $pw = $_POST['password'];
```

```
        checkPW($pw);
        break;
    default:
        echo "<form method=\"POST\"
            action=\"login.php?f=check\">
            <p>Enter Password to Login</p>
            <p><input type=\"text\"
            name=\"password\" size=\"20\">
            </p>
            <p><input type=\"submit\"
            value=\"login\"
            name=\"button1\"></p>
            </form>";
        break;
    }
?>

<!-- login_a.aophp --> before(): exec(checkPW($p)){
    $ip = $_SERVER['REMOTE_ADDR'];
    //This is Where you Log to File or Database
    echo "<i>$ip attempting login @ " . time() . "</i><br>";
} after(): exec(redirectMember($p)){
    $ip = $_SERVER['REMOTE_ADDR'];
    //This is Where you Log to or Database
    echo "<i>$ip logged in @ " . time()  . "</i><br>";
} after(): exec(invalidLogin($p)){
    $ip = $_SERVER['REMOTE_ADDR'];
    //This is Where you Log to or Database
    echo "<i>$ip failed @ " . time() . "</i><br>";
}
```

**Listing 19: A simple aoPHP example [76]**

Although not as prolific as PHP in the implementations of Aspect Oriented Programming facilities, Ruby also counts with some extensions: AspectR-Fork and Aquarium being the most popular and widespread. For the sake of exemplification, I will focus on the Aquarium platform for depicting AOP in Ruby. Aquarium provides a Domain Specific Language (DSL) with which you can express "aspectual" system behavior in a modular way. Aquarium is a toolkit for Aspect-Oriented Programming (AOP) whose goals include:

- A powerful "pointcut" language for specifying where to apply aspects, comparable to the pointcut language in AspectJ for Java.
- Management of concurrent aspects (i.e., those acting on the same "join points").
- Adding and removing aspects dynamically.
- A user-friendly DSL.
- Support for advising Java types through JRuby.

Ruby's metaprogramming facilities already provide some of the capabilities for which static-language AOP toolkits like AspectJ are typically used. With Ruby, you can easily add new methods and attributes to existing classes and objects. You can alias and redefine existing methods, which provides the method interception and "wrapping" needed to extend or modify existing behaviour [78].

However, what is missing in Ruby is an expressive language for describing systemic modifications, a so-called "pointcut language". If you have simple needs for method interception and wrapping, then Aquarium will be overkill. However, if you have system-wide concerns that cross the boundaries of many objects, then an AOP toolkit like Aquarium can help implement these concerns in a more modular way [78].

Aquarium is suitable for non-trivial and large-scale aspect-oriented components in systems. Aquarium will be most valuable for systems where aspects might be added and removed dynamically at runtime and systems where nontrivial pointcut descriptions are needed, requiring a full-featured pointcut language. For less demanding needs, the alternatives are lighter weight and hence may be more appropriate.

Listing 20 shows a couple of simple examples of APO using Aquarium. The first part is an example that traces invocations of all public instance methods (included inherited ones) of the classes or modules Foo and Bar. The second part is the same example using the convenience DSL that adds aspect methods to Object (available only if it is require **aquarium/dsl/object_dsl**, since other toolkits, like Rails, define similar methods on Object).

```
Part I:
require 'aquarium'
Aspect.new :around, :calls_to => :all_methods, :on_types => [Foo, Bar] do |join_point,
object, *args|
    p "Entering:  #{join_point.target_type.name}##{join_point.method_name}  for  object
#{object}"
    result = join_point.proceed
    p "Leaving:  #{join_point.target_type.name}##{join_point.method_name}  for  object
#{object}"
    result  # block needs to return the result of the "proceed"!
end
Part II:
require 'aquarium/dsl/object_dsl'
around :calls_to => :all_methods, :on_types => [Foo, Bar] do |join_point, object,
*args|
    p "Entering:  #{join_point.target_type.name}##{join_point.method_name}  for  object
#{object}"
    result = join_point.proceed
    p "Leaving:  #{join_point.target_type.name}##{join_point.method_name}  for  object
#{object}"
    result  # block needs to return the result of the "proceed"!
end
```

**Listing 20: Examples of AOP for Ruby Using Aquarium**

Again, I consider this to be equal ground for PHP and Ruby. Although not directly implemented on the standard for both languages, they count with extensions for Aspect Oriented Programming. It is difficult to see any strong difference between the two.

### 2.4.7 Functional Programming

PHP facilities are somewhat limited in the current version for functional programming. PHP, as it stands currently (pre 5.3), already has some support for higher-order programming by passing around the names of functions as strings. It also supports so-called anonymous functions[8], via the **create_function**[9] function, though PHP does give those functions a name (something like "**lambda_N**").

That's not to say that progress hasn't been done in the field of high order programming in PHP. *Fn.php*[10] is an attempt to define lots of useful higher-order functions to PHP, and fix some of the things that are inconsistent with the others. *Fn.php* already supports the things in PHP that already exist, but adds **foldr**, **compose**, **zip**, **andf**, **orf**, **not**, **any**, **every**, **curry**, **I**, **K**, **S**, **flip** and a new short hand way to define functions with strings [96].

---

In the article "The State of Functional Programming in PHP," by Troels Knak-Nielsen [95], the author discusses what defines functional programming, dynamic dispatch, the binding state, closures, currying, implementing currying and weighing the options, all within the scope of PHP. It is concluded that of the current options for functional programming in PHP, the only practical solution is currying using command objects. `curry`, given a function `g` and an argument `a`, returns a function that promises to call the function `g` with the argument `a` plus whatever arguments are passed to it. In effect, `curry`, delays the function call until more knowledge is known.

Another initiative is *Phuntional[11]*, a library of utility classes written by Alan Dipert that enables programmers to write streamlined array-handling code. It aims to implement lambdas, closures, and a rich assortment of methods for operating on lists, hashes, and arrays.

Ruby is not a functional programming language in the way that languages like Lips and Haskell are, but Ruby's blocks, procs, and lambdas lend themselves nicely to a functional programming style. Any time a block is used with an `Enumerable` iterator like `map` or `inject`, a functional programming style is used [97]. Listing 21 illustrates an example of this case.

```
#Compute the average and standard deviation of an #array of numbers
mean = a.inject {|x,y| x+y } / a.size
sumOfSquares = a.map{|x| (x-mean)**2 }.inject{|x,y| x+y }
standardDeviation = Math.sqrt{sumOfSquares/(a.size-1)}
```
**Listing 21: Example of use of map/inject in Ruby**

`map`[12] and `inject`[13] are two of the most important iterators defined by `Enumerable`. Each expects a block. If programs are to be written in a function-centric way, it may be needed methods in the functions that allow programmers to apply those functions to a specified `Enumerable` object.

The article "Ruby and Functional Programming" by Khaled alHabache [98], relates common functional programming concepts and how can they be accomplished in Ruby. Such concepts are pure functions, expressions, higher order functions, currying and partial functions, and iteration through recursion among other topics. Even though the author notes that Ruby lacks two important aspects for functional programming (pattern matching and lazy evaluation), facilities such as blocks, lambdas [99], and the fact that everything is evaluated as an expression support a functional programming style.

Ruby clearly offers more support than PHP for functional programming capabilities. Although not a purely FP language like Lisp or Haskell, Ruby offers some structures and methods that allow programmers to write functional code (with the four closure types: blocks, Procs, lambdas and Methods) but does not force it.

---

[11] http://alan.dipert.org/post/154093030/functional-kinda-php-with-phunctional

[12] **enum.map {| obj | block } => array** : returns a new array with the results of running *block* once for every element in *enum*.

[13] **enum.inject {| memo, obj | block } => obj** : combines the elements of *enum* by applying the block to an accumulator value (*memo*) and each element in turn. At each step, *memo* is set to the value returned by the block. The first form lets you supply an initial value for *memo*. The second form uses the first element of the collection as the initial value (and skips that element while iterating).

### 2.4.8 Declarative Programming

Foremost, PHP supports an imperative/procedural programming paradigm [54], which describes computation in terms of statements that change a program state. In other words, imperative programs define sequences of commands for the computer to perform. This goes in contrast to declarative programming, which expresses *what* needs to be done, without prescribing *how* to do it in terms of sequences of actions to be taken.

Although some research have been done in the declarative development field for PHP using annotations [69]; largely, it can be considered that PHP does not support a purely declarative programming paradigm unlike languages such as Prolog, Smalltalk, or Erlang.

As the same case as PHP, Ruby is at first an imperative Object-oriented programming language, although it has functional features, it is not declarative. Some articles have been written researching the possibilities of declarative programming in Ruby [101], but it can be considered that Ruby at its core is Object-oriented and imperative, which goes in opposition to the declarative programming paradigm.

Neither PHP nor Ruby support declarative programming at the design level.

### 2.4.9 Batch Scripting

It is possible to make a PHP script to run it without any server or browser. The PHP parser is the only thing needed to use the language this way. This type of usage is ideal for scripts regularly executed using cron (on *nix or Linux) or Task Scheduler (on Windows). These scripts can also be used for simple text processing tasks. PHP supports a CLI (Command Line Interpreter/Interface) SAPI (Server Application Programming Interface) as of PHP 4.3.0. The main focus of this SAPI is for developing shell applications with PHP, which is of utmost importance for Batch processing.

Batch processing is relatively straightforward. In most cases, it comes down to one of two workflows. The first is used in reporting; the script runs once a day to generate reports and send them out to a set group of people. The second is a batch job created in response to some request. For example, I log in to the Web application and ask it to send all the people registered in the system a message telling them about a great new feature. This action has to be done as a batch because there are 10,000 people on the system. Such a task will take PHP a while to complete, so it must be done by a job outside the browser.

In this second workflow, the Web application simply drops information in a location shared with the batch-processing application. That information specifies the nature of the job (for example, "Send this e-mail to all the people on the system"). The batch processor runs the job, and then deletes the job. Alternatively, the processor marks the job as completed. Either way, the job should be identified as completed so it doesn't run again.

In the [Appendix E] there is a comprehensive example of a dedicated mail queuing system. In this model, there's a table in the database with a list of e-mail messages that should be sent out to various people. The Web interface uses the **mailouts** class to add an e-mail to the queue. The e-mail processor uses the **mailouts** class to retrieve the pending e-mail, and then uses it again to delete the pending messages from the queue [79].

There are a few options to execute Batch jobs on Ruby, and also on the Rails framework. The Module `ActiveRecord::Batches::ClassMethods` allows processing large number of records and sports two primary methods to do batch scripting. Listing 22 shows a simple example for both functions.

- **find_each**`(options = {}) {|record| ...}`

Yields each record that was found by the find `options`. The find is performed by find_in_batches with a batch size of 1000 (or as specified by the `:batch_size` option). This method is only intended to use for batch processing of large amounts of records that wouldn't fit in memory all at once. If you just need to loop over less than 1000 records, it's probably better just to use the regular find methods.

- **find_in_batches**`(options = {}) {|records| ...}`

Yields each batch of records that was found by the find options as an array. The size of each batch is set by the :batch_size option; the default is 1000.

You can control the starting point for the batch processing by supplying the :start option. This is especially useful if you want multiple workers dealing with the same processing queue. You can make worker 1 handle all the records between id 0 and 10,000 and worker 2 handle from 10,000 and beyond (by setting the :start option on that worker).

It's not possible to set the order. That is automatically set to ascending on the primary key ("id ASC") to make the batch ordering work. This also means that this method only works with integer-based primary keys. You can't set the limit either, that's used to control the batch sizes [80].

```
Person.find_each(:conditions => "age > 21") do |person|
    person.party_all_night!
end

Person.find_in_batches(:conditions => "age > 21") do |group|
    sleep(50) # Make sure it doesn't get too crowded in there!
    group.each { |person| person.party_all_night! }
end
```
**Listing 22: Examples of the methods 'find_each' and 'find_in_batches' for Ruby**

Also for the Rails framework there are two known extensions for Batch processing. *BackgroundDRb*[14] is a Ruby job server and scheduler. Its main intent is to be used with Ruby on Rails applications for offloading long-running tasks. Since a Rails application blocks while serving a request it is best to move long-running tasks off into a background process that is divorced from http request/response cycle [81]. The latest version is also modular and can be used without Rails so that any Ruby program or framework can use it.

*RailsCron*, by Kyle Maxwell, is a way to run background tasks using the Ruby on Rails environment. It could arguably do whatever you want to do in the background via running a crontab that executes the script/runner, but RailsCron enables you to do it all from ruby code, and seeing as it is an ActiveRecord object, it can be manipulated as such from the application [82].

---

[14] http://backgroundrb.rubyforge.org/

In the article "Scheduling tasks in Ruby / Rails" [83], the author exposes three methods for Batch processing using the Ruby programming language: a thread based scheduler, the OpenWFEru[15] project, which implements a standalone-ready scheduler, and the already mentioned BackgroundDRb. Examples for the three methods exposed are given by the author.

Both programming languages offer features to Batch scripting. But I would say PHP has a slight advantage over Ruby for the more comprehensive command line scripting facility. Although Ruby has extensions made by the programming community and features of its own for Bash processing [80], PHP offers various methods for sharing data between the Web application front end and the batch-processing back end. In conclusion, with the use of conventional tools (cron, MySQL, standard object-oriented PHP, and Pear::DB) creating batch jobs in PHP applications is easy to do, easy to deploy, and easy to maintain [79].

### 2.4.10  UI Prototype Design

PHP being primary a server side scripting language for Web development the options for UI prototyping are rather constricted and limited. Given that usually PHP code is embedded into HTML, designing a UI for a PHP application that displays in HTML browsers has some constraints that a standard application UI does not have. For instance, in HTML you cannot overlap one control over each other if you want your application to display properly on all browsers, and is restricted to HTML form controls.

Efforts have been done by the PHP Group to enable the programming language to write desktop applications. PHP is probably not the very best language to create a desktop application with a graphical user interface, but if you know PHP very well, and would like to use some advanced PHP features in your client-side applications you can also use PHP-GTK[16] to write such programs. You also have the ability to write cross-platform applications this way. PHP-GTK is an extension to PHP, not available in the main distribution [66]. Figure 3 show a desktop application built using PHP-GTK2.

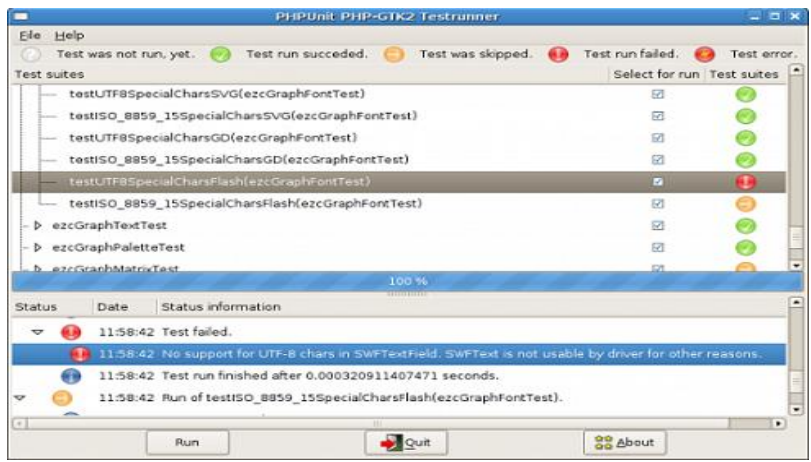

**Figure 15: Example of a desktop application using PHP-GTK**

There are various contributions to building UI components for PHP. One compilation of libraries can be found at [103]. In addition, some COTS solutions are available to PHP for UI controls like KoolPHPSuite[17].

Generally speaking, PHP would have to rely on HTML/JavaScript wizardry to build UI prototypes and emulate behavior. The purpose of prototyping is to explore possible solutions, experiment with components, layouts, menus, navigation, colors, fonts, graphics, etc. PHP servers to code the backend of the application, and has little or no purpose for presentation, it serves to process the data received and generates the output but has no control on how to display that information, that's HTML/CSS job.

Ruby is an object-oriented programming language. Until recently, its only usable implementation was an open-source interpreter. As with interpreters for comparable languages such as Perl and Python, Ruby is a command-line tool in the UNIX tradition. By default Ruby programs are not interactive; those that are accept text inputs in a terminal, and are limited to using text to provide feedback to users. Ruby programs can be developed in any editor; the standard interpreter ruby is not closely integrated with any particular program development environment.

A number of libraries exist to enable Ruby to provide graphical user interaction (GUI). These libraries enable a program to accept input and provide feedback using desktop computing interface elements and conventions, such as buttons, text boxes and windows. The standard distribution of Ruby includes the 'Tk' library[18] to do this; a number of other libraries are provided by third parties. Most of these are wrappers around toolkits written in C or C++, such as FxRuby[19], wxRuby[20], ruby-GNOME2[21], and QtRuby[22]; Shoes[23] is a graphical library which includes some interactive elements, implemented for Ruby alone. More recent alternate implementations of Ruby have their own GUI facilities provided by toolkits associated with the environment, such as Cocoa for MacRuby and Swing for JRuby [102].

There is a substantial overlap between the libraries in terms of their capabilities; they all enable display of interface elements, on-screen drawing, and handling of user interaction. There are also, however, considerable differences between the libraries in their aims, supported platforms, size, API style, range of widgets, aesthetics, licensing terms and supporting tools and documentation [101]. Figure 4 shows an example of a desktop application using wxRuby, this helps to illustrate that by using any of this GUI libraries it is possible to build user interface prototypes is a seamless and fast way. As it can be seen in the figure bellow, by using typical UI components a programmer or designer can built up a mock-up interface with some simulated or implemented behavior that can be easily modified as actual source code and data is integrated.

---

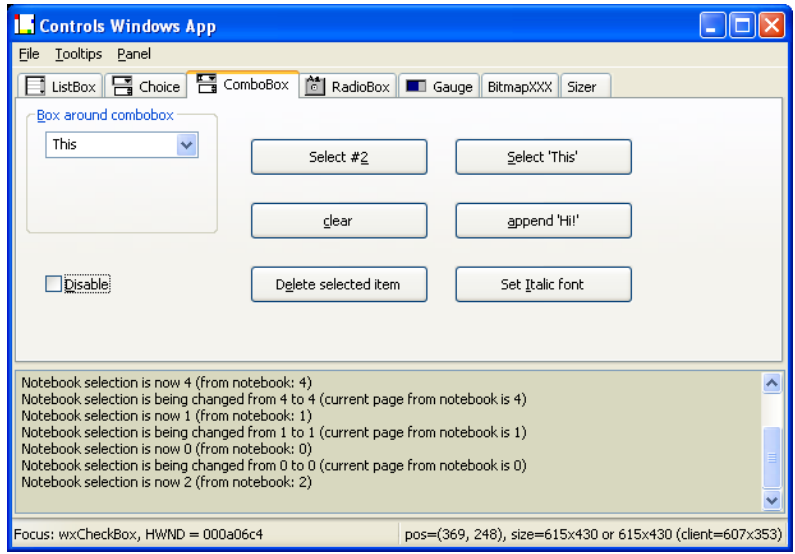

**Figure 16: Example of a desktop application using wxRuby**

Given the fact that Ruby has an extensive collection of implementations to provide GUI iteration, I think it holds the advantage over PHP in this area, which has to rely on HTML/JavaScript/CSS for prototyping User Interfaces.

## 2.5  Bash Scripts vs. Scheme

### 2.5.1  Default more secure programming practices

Shell environments were designed to automate certain process interactively with a user and were not meant to prevent a determined hacker to corrupt the shell script. Therefore, shell scripting is suitable for programs that do not need to be secure [189]. They can also be useful when they are executed with privileges by trusted users of the system.  In that respect, it becomes very difficult to write secured Bash scripts in an environment filled with malicious inputs because of various external factors/components affecting the shells [187]. Among those factor we can name environment variables (IFS, ENV…), filenames and file content (through piping), etc. Some common practices to use while coding Bash scripts especially when dealing with *setuid* or *setgid* script can be found in [190]. More information about secure shell scripts practices can also be found in [187], [188].

As far as Scheme is concerned, a valuable tutorial on how to write secure Scheme programs can be found in [236]. However, the application domain of those programming practices is restricted to web development, area where the most security is required over network applications to guarantee integrity of the data.

Given the details mentioned above and the element at our disposal that, the disjunction of domains of application of each language secures programming practices and the intrinsic nature of the latters determine whether to use or not to assess certain of application programming. That is to say, secure programming practices have nothing over Schemes once and vice-versa.

### 2.5.2  Web Application development

It is common knowledge that Bash scripts can be used to develop web applications if existing script functionality is needed over the Internet. This approach then comes handy. To use a Bash script via a network as a Web application we can for example [193]:

- use *cgi-bin* or enable *.cgi* files to run under Apache web server. It is setup through *httpd.conf*

- Invoke the Bash script through another script using PHP with a code such as:
    $view = exec ('/usr/local/bin/myscript.sh');

- use Server Side Includes by executing the following command within SSI
                    $view = exec ('/usr/local/bin/myscript.sh');

Below is a small Bash script "myscripit.sh" that could be triggered by the aforementioned second method:

```
#!/bin/Bash
# get kernel name and version
MSG="$(uname -r)"
#get Total memory
TOTMEM="$(free -m | egrep Mem | awk '{print $2}')"
# get available memory
FREEMEM="$(free -m | egrep Mem | awk '{print $4}')"
#Embedded HTML tags to port the outputs in a web browser format
echo "Content-type: text/html"
echo ""
echo"<html><head><title>Web Application Prototype</title></head><body>"
echo "My Web service name is: $0"
echo "My Kernel version is $MSG <br>"
echo "The current system total memory is: $TOTMEM bytes<br>"
echo "The current system total memory is: $FREEMEM bytes <br>"
echo "</body></html>"
```

**Listing 23: Example of Web service design in Bash**

On the other part, Web applications development in Scheme is a feasible task. Scheme can be used to code web application such as applets, or to generate dynamic web content using Scheme servlets and JScheme (a dialect of Scheme in Java) for both presentation and business logic purposes [216] or LAML [217] for presentation only. Also, one can use PLT-Scheme/Racket [217][218] which is also a language inspired by Scheme. In that case, Scheme web applications will be posted on PLT-Scheme web-server.

It appears that both Bash scripts and Scheme can be used to develop web applications. However, web application development with Bash essentially resides on automating or executing shell scripts over a network with a reasonable low cost. On the other side, Scheme language can be used to build high-end multi-tiered web applications with all the complexity that it entails through dialect languages such as JScheme or Racket (former PLT Scheme). This means an extended learning curve due to:

a) the new syntax coming into play with those new dialects
b) the deployment constraints such as web servers configuration
c) The lack of supporting and tutoring documentation on this subject [217].
d) tedious coding and debugging cycles compared to Bash scripts approach

### 2.5.3 Web services Design and Web services composition

Shell scripts and for instance Bash scripts can be wrapped as web service through various artifices and be ported from local accessibility to a wider scope. One of them is to use a WSO2's Web Services Application Server (WSAS) as explained in [197]. Bash scripts are even capable to consume web services [195]. Since BEPL processes are available as web services, it is then possible to invoke them using shell/Bash scripts [199]. They can also be deployed using Bash scripts [201]. Moreover, BPEL pipelining can be achieved through shell/Bash scripts [200].

In Scheme this can be achieved using one of his dialects called Racket formerly PLT/Scheme [221] with the affiliated software dependencies for example. As for web services composition, we did not find much tutorial to assess how difficult it is to achieve such a task but it is theoretically possible to orchestrate Scheme web services via a BEPL process which in its turn can be called via a Scheme scripts[216][217][218].

In summary, Bash and Scheme programming can used for web services design. Each of them having specific back-end technologies and sometimes needed extensions (Scheme) to perform the task. However, web service composition seems to be easier in Bash scripts in respect with their scope compared to a Scheme counter-part which could be implemented in Racket/PLT Scheme.

### 2.5.4  Object Oriented based abstractions

To perform object oriented programming in Bash scripts is quite impossible unless building a complex system of allocation of objects [202]. Also, Kragen Sitaker attempted an implementation of object oriented Bash scripts based on a Martin Hinsch's idea [203]. Among features he tried to implement we have: transparent persistence (but terrible memory management!), multiple inheritance, and local-network distributed objects. As mentioned by the author himself, those features do not make the shell a general purpose programming language.  However, Kragen Sitaker workaround is able to mimic OO-based abstractions in Bash scripts to some extents.

Regarding Scheme, is nearly an oriented programming language. To add an Object Oriented paradigm while keeping its minimalistic semantics and few forms of expression, a solution can be articulated around the following points [222]:

- add  new disjoint data type to Scheme to permit generic operations to all Scheme objects
- ensure that anonymous generic operations supports modularity
- use the terms *object* and *instance* idiosyncratically
- Generic function (operation) should be thought of as procedure whose definition propagates through the diverse objects it operates on.

Thus the addition of a single new, but simple primitive type allows creating an object system with relatively small addition to the language.

Overall, to implement an object oriented paradigm is easier in Scheme than Bash script. Indeed, Scheme offers by design a lot more mechanisms than Bash  to design Object Oriented abstraction behaviors and seamlessly integrate them into the Scheme minimalistic semantic.  On the other side, Bash scripts can be augmented with some object oriented features but maintainability and flexibility quality features of the code costs are high.

### 2.5.5  Reflection

Iterators are one the fundamental criteria to provide reflection mechanism to an imperative programming language [205]. Bash does not support them natively, reinforcing the fact that it is not a general purpose programming language. However, Bash scripts are capable of meta-programming [228] and therefore can materialize the notion of reification [229]. The process of evaluating the data from reification process (es) is called reflection [224]. Below, there is an attempt to mimic reflection in a Bash script given those 2 definitions (i.e. reification and reflection):

```bash
#!/bin/bash
#Metaprogramming and reflection simulation script

#Reification
echo "#!/bin/bash" > myprog
RANGE=60
number=$RANDOM
let "number %= $RANGE"
for ((I=0;I<=$number;I++)) do
        line="Line"
        sentence=$line" "$(($I+1))
        echo "echo $sentence" >> myprog
done
chmod +x myprog

#Evaluation
numLines=$(wc -l < myprog)
#echo $numLines
MAXLINES=20
if [A $numLines -lt $MAXLINES ]
then
    echo "This is a small program"
else
    echo "Renaming old program to \"bigprog\""
    mv myprog bigprog
    echo "echo  \"Reflective line appending\"" >> bigprog
fi
```

Listing 24: Example of reflection mechanism in Bash

Scheme as a dialect of LISP embeds the required features to perform reification and evaluation. Indeed, it is possible to represent programs using *quote function* and *list constructors.* It is then possible to execute then dynamically using the *"eval"* procedure [224].

As above, Bash scripts do not incorporate reflection mechanism by design. Since shell scripts are meta-programming capable, the existence of reflection would have been a valuable asset to facilitate meta-program development. However, reflective-style automation could be executed in Bash aiding with its reification facilities. On Scheme counterpart, built-in mechanisms are available to perform reflection. Hence, to develop strict reflective programs, Scheme is a better option over Bash scripts.

## 2.5.6  Aspect Orientation

Bash scripts like many others programming languages implement certain level of separation of concerns especially when dealing with modules function etc [202]. But do Bash scripts offer the native capabilities found in AOP? As we know on the main purpose of shell scripting is the automation of tasks and the execution of external programs.  So when it comes to compile JAVA or C code, through an automated script, will a Bash scripting be flexible enough to derive the target program (Aspect) for example?

The answer is likely to be negative because of the multiple dependencies that exist during the process [206]. Dependencies imply cross-cutting contexts which will be hectic to handle using a Bash script. Indeed unnecessary dependencies are always recompiled due to the nature of the shell interpreter [206]. A solution is to use "Makefiles" to separate aspect concerns. "Makefiles" might embed shell commands in their structure or shell scripts [207].

Meanwhile, performing pure functional-style AOP using Scheme is possible but a relatively experimental [231]. Indeed, the goals of the authors were to add AspectJ-style aspects to a functional language like such as Scheme. The main difficulties encountered were to specify pointcuts and advice. Overall they demonstrate it is possible to support several aspects of AOP in functional languages by adding three language constructs. Those constructs were then implemented as syntactic extensions of the Scheme language. But to do so, they have to use PLT Scheme facilities to create a new language, and default Scheme macro system. They also deliberately neglected type systems.

Based on the aforementioned points, Bash scripts and Scheme do not implement by design aspect orientation paradigm, However both languages offer syntactically derived workarounds, if one wish to use look alike language(s). Bash offers "makefiles" programming whereas Scheme offers a much heavier solution in terms of learning curve and complexity. Between the two, the best choice will depend on the aspect oriented task to be completed. For example "makefiles" will be more suitable to describe compilation processes than Scheme in terms of usability and maintainability.

### 2.5.7 Functional Programming

Bash scripts are known to be procedural program by essence [210]. Moreover, they also implement the notion of imperative functions and modules and also recursive functions [202]. However, being procedural, Bash scripts can generate side effects. Is it then possible to perform pure functional programming in Bash scripts?

One important foundation for writing functional-style programs is the presence of iterators. [209] Bash does not implement iterators since they are it shes not conceptually object oriented. Thus, Bash scripts are unlikely to mimic pure functional programming even though they implement some of the functional concept like recursion.

On the other side, Scheme is designed as pure functional language with all what it entails. Below is a code snippet showing how to compute Fibonacci numbers on pure functional basis, avoiding therefore side effects. The general definition of a Fibonacci is as described as follow:

$$F(n) = \begin{cases} 0 & n = 0 \\ 1 & n = 1 \\ F(n-1) + F(n-2) & n > 1 \end{cases}.$$

This definition can directly be translated to Scheme as:

```
(define (myfibo N)
 (cond ((< N 2) N)
       (else (fibup N 2 1 0))))
(define (fibup MAX count n-1 n-2)
 (cond ((= MAX count) (+ n-1 n-2))
       (else (fibup MAX (+ 1 count) (+ n-1 n-2) n-1))))
```

**Listing 25: Fibonacci computation in Scheme**

As we see, the computation of the numbers only deals with functions *(fibup())* and parameters(N). There is no need to understand inside mechanisms of those functions as the goal is to focus on the results only. Another task which can be completed easily in scheme as opposed to Bash Scripts for example, is the matrix cross-product:

```
(define v (vector 1 2 3))
(define (dot-product a b)
  (apply+(map*ab)))
display (dot-product v v)
```

**Listing 26: Matrix cross product in Scheme**

As conclusion, it is obvious that Scheme as opposed to Bash scripts should be the preferred language to perform pure functional computation regarding all its built-in mechanisms and abstractions in order to avoid side effects programming.

## 2.5.8 Declarative Programming

As mentioned in 2.5.7, Bash scripts follow procedural paradigm and does not natively support the features of declarative programming. However, "makefiles" (hybrid language) which can often embed shell scripts, specify dependencies in a declarative form [212], but also include an imperative list of actions (shell scripts, etc) are the closest thing a Bash script could be regarding declarative programming.

In Scheme however, you can use the macro facility to define nearly any sort of domain-specific languages you want [230]. It is known that meta-programming is a form of declarative programming. Thus, Scheme embodies powerful meta-programming facilities which allow to:

- Making the syntax nicer
- Automating boilerplate generation
- Writing declarative sub-programs

Thus, both Scheme and Bash Scripts can be used to do declarative programming. In bash scripts, not only built-in functionalities but also hybrid "makefiles" scripting. In Scheme however, its powerful system of macro will do the job. Thus, the choice of the language relies essentially on the domain of application rather than a limitation of both languages and/or their dialects.

### 2.5.9  Batch Scripting

Shell scripts are by conception a form of batch scripting under Unix-like operating systems. They allow several commands that would be entered manually at a command line interface to be executed in lot and without having to wait for a user to trigger each stage of the sequence (automation). Bash script are able to launch external program with the command "*exec".*

More generally, all scripting languages in their essence are able to execute series of command in lot. However, is it possible to automate process execution with Scheme? A positive answer can be found in [237] where to interact with Fluent Journal files via parameters it is necessary to use SCHEME

Thus, by design, Bash scripts are likely to be more powerful than Scheme programs to perform repetitive and automated tasks. Indeed, pipe programming and built-in process execution in Bash, gives it a net advantage over Scheme in terms of repetitive automation of external processes.

### 2.5.10  UI Prototyping

Shell and for instance shell scripts are designed around CLI. Indeed, CLIs are often used by programmers and system administrators, in engineering and scientific environments, and by technically advanced personal computer users. CLIs are also popular among people with visual disability; since the commands and feedbacks can be displayed using Refreshable Braille displays [234].

Hence, Bash scripts can be mostly used to prototype user interactions with a Unix-like OS via the CLI to perform specific tasks or trigger automation process (es). In addition, thanks to their external process execution capabilities such as "***echo**-ing*", they can generate WUI if declarative syntax code like HTML or XML is embedded inside the very script.

Scheme can be used to define command line interactions using implementations like Gambit, Chicken or Bigloo [214]. Moreover, using specific libraries like Racket Graphic Toolkit, extension of Racket one can achieve Scheme-like GUI prototyping [235].

Overall, both Bash scripts and Scheme are capable of user interface prototyping. Bash scripts are likely to focus on CLIs prototyping even if they can be involve in hybrid GUI-CLIs. Scheme is as capable as Bash script in that matter provided required extension or specific Scheme dialects are used.

## 3. Consolidated Analysis and Synthesis of the Results

Here is a point overview of the criteria out of 10 points.  It is a quick overview of criteria compatilility for each language.

|  | Java | Haskell | C++ | Perl | AspectJ | COBOL | Ruby | PHP | Bash Scripts | Scheme |
|---|---|---|---|---|---|---|---|---|---|---|
| Security | 8 | 9 | 6 | 7 | 9 | 5 | 8 | 6 | 2 | 5 |
| Web Development | 9 | 7 | 7 | 7 | 7 | 1 | 8 | 10 | 4 | 5 |
| Web Services | 9 | 7 | 8 | 7 | 7 | 1 | 7 | 9 | 3 | 5 |
| OO-base abstraction | 10 | 3 | 9 | 6 | 9 | 9 | 10 | 8 | 1 | 7 |
| Reflection | 8 | 1 | 3 | 9 | 9 | 0 | 7 | 7 | 1 | 8 |
| Aspect-Orientation | 5 | 0 | 4 | 6 | 10 | 5 | 6 | 6 | 2 | 5 |
| Functional Programming | 5 | 10 | 8 | 8 | 0 | 0 | 8 | 5 | 3 | 10 |
| Declarative Programming | 6 | 10 | 6 | 7 | 5 | 0 | 0 | 0 | 3 | 10 |
| Batch Scripting | 3 | 9 | 4 | 10 | 7 | 9 | 5 | 8 | 10 | 4 |
| UI prototype design | 8 | 10 | 8 | 5 | 5 | 8 | 7 | 5 | 10 | 5 |

## 3.1  Criteria 1: Default more secure programming practices

In terms of security the most secure languages from the set we compared are Haskell, Java and its aspect-oriented flavor, AspectJ.  While Haskell will not execute any insecure code, AspectJ and Java benefit from the sandbox model with bytecode verifiers to prevent security breaches. Other languages that ranked as good to acceptable were Ruby, Perl, C++ and PHP due to a number of known and exploitable vulnerabilities. The worst security scores got COBOL with little security features, Scheme with some features and Bash with almost non-existant security relying on the operating system.

## 3.2  Criteria 2: Web Application development

The best picked languages to develop web applications in are Java, PHP and Ruby due to a large availability of modules that support web frameworks, large existing support libraries and web integration as part of the premise of the language. The languages in which web development is possible but not as prominent are C++, Haskell, Perl and AspectJ which mostly rely on other third party frameworks to which they bind their code to fulfill their work. Lastly the lowest level of support for web applications due to low quality or availability is in Scheme, Bash and COBOL in which web application are almost non-existent.

## 3.3  Criteria 3: Web services design and composition

In the category of web service design the best languages to use are Java, C++ and PHP.  They have the most compatibility with the most amount of web service types offered and widest array of libraries as well as best performance.  Second to them come Haskell, AspectJ, Ruby and Perl where the availability of libraries or performance is still decent but inferior to languages listed before.  The languages which should not be used for web service design due to either lack of capacity or difficulties implementing solutions properly are Scheme, Bash and COBOL.

## 3.4  Criteria 4: OO-base abstraction

Object Oriented Based Abstraction:  according to our papers, we realized that all of the 10 languages have some support of the object oriented concepts but in different levels. Ruby in Java got the highest mark in criterion because they are pure object oriented languages, and then we find C++, AspectJ, Object Oriented COBOL.  These languages are either not pure object oriented (like C++ and OO COBOL, PHP) or they have some restriction regarding some OO concepts (AspectJ).  We find in the third place languages that are not object oriented but the support some features object oriented paradigm like Scheme, Perl. Haskell and Bash Script have a very poor support for object oriented features because it is a functional language and Bash Script which is mainly a scripting. For this criterion we took into consideration the Object Oriented COBOL.

## 3.5  Criteria 5: Reflection

Reflection:  As we can see Scheme, AspectJ and Perl are highly reflective languages, the same thing applies for Java, Ruby and PHP but with less efficiency than the first set of languages. Bash Script, C++ have a poor support for reflection while Haskell and COBOL don't have any mechanisms for reflection.

## 3.6  Criteria 6: Aspect orientation

Aspect Orientation: Not surprisingly, AspectJ is the top ranked in this criterion, while all the other languages have a fair support of aspect oriented paradigm. Surprisingly COBOL has some basic AOP features. It is noteworthy that Haskell does not support aspect oriented programming because it is a functional programming language.

## 3.7  Criteria 7: Functional programming

We have found out that the programming languages Haskell and Scheme are particularly well suited for the functional programming paradigm (if not the best choices among the 10 languages) , followed by C++, Perl, and Ruby. Haskell and Scheme are purely functional programming languages from the design which means that in general, functions in Haskell and Scheme do not have side effects; this is illustrated in the references and code examples given in this document. Behind Scheme and Haskell, C++, Perl, and Ruby also offer some high-order programming and meta-programming capabilities for FP. Perl and Ruby support a functional programming style and C++ counts with different libraries in which FC++ is the most popular and widespread. In conclusion, Haskell and Scheme are the languages to go to when one wants to do pure functional computation regarding all its built-in mechanisms and abstractions in order to avoid side effects on the applications.

## 3.8  Criteria 8: Declarative programming

Then again, Haskell and Scheme are the most suited for Declarative programming among our selection of languages, followed far behind by Perl, C++, and Java. In Scheme, you can use the macro facility to define nearly any sort of domain-specific languages you want. It is known that meta-programming is a form of declarative programming; thus, Scheme embodies powerful meta-programming facilities, same is the case with Haskell. This is all supported with the cited references and examples disclosed in this document.

## 3.9  Criteria 9: Batch scripting

In the area of batch scripting the languages that stand out are Bash, Perl, COBOL as most adapted to processing large quantities of data and being the most versatile for processing files in different manners. Languages that are also suitable are Haskell and PHP. The remaining languages do an ok to poor job at batch processing lacking it as their native functionality.  Languages such as C++ or Java-based languages do processing well due to their flexible architecture, but need to be compiled to work efficiently.

## 3.10  Criteria 10: UI prototype design

The best languages to do UI prototyping are Java, C++, COBOL and Ruby.  They have the best and most versatile tools and libraries for user interface purposes.  The remaining languages perform adequately depending on the availability of features and quality of bindings between the language and libraries used, with the exception of Bash which does not have user interface libraries and can only do text or web prototyping.

## 3.11  Criteria summary table.

Below is the criteria summary table split up into four tables that illustrate the junction of all the languages and all criteria .

| | Java | Haskell | C++ | Perl | AspectJ |
|---|---|---|---|---|---|
| **Default Secure programming practices** | - Java puts a lot of conformations on early checking for possible problems, later dynamic checking, and eliminating situations that are possibly causing error.<br>- Robustness and good memory management.<br>- No pointers | - Concurrent threads supported by GHC Haskell dialect, which these threads implicitly scheduled and allocated by underlying run time system<br>- Memory allocation and de-allocation can be done in safe and implicit manner via garbage-collector heap.<br>- lazy evaluation property in Haskell will take care of making sure that the data dependency order is respected so this gives programmers an opportunity to not worry about the order of the execution, even within single thread. | - main security issues are memory safert issues (buffer overflows, dangling pointers, double-free)<br>- some functions in the standard library contribute to the memory problems<br>- race conditions are present as well<br>- new libraries, static and dynamic toos available for analysis | - memory problems not present due to dynamic allocation<br>- main issue is input validation due to the vast array of input methods available for files, shell commands and web request parameters<br>- race conditions and privilege escalation present<br>- source code review tools and taint mode available to perform security review | - Strong typing<br>- Automatic memory management<br>- Not pointers<br>- Exception handling |
| **Web applications development** | - Java has strong presence between client side and server side.<br>- Servlet API leverages the advantages of Java platform to connect and solve the issues of APIs and CGI.<br>- Java web applications can be run anywhere and all you need is a Java Virtual Machine (JVM) running on client's side<br>- The problem with Servlet is bit complicated and you have to write, compile and deploy life cycle, from this reason Java Server Pages (JSP) considered to be the third generation solution | - It has some implemented projects in this field.<br>- Happstack is another common example and new name for HApps. It is a Haskell web framework. The developers in this web framework can prototype quickly, scale massively, change easily, and operate reliably.<br>- It has ability to support OS X, Windows, FreeBSD, and GNU/Linux environments.<br>- Flexible Routing and Request Processing DSL, Flexible Templating Options, and Type Safty. | - web applications integrated since early stages such as CGI<br>- integration continued on many platforms and web services with improved performance<br>- has some nice framworks with bindings to C++ that allows for some good web development | - has been used for web development since its beginnings<br>- mod_perl Apache plugin allowed for tight web application integration with web server<br>- Catalyst framework the current best solution for making web services as A decent framework with A lot of popular solutions integrated | - Run on web server fine<br>- Could be used to manage crosscutting concerns in web application like monitoring, logging... |
| **Web service design and composition** | - In web service process composition Haskell usage in valid XML generation has been discussed in [15, 16]. HaXML is a domain specific language for parsing, filtering, transforming, and generating XML documents. HaXML utilities treat a Document Type Definition (DTD) as a series of type declarations in Haskell, thus conflating validation with type checking. | - Metro stack project consist of Web Services Interoperability Technologies (WIST), Java Architecture for XML Binding (JAXB) and Java API for XML-Based Web Services (JAX-WS), which enable you to freely create and deploy reliable, interoperable, transactional, and secure Web services and clients.<br>- in java web services devided into Core Web Services, Enhanced Web Services, Secure Web Services, Legacy Web Services. | - web services started with sockets and evolved through an array of protocols such as CORBA and DCOM.<br>- having limited support for standardized RPC and early standardized web services<br>- some decent toolkits developped recently for SOAP and other standards such as WSO2 and .Net Framework | - support for simple web services rather limited<br>- good and well used support for standardized RPC methodologies, but mediocre support for SOAP, WSDL and REST<br>- WSF/Perl currently one of the best framworks to use having Perl bindings | - Could be used for failures handling, statistics gathering and monitoring web services<br>- Could be used to implement very common concerns in distributed computing like persistence and distribution concerns |
| **OO-based abstraction** | - Java is an Object-Oriented programming language<br>- Objects in Java offer several benefits such as modularity, hiding information, reusability, and debugging made simple. | - Functional languages are not object-oriented languages because functional programming is considered to be an alternative to object-oriented<br>- Haskell as functional language provide strong module systems with excellent and flexible security.<br>- it is designed to avoid the use of state, encapsulates it via monads when necessary; and provides a mechanism called type classes which provides a great mechanism for polymorphism | - supports object orientation with classes, instances, methods, message calling though procedure calls, inheritance including multiple inheritance, abstractions, encapsulation, polymorphism and reduced coupling<br>- falls short because primitive types are not objects and applications are required to be written in a procedure | - supports object orientation to a degree by allowing a hash reference to have methods and attributes, but is weak in areas of encapsulation and abstraction.<br>- it does support all features to an extent and with some tricks needed<br>- internal types are not objects and requires to run in a procedure | - Support the main concepts of OOP<br>- Encapsulation, abstraction, message passing, Inheritance, Inner aspects...<br>- Some restriction regarding inheritance (Serializable and Cloneable)<br>- An advice cannot have a visibility modifier |
| **Reflection** | - Reflection is a functional extension to OOP paradigm<br>- This property let java programs are able to inspect or examine upon themselves, and manipulates internal properties of the program<br>- java reflection quite powerful and can be useful for instance, with Java reflection the class can obtain the name of all its members and display them.<br>- With Java reflection, it is also possible to instantiate objects, get/set filed values and invokes methods. | - Haskell doesn't support reflection in the same way as many other languages do, but there are idiomatic ways to accomplish the same things when the developer would like to use reflection for. | - does not natively support reflection due to the additional overhead it may incurr<br>- has several methods of implementing reflection though templates, modified compilers and external libraries that analyze the source code | - supports reflection very well dynamically finding instantiating modules, calling methods and accessing attributes | - AspectJ is a reflective language, the method getSignature returns meta information about a Java program at runtime<br>- AspectJ has API contains the package org.aspectj.lang.reflect which could be used to take advantage of the reflective aspect of AspectJ |

| | COBOL | Ruby | PHP | Bash | Scheme |
|---|---|---|---|---|---|
| **Default Secure programming practices** | - Typing is very weak<br>- Possibility to lose of information during type conversion<br>- No automatic memory management (all the variable are allocated once the program is loaded ), but new COBOL provides mechanisms to manage the memory by the ALLOCATE and FREE<br>- Single thread of control<br>- No exception Handling<br>- PERFORM-THRU and GO TO statement that makes the program hard to understand. | - Supports taint checking<br>- Facilities for validation, authentication, authorization, data protection<br>- Ruby on Rails has some clever helper methods, for example against SQL injection<br>- Rails also has vulnerabilities in some areas such as file permission, sessions, input validation, and escape output | - Most of these PHP-related vulnerabilities can be exploited remotely: they allow crackers to steal or destroy data from data sources linked to the Web server (such as an SQL database), send spam or contribute to Denial of Service attacks using malware, which itself can be installed on the vulnerable servers<br>- As PHP is easy and popular, more bugs are written with it<br>- PHP appliion developers do not make a good job at securing their code: they do not care, or they do not know how, or they try but fail. | - Shell environments were designed to automate certain process interactively with a user and were not meant to prevent a determined hacker to corrupt the shell script<br>- Shell scripting is suitable for programs that do not need to be secure<br>- They can also be useful when they are executed with privileges by trusted users of the system<br>- It becomes very difficult to write secured Bash scripts in an environment filled with malicious inputs because of various external factors/components affecting the shells | - Fairly secure, no real memory issues<br>- Vulnerable to bad input |
| **Web applications development** | - Some web server support COBOL in order to create CGI scripts<br>- Natively, COBOL does not talk the language<br>- COBOL programs running on mainframe could be used but using middleware | - Primary done with the Ruby on Rails framework<br>- Rails is an open source Ruby framework for developing database-backed web applications<br>- Rails scaffolding is a quick way to generate some of the major pieces of an application<br>- Scaffold allows the programmer to generate code that permits a user to create, read, update, and delete (CRUD) data in the database | - PHP is a general-purpose scripting language, but is especially suited for server-side web development<br>- PHP is mainly focused on server-side scripting, so it can do anything any other CGI program can do, such as collect form data, generate dynamic page content, or send and receive cookies<br>- PHP can be deployed on most web servers, many operating systems and platforms (Windows, Linux, Unix, Mac OS X, etc.), and can be used with many relational database management systems (MySQL, PostgreSQL, IBM DB2, SQLite, Oracle, etc.). Also, it is compatible with many Web servers like Apache, Microsoft Internet Information Server, Personal Web Server, Netscape and iPlanet servers, and many others | - It is common knowledge that Bash scripts can be used to develop web applications if existing script functionality is needed over the Internet<br>- Web application development with Bash essentially resides on automating or executing shell scripts over a network with a reasonable low cost | - Web applications development in Scheme is a feasible task<br>- Scheme can be used to code web application such as applets, or to generate dynamic web content using Scheme servlets and JScheme (a dialect of Scheme in Java) for both presentation and business logic purposes or LAML for presentation only<br>- One can use PLT-Scheme/Racket which is also a language inspired by Scheme. In that case, Scheme web applications will be posted on PLT-Scheme web-server |
| **Web service design and composition** | - COBOL does not process or generate XML which is a common standard for web services<br>- COBOL could provide web services using middleware | - Web services in Ruby on Rails are implemented using the REST architecture, specifically in the Active Resource component of Rails<br>- Active Resource provides the tools to quickly and easily consume REST based web services conforming to the Rails RESTful URI structure and protocol conventions<br>- Active Resource automatically maps the response from any conforming service to rich Ruby objects. Active Resource also provides all the lifecycle methods needed to easily find, create, update, and delete resources without having to write any code | - PHP sports various extensions for Web services functionalities, among them are OAuth (Open Authorization), SCA (Service Component Architecture), SOAP, XML-RPC, and REST<br>- Web application frameworks usually (if not always) offer features to support Web services facilities. Symfony, Zend Framework, CakePHP, CodeIgniter, and Sapphire offer implementations for REST Web services.<br>- PHP has implementations of SOAP (e.g. PEAR::SOAP, PHP-SOAP, NuSOAP) | - Shell scripts and for instance Bash scripts can be wrapped as web service through various artifces and be ported from local accessibility to a wider scope<br>- One of them is to use a WSO2's Web Services Application Server (WSAS)<br>- Since BEPL processes are available as web services, it is then possible to invoke them using shell/Bash scripts | - In Scheme this can be achieved using one of his dialects called Racket formerly PLT/Scheme with the affiliated software dependencies<br>- It is theoretically possible to orchestrate Scheme web services via a BEPL process which in its turn can be called via a Scheme scripts |
| **OO-based abstraction** | - Object Oriented COBOL support OOP concepts<br>- OO COBOL programs are harder to understand if compared to procedural COBOL programs | - Ruby is a fully Object-oriented programming language, and thus, it provides extensive support for developing Object-oriented applications along with the features to cover the concepts of OOP<br>- Ruby is considered a pure Object-Oriented language, because everything appears, to Ruby, as an object. All Ruby data consists of objects that are instances of some class. Even a class itself is an object that is an instance of the Class class<br>- Inheritance, Composition, Polymorphism are all possible with Ruby Object-Oriented model. | - Starting with PHP 5, the object model was rewritten to allow for better performance and more features<br>- Among the features in PHP 5 are the inclusions of visibility, abstract and final classes and methods, additional magic methods, interfaces, cloning and typehinting<br>- PHP 5 introduces abstract classes and methods. It is not allowed to create an instance of a class that has been defined as abstract. Any class that contains at least one abstract method must also be abstract | - To perform object oriented programming in Bash scripts is quite impossible unless building a complex system of allocation of objects<br>- Among features tried to implement we have: transparent persistence (terrible memory management), multiple inheritance, and local-network distributed objects.<br>- those features do not make the shell a general purpose programming language. | - Scheme is nearly an oriented programming language. To add an object oriented paradigm while keeping its minimalistic semantics and few forms of expression, A solution can be articulated<br>- the addition of A single new, but simple primitive type allows creating an object system with relatively small addition to the language |
| **Reflection** | - COBOL does not have reflection | - Ruby also comes with an extensive reflection API. One of the many advantages of dynamic scripting languages such as Ruby is the ability to introspect—to examine aspects of the program from within the program itself<br>- Reflection in Ruby can be used to discover about a program what object it contains, the current class hierarchy, the contents and behavior of objects, information on methods among others | - A reflection API has been included in the standard PHP 5 distribution that adds the ability to reverse-engineer classes, interfaces, functions, methods and extensions. It provides an API to obtain information about classes, including all of their methods and properties, without the need to parse any source code. Simple functions, runtime objects and PHP extensions can be inspected as well. It is also possible to modify properties of objects, invoke arbitrary methods or functions, and instantiate objects, i.e. doing some basic intercession | - Bash does not support iterators them natively, reinforcing the fact that it is not a general purpose programming language. However, Bash scripts are capable of meta-programming and therefore can materialize the notion of reification<br>- The process of evaluating the data from reification process(es) is called reflection | - Indeed, it is possible to represent programs using quote function and list constructors<br>- It is then possible to execute them dynamically using the "eval" procedure |

| | **Java** | **Haskell** | **C++** | **Perl** | **AspectJ** |
|---|---|---|---|---|---|
| **Aspect-Orientation** | - In java there are open source aspect-oriented frameworks such as AspectWokz, JAC(Java Aspect Components), dynop, DynmicAspects, Nanning, JBossAOP, CASER, EAOP, Colt, CALI, PROSE, and Azuki Framework. | - Haskell as pure functional language doesn't support AOP from its design phase principled | - does not natively support aspect oriented programming<br>- ways exist that enable creating aspects either though templates or through external libraries which incur an overhead for hooking in the aspects at join points | - being a flexibly typed language with the use of aspect oriented library it is possible to hook in aspects to Perl code without a big overhead | - Not surprisingly, ASpectJ is an aspect oriented language |
| **Functional programming** | - Developers tried to embed some libraries developed in Java and using FP patterns.<br>- One example of the developed libraries in this manner (FP) is FunctionalJ. This library makes it easy to use functional programming constructs in java code and this library provides many features such as easily representing functions as objects, uses parameter binding, replace procedural code with functional code, and no need to deal with exception if you don't need. | - Haskell is categorized from family of pure functional language paradigm.<br>- does not have side effect<br>- Another interesting feature of Haskell is its lack of any loop construct. There is no for and no while. There is no GOTO or branch or jmp or break. | - does not natively support functional programming, but there exists a well known library FC++ that does a very good job at adapting C++ to act like a functional language | - supports functional programming to a degree having list manipulation functions<br>- functional programming can be extended to a great extent through a library that supports a lot of functions that imitate Haskell | - Functional programming is absent in AspectJ |
| **Declarative programming** | - Java Tiger release (JDK 1.5) adds a new language construct called annotation (proposed by JSR-175).<br>- Enterprise JavaBean (EJB) has been designed by declarative model<br>- Annotation is a generic mechanism for associating metadata and combining the information (declarative information) with program elements such as classes, methods, fields, parameters, local variables, and packages. | - Haskell is declarative programming based on the brief details in Criteria 7. | - having FC++ for functional programming, it can be used for declarative programming<br>- logic programming is available through LC++ and other types of declarative programming generally supported to a degree with other libraries | - is fairly competent in declarative programming style being suitable for functional and other types due to the flexible syntax of the language.<br>- libraries enable usage of Perl in different types of delcarative programming purposes | - If we consider annotations based programming as declarative, we can say that AspectJ has some support for declarative programming<br>- Many annotations are available is AspectJ like @Aspect @pointCut.... |
| **Batch Scripting** | - Java provides some libraries that satisfied characteristics of batch scripting such as executing external commands and automation.<br>- This property involves the use of two java classes Runtime class and Process class. | - Haskell is supporting batch scripting techniques by some ways. Library called System (System :: String -> IO ExitCode) used in Haskell to perform and executing external commands such as external Linux command 'ls'. | - does not natively support batch scripting, although due to fast compilation time is well suited for processing<br>- has a interpreter that can interpret typed code and execute it without compilation | - by nature a great scripting language with easy access to file reading, system commands, shell and text processing through regular expressions<br>- has lots of libraries that enahce its processing capabilities | - Even if AspectJ is not designed do batch scripting, it could be integrated to a Java program doing batch scripting |
| **UI prototype design** | - The most famous and known GUI library in java is Swing<br>- In Java it is easy to build GUI but the problem is hard to get something easy to build and maintain, and look fancy.<br>- In Java GUI there are two separation concerns "BAD" and "GOOD". First, BAD is resulted by mixing logic and interface, and this style for small student program. Second, GOOD is separating GUI from logic. When the program grows larger it is suppose to separate GUI processing from logic. | - Haskell has many benefits tools and it gives a good combination of robustness and simplicity for the graphical user interface development model. Haskell has many of toolkits for programming graphical user interface (UGI) such as wxHaskell, Gtk2Hs, hoc, and qtHaskell. | - has a good array of user interface development tools such VisualStudio, Qt Creator and wxWidgets Form Designer that allow for drag-and-drop visual design<br>- the underlying event functionality is not as easy to hookup in complex applications and usually requires manual tweaking | - has some tools like QtCreator and wxWidgets Form Designer but is not meant for graphical user design<br>- has an easier time to design textual user interfaces than others due to its scripting nature | - AspectJ could integrated to user interfaces if there are some crosscutting concerns, but in genrally we can say that AspectJ is not designed for user interface prototyping |

| | COBOL | Ruby | PHP | Bash | Scheme |
|---|---|---|---|---|---|
| **Aspect-Orientation** | - Surprisingly, COBOL has some support for pointcuts and advices, but this support is very poor<br>- The is a prototypical language AspectCOBOL to design an extension to COBOL for aspect oriented programming | - Ruby also counts with some extensions: AspectR-Fork and Aquarium being the most popular and widespread<br>- Aquarium provides a Domain Specific Language (DSL) with which you can express "aspectual" system behavior in a modular way. Aquarium is a toolkit for Aspect-Oriented Programming (AOP)<br>- Ruby's metaprogramming facilities already provide some of the capabilities for which static-language AOP toolkits like AspectJ are typically used. | - There exist various PHP implementations for Aspect Oriented Programming: PHPaspect, Aspect-Oriented PHP (aoPHP), Seasar.PHP, PHP-AOP, and FLOW3 to name a few<br>- aoPHP is an extension of PHP that supports AOP in the Web development context. Currently, functions are the first class of joint points supported in aoPHP, so the pointcut descriptors are defined based on function calls. | - Bash scripts has many others programming languages implement certain level of separation of concerns especially when dealing with modules function<br>- Aquarium provides a Domain Specific Language (DSL) with which you can express "aspectual" system behavior in a modular way. Aquarium is a toolkit for Aspect-Oriented Programming (AOP)<br>- Bash's metaprogramming facilities already provide some of the capabilities for which static-language AOP toolkits like AspectJ are typically used. | - Performing pure functional-style AOP using Scheme is possible but a relatively experimental<br>- It is possible to support several aspects of AOP in functional languages by adding three language constructs. Those constructs were then implemented as syntactic extensions of the Scheme language<br>- To do so, they have to use PLT Scheme facilities to create a new language |
| **Functional programming** | - COBOL cannot do functional programming | - Ruby's blocks, procs, and lambdas lend themselves nicely to a functional programming style.<br>- Ruby lacks two important aspects for functional programming (pattern matching and lazy evaluation), but its facilities such as blocks, lambdas, and the fact that everything is evaluated as an expression support a functional programming style | - PHP facilities are somewhat limited in the current version for functional programming. PHP, as it stands currently (pre 5.3), already has some support for higher-order programming by passing around the names of functions as strings<br>- Fn.php is an attempt to define lots of useful higher-order functions to PHP, and fix some of the things that are inconsistent with the others.<br>- Another initiative is Phuntional, a library of utility classes written by Alan Dipert that enables programmers to write streamlined array-handling code. | - Bash is known to be procedural programs by essence<br>- Being procedural, Bash scripts can generate side effects<br>- One important foundation for writing functional-style programs is the presence of iterators. Bash does not implement iterators. Thus, Bash scripts are unlikely to mimic pure functional programming even though they implement some of the functional concepts like recursion | - Scheme is designed as pure functional language with all what it entails |
| **Declarative programming** | - COBOL is mainly an imperative language, we cannot find any clue about declarative programming in COBOL | - Ruby is at first an imperative Object-oriented programming language, although it has functional features, it is not declarative. Some articles have been written researching the possibilities of declarative programming in Ruby, but it can be considered that Ruby at its core is Object-oriented and imperative, which goes in opposition to the declarative programming paradigm | - Foremost, PHP supports an imperative/procedural programming paradigm, which describes computation in terms of statements that change a program state. In other words, imperative programs define sequences of commands for the computer to perform. This goes in contrast to declarative programming, which expresses what needs to be done, without prescribing how to do it in terms of sequences of actions to be taken | - Bash scripts follow procedural paradigm and does not natively support the features of declarative programming<br>- However, "makefiles" (hybrid language) which can often embed shell scripts, specify dependencies in a declarative form, but also include an imperative list of actions (shell scripts, etc) are the closest thing a Bash script could be regarding declarative programming | - In Scheme, you can use the macro facility to define nearly any sort of domain-specific languages you want. It is known that meta-programming is a form of declarative programming. Thus, Scheme embodies powerful meta-programming facilities |
| **Batch Scripting** | - Very known in batch scripting, COBOL can automate tasks by calling external programs like data base driver.<br>- Very strong in automating tasks related to file system<br>- COBOL programs could be called by another scripting language which is JCL | - There are a few options to execute Batch jobs on Ruby, and also on the Rails framework. The Module ActiveRecord::Batches::ClassMethods allows processing large number of records and spares two primary methods to do batch scripting<br>- Also for the Rails framework there are two known extensions for Batch processing. BackgrounDRb is a Ruby job server and scheduler. RailsCron, by Kyle Maxwell, is a way to run background tasks using the Ruby on Rails environment. | - It is possible to make a PHP script to run it without any server or browser. The PHP parser is the only thing needed to use the language this way<br>- PHP supports a CLI SAPI as of PHP 4.3.0. The main focus of this SAPI is for developing shell applications with PHP, which is of utmost importance for Batch processing | - Shell scripts are by conception a form of batch scripting under Unix-like operating systems.<br>- They allow several commands that would be entered manually at a command line interface to be executed in lot and without having to wait for a user to trigger each stage of the sequence (automation). Bash script are able to launch external program with the command "exec" | - All scripting languages in their essence are able to execute series of command in lot. |
| **UI prototype design** | - Generally COBOL programs are run through the command line, but there are some tools available to create graphical interfaces (they could be used for user interface prototyping).<br>- The tools for creating graphical user interface are generally operating system dependant. | - A number of libraries exist to enable Ruby to provide graphical user interaction (GUI).<br>- The standard distribution of Ruby includes the 'Tk' library. A number of other libraries are provided by third parties. Most of these are wrappers around toolkits written in C or C++, such as FxRuby, wxRuby, ruby-GNOME2, and QtRuby; Shoes is a graphical library which includes some interactive elements, implemented for Ruby alone<br>- More recent alternate implementations of Ruby have their own GUI facilities provided by toolkits associated with the environment, such as Cocoa for MacRuby and Swing for JRuby | - PHP being primary a server side scripting language for Web development the options for UI prototyping are rather constricted and limited. Given that usually PHP code is embedded into HTML, designing a UI for a PHP application that displays in HTML browsers has some constraints that a standard application UI does not have<br>- Generally speaking, PHP would have to rely on HTML/JavaScript wizardry to build UI prototypes and emulate behavior. | - Bash and for instance scripts are designed around CLI.<br>- Hence, Bash scripts can be mostly used to prototype user interactions with a Unix-like OS via the CLI to perform specific tasks or trigger automation process(es). In addition, thanks to their external process execution capabilities such as "echo-ing", they can generate WUI if declarative syntax code like HTML or XML is embedded inside the very script | - Scheme can be used to define command line interactions using implementations like Gambit, Chicken or Bigloo<br>- Using specific libraries like Racket Graphic Toolkit, extension of Racket one can achieve Scheme-like GUI prototyping |

# 4. Conclusion

It is important to make technology decisions at the right time and for the right reasons. Good business decisions provide good people with appropriate supporting tools so they can produce good products. When it comes to software development, dealing with tough language issues head-on is one requirement for today's visionary manager. When combined with other software engineering considerations, a good language decision can support the development of cost-effective software systems that, in turn, provide valuable, reliable business support. As someone once said: *"Selecting the best tool for the job."*

## 4.1 Future Work

We'd like to refine our analysis of our chosen languages within the stated criteria further as we became more familiar with them over time. We also plan on expanding our analysis onto other criteria and other languages also provide more programming snippets as proof-of-concept illustration.

## 4.2 Acknowledgments

We would like to acknowledge the following people and entities who made this work possible:

- Dr. Serguei A. Mokhov for his suggestions, advises, and supervision in this paper.
- Faculty of Engineering and Computer Science, Concordia University, Montreal, Canada.
- Concordia University Libraries for access to the invaluable digital libraries of ACM, IEEE, Springer and others to do search.
- Our deepest gratitude goes to our families for their support and unflagging love.
- Finally, our thanks goes to our Bosses and clleouges also.

## [Appendix A] - Java, Haskell, C++ and Perl codes examples

Using Haskell in web services:

```haskell
{-
  Demo Application using HaXml 1.02
  Copyright (c) 2001 by Michael Neumann
  $Id: haxml-demo.hs,v 1.1 2001/10/17 22:41:17 michael Exp $
  HaXml: http://www.cs.york.ac.uk/fp/HaXml
  Compile with: hmake -nhc98 haxml-demo
  For sample XML file, see below.
-}
module Main where
import XmlLib
import XmlCombinators

main = processXMLwith
    ( mkElem "HTML"
        [ mkElem "BODY"
            [ message `o` (tag "messages" /> tag "message") ]
        ]
    )

message =
    mkElem "DIV"
    [
    -- Message Title
    mkElem "DIV"
        [ mkElem "B"
            [ ("["!) , tag "message" /> tag "category" /> txt , ("] "!) ]
        , mkElem "EM"
            [ txt `o` children `o` tag "title" `o` children `o` tag
"message" ]
        , literal " at "
        , tag "message" /> tag "date" /> txt
        , literal " sent by "
        , tag "message" /> tag "author"
            ?> mkElemAttr "A"
            [ ("HREF", (tag "message" /> tag "author" /> tag "email"
/> txt) ) ]
                [ tag "message" /> tag "author" /> tag "name" /> txt ]
            :> literal "Unknown Person"
        ]
    ,
    -- Message Body
    mkElem "DIV"
```

```
      [ keep `o` (tag "message" /> tag "text") ]
  ,
  mkElem "BR" []
  ]

{- Sample XML file follows:
<?xml version="1.0"?>
<messages>
  <message id="1">
    <date>2001-12-31</date>
    <category>Admin</category>
    <title>CVS Updated</title>
    <text>CVS has been updated.</text>
  </message>

  <message id="2">
    <author>
      <name>Michael</name>
      <email>neumann@s-direktnet.de</email>
    </author>
    <date>2001-12-30</date>
    <category>Programming</category>
    <title>New function</title>
    <text>New method <font color="red">set_parser</font> in module
XXX.</text>
  </message>
</messages>
-}
```

Java Reflection example:

```
/*
 * To change this template, choose Tools | Templates
 * and open the template in the editor.
 */
package javaapplication7;
/**
 *
 * @author Sultan
 */
import java.lang.reflect.*;

 class Reflect {
    public static Constructor [] constructors;
    public static Method [] methods;
    public static Field [] fields;
```

```
    public static Class thisClass;

    static{
        try{
            // dissect and see this class itself
            thisClass = Class.forName("reflect");
        }
        catch(ClassNotFoundException cnfe){
            System.out.println("class doesn't exist"+cnfe);
            System.exit(1);
        }
        constructors = thisClass.getConstructors();
        methods = thisClass.getMethods();
        fields = thisClass.getFields();
    }

    public static void main(String [] args){
        System.out.println("This class is "+thisClass);
        for(int i = 0; i < constructors.length; i++)
            System.out.println("Ctor"+i+""+constructors[i]);
        for(int i = 0; i < methods.length; i++)
            System.out.println("Ctor"+i+""+methods[i]);
        for(int i = 0; i < fields.length; i++)
            System.out.println("Ctor"+i+""+fields[i]);
    }
}
```

Declarative Programming example in Java:

```
package njunit;

import java.lang.reflect.*;
import njunit.annotation.*;

public class TestRunner {
    static void executeUnitTests(String className) {
        try {
            Object testObject =
                Class.forName(className).newInstance();
            Method [] methods =
                testObject.getClass().getDeclaredMethods();
            for(Method amethod : methods) {
            UnitTest utAnnotation =
                amethod.getAnnotation(UnitTest.class);
            if(utAnnotation!=null) {
                System.out.print(utAnnotation.value() +
```

```
                        " : " );
            String result =
                invoke(amethod, testObject);
            System.out.println(result);
            }
        }
    }catch(Exception x) {
        x.printStackTrace();
    }
}

static String invoke(Method m, Object o) {
    String result = "passed";
    try{
        m.invoke(o,null);
    } catch(Exception x) {
        result = "failed";
    }
    return result;
}

public static void main(String [] args) {
    executeUnitTests(args[0]);
}
}
```

Criteria 2: C++ and CGI "Hello World"

```
#include <stdio.h>   // include standard IO library
int main ( void ) {  // program main
  // print the HTML response header
  printf("Content-Type: text/plain;charset=us-ascii\n\n");
  // print the plain text to the page
  printf("Hello world\n\n");
  // return success
  return 0;
}
```

Criteria 2: C++ and ISAPI [241]

```
void CHelloWorldExtension::Default(CHttpServerContext* pCtxt)
{
    StartContent(pCtxt); // open the response for writing
    WriteTitle(pCtxt);        // write the title of the response
    *pCtxt << T("<H1>Hello World!</H1>.\r\n"); // write HTML content
    EndContent(pCtxt);        // close the response
```

```
}
```

```
#include <boost/date_time/posix_time/posix_time.hpp>
#include <fstream>
#include <fastcgi++/request.hpp>
#include <fastcgi++/manager.hpp>

void error_log(const char* msg)
{
    using namespace std;
    using namespace boost;
    static ofstream error;
    if(!error.is_open())
    {
        error.open("/tmp/errlog", ios_base::out | ios_base::app);
        error.imbue(locale(error.getloc(), new
posix_time::time_facet()));
    }

    error << '[' << posix_time::second_clock::local_time() << "] " <<
msg << endl;
}

class HelloWorld: public Fastcgipp::Request<wchar_t>
{
    bool response()
    {
        wchar_t russian[]={ 0x041f, 0x0440, 0x0438, 0x0432, 0x0435,
0x0442, 0x0020, 0x043c, 0x0438, 0x0440, 0x0000 };
        wchar_t chinese[]={ 0x4e16, 0x754c, 0x60a8, 0x597d, 0x0000 };
        wchar_t greek[]={ 0x0393, 0x03b5, 0x03b9, 0x03b1, 0x0020,
0x03c3, 0x03b1, 0x03c2, 0x0020, 0x03ba, 0x03cc, 0x03c3, 0x03bc,
0x03bf, 0x0000 };
        wchar_t japanese[]={ 0x4eca, 0x65e5, 0x306f, 0x4e16, 0x754c,
0x0000 };
        wchar_t runic[]={ 0x16ba, 0x16d6, 0x16da, 0x16df, 0x0020,
0x16b9, 0x16df, 0x16c9, 0x16da, 0x16de, 0x0000 };

        out << "Content-Type: text/html; charset=utf-8\r\n\r\n";

        out << "<html><head><meta http-equiv='Content-Type'
content='text/html; charset=utf-8' />";
        out << "<title>fastcgi++: Hello World in UTF-
8</title></head><body>";
```

```
        out << "English: Hello World<br />";
        out << "Russian: " << russian << "<br />";
        out << "Greek: " << greek << "<br />";
        out << "Chinese: " << chinese << "<br />";
        out << "Japanese: " << japanese << "<br />";
        out << "Runic English?: " << runic << "<br />";
        out << "</body></html>";

        err << "Hello apache error log";

        return true;
    }
};

int main()
{
    try
    {
        Fastcgipp::Manager<HelloWorld> fcgi;
        fcgi.handler();
    }
    catch(std::exception& e)
    {
        error_log(e.what());
    }
}
```

Criteria 2: C++ and Wt [242]

```
#include <WApplication>
#include <WText>
#include <WPushButton>

int wmain(int argc, char **argv)
{
WApplication appl(argc, argv);
// Widgets can be added to a parent by calling addWidget() ...
appl.root()->addWidget(new Wtext("<h1>Hello, World!</h1>"));
// ... or by specifying a parent at construction time
WPushButton *Button = new WPushButton("Quit", appl.root());
Button->clicked.connect(SLOT(&appl, Wapplication::quit));
return appl.exec();
}
```

Criteria 2: Perl and CGI

```
#!/usr/bin/perl -Tw
print "Content-type: text/plain\n\n";
print "Hello world!\n";
```

Criteria 2: Perl and CGI Library

```
#!/usr/bin/perl
use CGI qw(:standard);
my $name = param ( "name" );
print header ( "text/plain" );
print "Hello $name!\n\n";
```

Criteria 2: Perl and FastCGI [144]

```
#!/usr/local/bin/perl # must be a FastCGI version of perl!
use CGI::Fast;
# do some initialization
my $x = 0;
$ENV{FCGI_SOCKET_PATH} = "sputnik:8888";
$ENV{FCGI_LISTEN_QUEUE} = 100;
while ($q = new CGI::Fast) {
# process request $q
$x++;
print $x;
}
```

Criteria 2: Perl and mod perl
This is an example file ran from a Location by Apache as a module:

```
ModPerl/HelloWorld.pm
  ----------------
  package ModPerl::HelloWorld;
  use Apache::Constants qw(:common);

  sub handler { # 'handler' subroutine gets called when a request
comes
      my ($request) = @_;
      $request ->send_http_header( 'text/plain' );
      $request ->print( "Hello World!\n" );
      return Apache::OK; # Response OK
  }

  1; # Modules return 1
```

Criteria 5: Reflection in Perl

```
$ ./reflection.pl
Name: Test Reflection Class
```

```
Perl internal literals:
Package: Test::Reflection
File:Test/Reflection.pm
Line:23
Internal field [uncle] : Ben
itk@netqb ~/epawpie $ cat reflection.pl
#!/usr/bin/perl

use strict;
use lib '.';  # library located in local dir
use Test::Reflection;

# set the variables for reflection
my ($class, $method, $field) = ("Test::Reflection", "print_info",
"uncle");
# create an instance
my $instance = $class -> new();
# call the method
$instance -> $method();
# get the field
print "Internal field [$field] : " .  $instance -> {$field} . "\n";
```

**$ cat ./Test/Reflection.pm**

```
package Test::Reflection;
# intercal class fiends
use fields qw(uncle aunt);
# a static public variable
our $CanonicalName = "Test Reflection Class";
# constructor
sub new {
        # get the class name
        my $class = shift;
        # create a pseudo-hash with fields and functions
        my $self = bless {};
        # assign internal varialbes
        $self -> {'uncle'} = 'Ben';
        $self -> {'aunt'} = 'Lucy';
        # return the instance
        return $self;
};
# a method
sub print_info {
        # print statick varaible
        print "Name: $CanonicalName\n";
        # displays the interlan information of the module and code
```

```
        print "Perl internal literals:\nPackage:
".__PACKAGE__."\nFile:".__FILE__."\nLine:".__LINE__."\n";
};
# all modules in Perl have to finish with 1;
1;
```

Criteria 7: Functional programming in C++

```
#include <assert.h>
#include <string>
#include "prelude.h"

using namespace fcpp;
int main() {
        int x=1, y=2, z=3;
        std::string s="foo", t="bar", u="qux";
        List<int> li = cons(x,cons(y,cons(z,NIL)));
        List<std::string> ls = cons(s,cons(t,cons(u,NIL)));
        assert( head(li) == 1 );
        // list_with() makes short lists
        assert( tail(li) == list_with(2,3) );
        ls = compose(tail,tail)(ls);
        assert( head( ls ) == "qux" );
        assert( tail( ls ) == NIL );
}
```

Criteria 10: GTK+ with C++ - plain Hello World window [243]

```
#include <gtk/gtk.h> // include gtk header

int main( int argc, char *argv[])
{
  GtkWidget *window; // create the window handle
  GtkWidget *label;  // a text label

  gtk_init(&argc, &argv);   // initialize gtk with command line options

  window = gtk_window_new(GTK_WINDOW_TOPLEVEL);  // create a new window

  label = gtk_label_new ("Hello, World");  // set label text to Hello
World

  gtk_container_add (GTK_CONTAINER (window), label); // add label to
window

  gtk_widget_show(window);  // set window to visible
```

```
  gtk_main(); // enter main event loop

  return 0;   // exit application - success
}
```

# PHP Example of HTML form processing

L. Beighley and M. Morrison. *Head First PHP & MySQL*. Head First Labs from O'Reilly Media, Inc. 2008.

- How it looks:

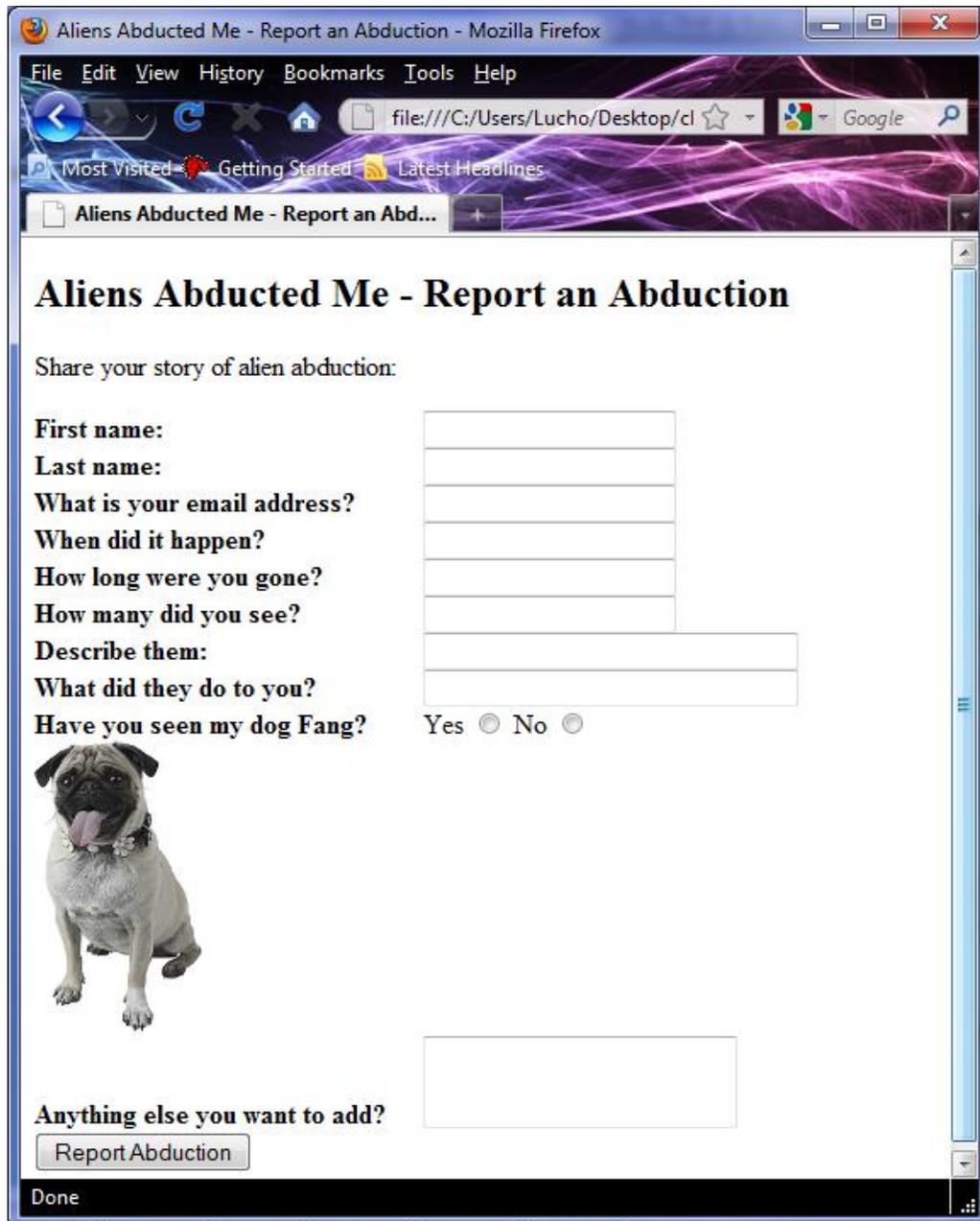

- The HTML file containing the form:

```
<!DOCTYPE html PUBLIC "-//W3C//DTD XHTML 1.0 Transitional//EN"
  "http://www.w3.org/TR/xhtml1/DTD/xhtml1-transitional.dtd">
<html xmlns="http://www.w3.org/1999/xhtml" xml:lang="en" lang="en">
<head>
  <meta  http-equiv="Content-Type"  content="text/html;  charset=utf-8"
```

```
  />
    <title>Aliens Abducted Me - Report an Abduction</title>
    <link rel="stylesheet" type="text/css" href="style.css" />
  </head>
  <body>
    <h2>Aliens Abducted Me - Report an Abduction</h2>

    <p>Share your story of alien abduction:</p>
    <form method="post" action="report.php">
      <label for="firstname">First name:</label>
      <input type="text" id="firstname" name="firstname" /><br />
      <label for="lastname">Last name:</label>
      <input type="text" id="lastname" name="lastname" /><br />
      <label for="email">What is your email address?</label>
      <input type="text" id="email" name="email" /><br />
      <label for="whenithappened">When did it happen?</label>
      <input type="text" id="whenithappened" name="whenithappened" /><br
  />
      <label for="howlong">How long were you gone?</label>
      <input type="text" id="howlong" name="howlong" /><br />
      <label for="howmany">How many did you see?</label>
      <input type="text" id="howmany" name="howmany" /><br />
      <label for="aliendescription">Describe them:</label>
      <input  type="text"  id="aliendescription"  name="aliendescription"
  size="32" /><br />
      <label for="whattheydid">What did they do to you?</label>
      <input  type="text"  id="whattheydid"  name="whattheydid"  size="32"
  /><br />
      <label for="fangspotted">Have you seen my dog Fang?</label>
      Yes  <input   id="fangspotted"   name="fangspotted"   type="radio"
  value="yes" />
      No  <input   id="fangspotted"   name="fangspotted"   type="radio"
  value="no" /><br />
      <img src="fang.jpg" width="100" height="175"
        alt="My abducted dog Fang." /><br />
      <label for="other">Anything else you want to add?</label>
      <textarea id="other" name="other"></textarea><br />
      <input type="submit" value="Report Abduction" name="submit" />
    </form>
  </body>
</html>
```

- The PHP file for processing the form data:

```
<!DOCTYPE html PUBLIC "-//W3C//DTD XHTML 1.0 Transitional//EN"
  "http://www.w3.org/TR/xhtml1/DTD/xhtml1-transitional.dtd">
<html xmlns="http://www.w3.org/1999/xhtml" xml:lang="en" lang="en">
<head>
  <meta http-equiv="Content-Type" content="text/html; charset=utf-8"
/>
  <title>Aliens Abducted Me - Report an Abduction</title>
</head>
```

```
<body>
  <h2>Aliens Abducted Me - Report an Abduction</h2>

<?php
  $name = $_POST['firstname'] . ' ' . $_POST['lastname'];
  $when_it_happened = $_POST['whenithappened'];
  $how_long = $_POST['howlong'];
  $how_many = $_POST['howmany'];
  $alien_description = $_POST['aliendescription'];
  $what_they_did = $_POST['whattheydid'];
  $fang_spotted = $_POST['fangspotted'];
  $email = $_POST['email'];
  $other = $_POST['other'];

  $to = 'owen@aliensabductedme.com';
  $subject = 'Aliens Abducted Me - Abduction Report';
  $msg = "$name was abducted $when_it_happened and was gone for
$how_long.\n" .
    "Number of aliens: $how_many\n" .
    "Alien description: $alien_description\n" .
    "What they did: $what_they_did\n" .
    "Fang spotted: $fang_spotted\n" .
    "Other comments: $other";
  mail($to, $subject, $msg, 'From:' . $email);

  echo 'Thanks for submitting the form.<br />';
  echo 'You were abducted ' . $when_it_happened;
  echo ' and were gone for ' . $how_long . '<br />';
  echo 'Number of aliens: ' . $how_many . '<br />';
  echo 'Describe them: ' . $alien_description . '<br />';
  echo 'The aliens did this: ' . $what_they_did . '<br />';
  echo 'Was Fang there? ' . $fang_spotted . '<br />';
  echo 'Other comments: ' . $other . '<br />';
  echo 'Your email address is ' . $email;
?>

</body>
</html>
```

## [Appendix B] – Rails example files for a basic online ads system built using scaffolding

D. Griffiths. *Head First Rails*. Head First Labs from O'Reilly Media, Inc. 2008.

- How it looks like:

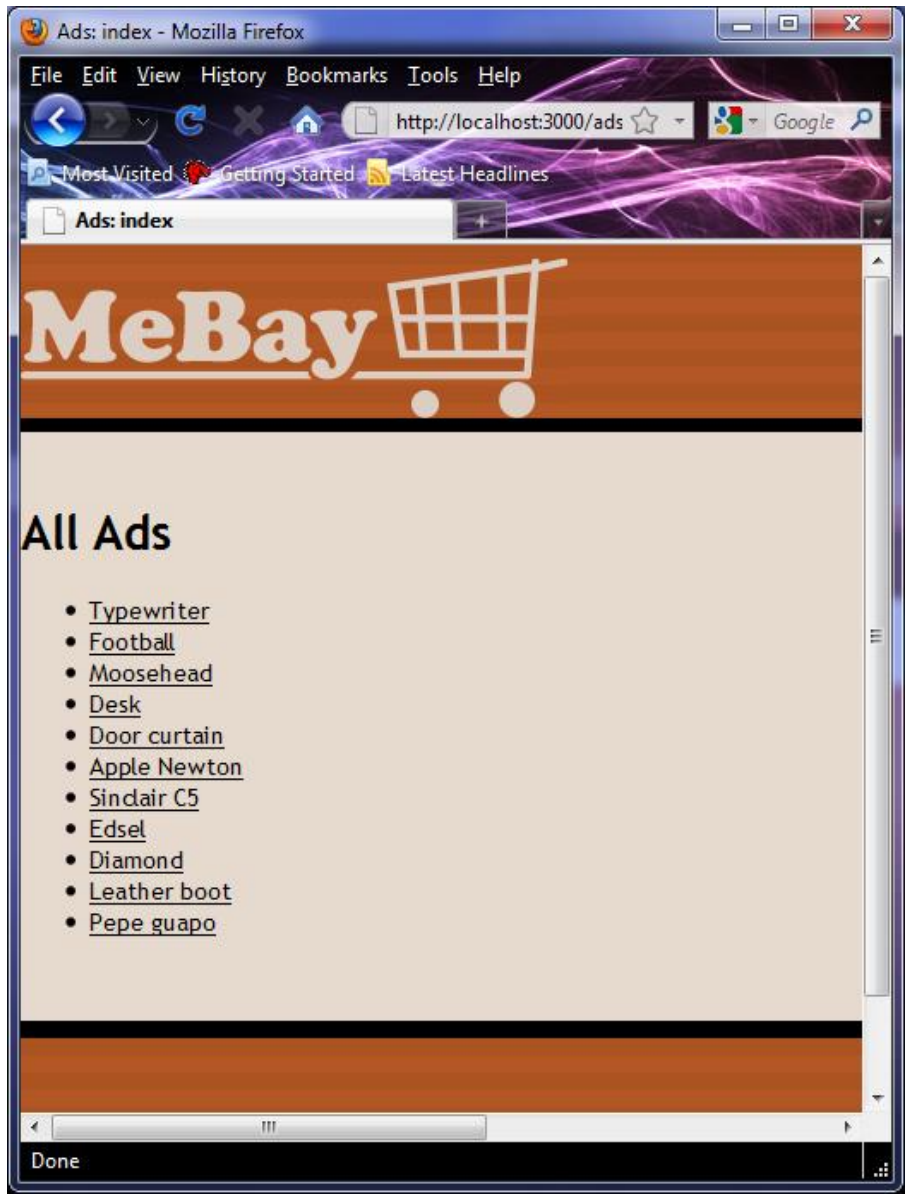

- Controller file (`ads_controller.rb`):

```
class AdsController < ApplicationController
  def new
    @ad = Ad.new
  end

  def index
    @ads = Ad.find(:all)
```

```
  end

  def show
    @ad = Ad.find(params[:id])
  end

  def create
    @ad = Ad.new(params[:ad])
    @ad.save
    redirect_to "/ads/#{ @ad.id }"
  end

  def edit
    @ad = Ad.find(params[:id])
  end

  def update
    @ad = Ad.find(params[:id])
    @ad.update_attributes(params[:ad])
    redirect_to "/ads/#{ @ad.id }"
  end

end
```

- Template file for the application (**ads.html.erb**):

```
<!DOCTYPE html PUBLIC "-//W3C//DTD XHTML 1.0 Transitional//EN"
      "http://www.w3.org/TR/xhtml1/DTD/xhtml1-transitional.dtd">
<html xmlns="http://www.w3.org/1999/xhtml">
<head>
   <title>Ads: <%= controller.action_name %></title>
   <%= stylesheet_link_tag 'default.css' %>
</head>
<body>
   <div id="wrapper">
            <div id="header">
                  <div>

                        <h1>MeBay</h1>
                        <ul id="nav">
                              <li><a href="/ads/">All Ads</a></li>
                        </ul>
                  </div>
            </div>

            <div id="content">
             <%= yield %>
            </div>
            <div id="clearfooter"></div>
      </div>
```

```
        <div id="footer"></div>
</body>
</html>
```

- File to display all ads in the database (index.html.erb):

```
<h1>All Ads</h1>
<ul>
<% for ad in @ads %>
  <li><a href="/ads/<%= ad.id %>"><%= ad.name %></a></li>
<% end %>
</ul>
```

- File to create a new add (**new.html.erb**):

```
<h1>New ad</h1>

<% form_for(@ad,:url=>{:action=>'create'}) do |f| %>

<p><b>Name</b><br /><%= f.text_field :name %></p>

<p><b>Description</b><br /><%= f.text_area :description %></p>

<p><b>Price</b><br /><%= f.text_field :price %></p>

<p><b>Seller</b><br /><%= f.text_field :seller_id %></p>

<p><b>Email</b><br /><%= f.text_field :email %></p>

<p><b>Img url</b><br /><%= f.text_field :img_url %></p>

<p><%= f.submit "Create" %></p>

<% end %>
```

# [Appendix C] – PHP Web service example using SOAP Client

A. Trachtenberg. *Web Services in PHP*. O'Reilly Open Source Conference July 9, 2003. Portland, OR. Available at http://talks.php.net/show/oscon-webservices/.

- oad in the SOAP Client

```php
<?php require 'SOAP/Client.php'; ?>
```

- Generate the Client proxy (from Amazon.com)

```php
<?php
// We have human readable explanation of the API.
$wsdl_url = 'http://soap.amazon.com/schemas3/AmazonWebServices.wsdl';
$WSDL = new SOAP_WSDL($wsdl_url);
$client = $WSDL->getProxy();
?>
```

- Make ManufacturerSearchRequest for OXO Kitchen Supplies

```php
<?php
// I love OXO kitchen products
$params = array('manufacturer' => 'oxo',
                'mode'         => 'kitchen',
                'page'         => 1,
                'type'         => 'lite',
                'tag'          => 'trachtenberg-20',
                'devtag'       => 'XXXXXX'
               );

$hits = $client->ManufacturerSearchRequest($params);
?>
```

- What we get

```xml
<?xml version="1.0" encoding="UTF-8"?>
    <SOAP-ENV:Envelope xmlns:SOAP-
ENC="http://schemas.xmlsoap.org/soap/encoding/" SOAP-
ENV:encodingStyle="http://schemas.xmlsoap.org/soap/encoding/" xmlns:xs
i="http://www.w3.org/2001/XMLSchema-instance" xmlns:SOAP-
ENV="http://schemas.xmlsoap.org/soap/envelope/" xmlns:xsd="http://www.
w3.org/2001/XMLSchema" xmlns:amazon="http://soap.amazon.com">
  <SOAP-ENV:Body>
  <namesp37:ManufacturerSearchRequestResponse xmlns:namesp37="http://s
oap.amazon.com">
  <return xsi:type="amazon:ProductInfo">
  <TotalResults xsi:type="xsd:string">165</TotalResults>
  <Details SOAP-ENC:arrayType="amazon:Details[10]" xsi:type="SOAP-
ENC:Array">
   <Details xsi:type="amazon:Details">
```

```
        <Url xsi:type="xsd:string">http://www.amazon.com/...</Url>
        <Asin xsi:type="xsd:string">B00004OCKR</Asin>
        <ProductName xsi:type="xsd:string">OXO Good Grips Salad Spinner<
/ProductName>
        <Catalog xsi:type="xsd:string">Kitchen</Catalog>
        <Manufacturer xsi:type="xsd:string">OXO</Manufacturer>
        <ImageUrlSmall xsi:type="xsd:string">http://images.amazon.com/..
.</ImageUrlSmall>
        <ImageUrlMedium xsi:type="xsd:string">http://images.amazon.com/.
..</ImageUrlMedium>
        <ImageUrlLarge xsi:type="xsd:string">http://images.amazon.com/..
.</ImageUrlLarge>
        <ListPrice xsi:type="xsd:string">$35.00</ListPrice>
        <OurPrice xsi:type="xsd:string">$24.95</OurPrice>
    </Details>
    </Details>
</return>
</namesp37:ManufacturerSearchRequestResponse>
</SOAP-ENV:Body>
</SOAP-ENV:Envelope>
```

- What we see

```
stdClass Object
(
    [TotalResults] => 165
    [Details] => Array
        (
            [0] => stdClass Object
                (
                    [Url] => http://www.amazon.com/...
                    [Asin] => B00004OCKR
                    [ProductName] => OXO Good Grips Salad Spinner
                    [Catalog] => Kitchen
                    [Manufacturer] => OXO
                    [ImageUrlSmall] => http://images.amazon.com/...
                    [ImageUrlMedium] => http://images.amazon.com/...
                    [ImageUrlLarge] => http://images.amazon.com/...
                    [ListPrice] => $35.00
                    [OurPrice] => $24.95
                )
        )
)
```

- Parse the results

```
<?php
foreach ($hits->Details as $hit) {
    printf('<p style="clear:both"><img src="%s" alt="%s"
        align="left" /><a href="%s">%s</a><br/>%s</p>',
            $hit->ImageUrlSmall,
```

```
            htmlspecialchars($hit->ProductName),
            $this->Url, htmlspecialchars($hit->ProductName),
            htmlspecialchars($hit->OurPrice)
        );
}
?>
```

- Output

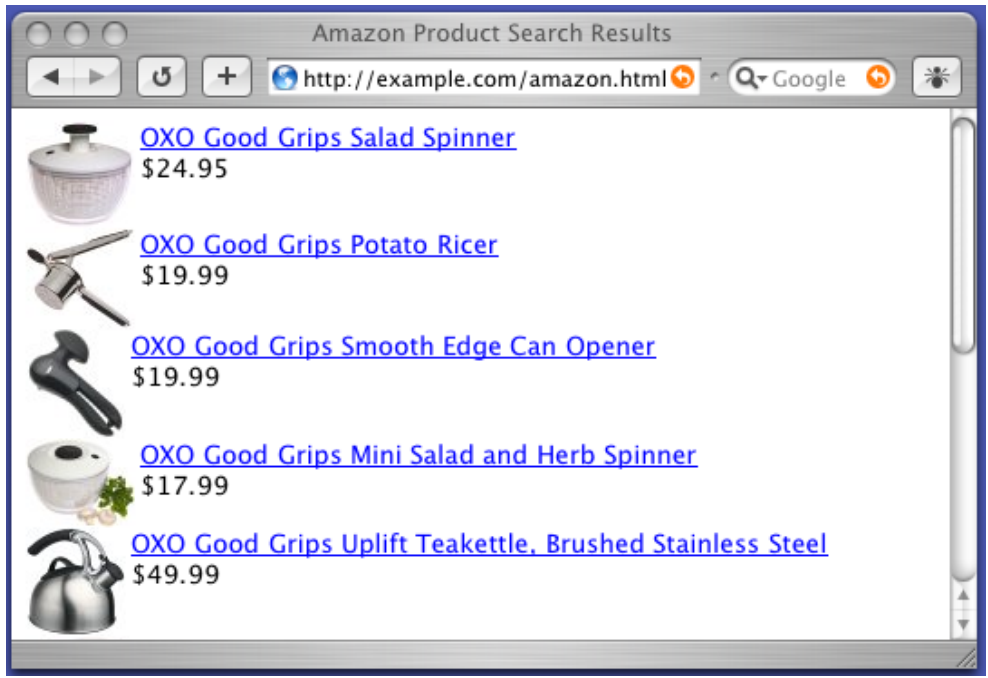

## [Appendix D] – Ruby on Rails Web service example using RESTful

the { buckblogs :here }. *Web services, Rails-style*. Available at
http://weblog.jamisbuck.org/2006/3/27/web-services-rails-style.

Consider: if you have a page in your app that displays a list of people, it might look something like this, without web-service support:

```
def list
  @people = Person.find(:all)
end
```

Here's the same action, with web-service support baked in:

```
def list
  @people = Person.find(:all)

  respond_to do |wants|
    wants.html
    wants.xml { render :xml => @people.to_xml }
  end
end
```

What that says is, "if the client wants HTML in response to this action, just respond as we would have before, but if the client wants XML, return them the list of people in XML format." (Rails determines the desired response format from the HTTP Accept header submitted by the client.)

Now, let's suppose you have an action that adds a new person, optionally creating their company (by name) if it does not already exist. Without web-services, it might look like this:

```
def add
  @company = Company.find_or_create_by_name(params[:company][:name])
  @person  = @company.people.create(params[:person])

  redirect_to(person_list_url)
end
```

Here's the same action, with web-service support baked in:

```
def add
  company = params[:person].delete(:company)
  @company = Company.find_or_create_by_name(company[:name])
  @person  = @company.people.create(params[:person])

  respond_to do |wants|
    wants.html { redirect_to(person_list_url) }
    wants.js
    wants.xml  { render :xml => @person.to_xml(:include => @company) }
  end
end
```

It was simple enough that I also added RJS support here. If the client wants HTML, we just redirect them back to the person list. If they want Javascript (wants.js), then it is an RJS request and we render the RJS template associated with this action. Lastly, if the client wants XML, we render the created person as XML, but with a twist: we also include the person's company in the rendered XML.

Note that you can define your own XML parameter parser which would allow you to describe multiple entities in a single request (i.e., by wrapping them all in a single root note), but if you just go with the flow and accept Rails' defaults, life will be much easier.

## [Appendix E] – Mail queuing system using PHP Batch scripting

This example was taken from [48].

- mailout.sql

```
DROP TABLE IF EXISTS mailouts;
CREATE TABLE mailouts (
  id MEDIUMINT NOT NULL AUTO_INCREMENT,
  from_address TEXT NOT NULL,
  to_address TEXT NOT NULL,
  subject TEXT NOT NULL,
  content TEXT NOT NULL,
  PRIMARY KEY ( id )
);
```

This schema is pretty simple. Each row has a from and a to address, along with a subject and the content of the e-mail.

Wrapped around the mailouts table in the database is the PHP mailouts class.

- mailouts.php

```php
<?php
require_once('DB.php');

class Mailouts
{
  public static function get_db()
  {
    $dsn = 'mysql://root:@localhost/mailout';
    $db =& DB::Connect( $dsn, array() );
    if (PEAR::isError($db)) { die($db->getMessage()); }
    return $db;
  }
  public static function delete( $id )
  {
    $db = Mailouts::get_db();
    $sth = $db->prepare( 'DELETE FROM mailouts WHERE id=?' );
    $db->execute( $sth, $id );
    return true;
  }
  public static function add( $from, $to, $subject, $content )
  {
    $db = Mailouts::get_db();
    $sth = $db->prepare( 'INSERT INTO mailouts VALUES (null,?,?,?,?)'
);
    $db->execute( $sth, array( $from, $to, $subject, $content ) );
    return true;
  }
  public static function get_all()
```

```
    {
        $db = Mailouts::get_db();
        $res = $db->query( "SELECT * FROM mailouts" );
        $rows = array();
        while( $res->fetchInto( $row ) ) { $rows []= $row; }
        return $rows;
    }
}
?>
```

The script includes the Pear::DB database access class. Then it defines a mailouts class with three central static functions: add, delete, and get_all. The add() method adds an e-mail to the queue and is meant to be used by the front end. The get_all() method returns all the data from the table. The delete() method deletes an individual method.

You might ask why I don't just have a delete_all() method that would be called at the end of the script. Such a method doesn't exist for two reasons: If I delete each message after I send it, there's no possibility that a message could be sent twice if a script is rerun after a problem; and new messages could have been added between the start of the batch job and its completion.

The next step is to write a simple test script that adds an entry to the queue.

- mailout_test_add.php

```
<?php
require 'mailout.php';

Mailouts::add( 'donotreply@mydomain.com',
    'molly@nocompany.com.org',
    'Test Subject',
    'This is a test of the batch mail sendout' );
?>
```

In this case, I'm adding a mailout to Molly at some company, with a test subject and e-mail body. I can run this script on the command line: php mailout_test_add.php.

To send the e-mail, I need another script that acts as my job processor.

- mailout_send.php

```
<?php
require_once 'mailout.php';

function process( $from, $to, $subject, $email ) {
    mail( $to, $subject, $email, "From: $from" );
}

$messages = Mailouts::get_all();
```

```
foreach( $messages as $msg ) {
  process( $msg[1], $msg[2], $msg[3], $msg[4] );
  Mailouts::delete( $msg[0] );
}
?>
```

This script uses the get_all() method to retrieve all the e-mail messages, then uses PHP's mail() method to send out the messages one by one. After each is successfully sent, the delete() method removes that individual record from the queue.

This script would run at periodic intervals using the cron daemon. How often the script runs is up to you and the needs of your application.

# [Appendix F] – Comparison of reported vulnerabilities in PHP and Ruby

From: National Vulnerability Database (NVD) CVE Statistics. Available at
http://web.nvd.nist.gov/view/vuln/statistics.

- **PHP**

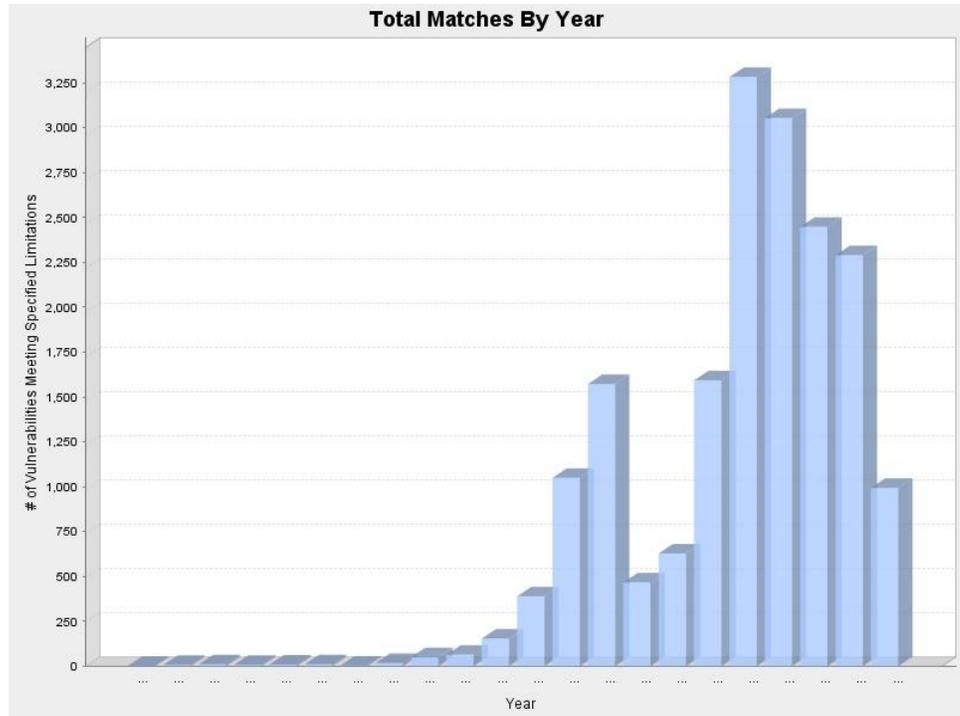

| Statistical Data | | |
|---|---|---|
| **Year** | **# of Vulns** | **% of Total** |
| **1995** | 3 | 12.00 |
| **1996** | 18 | 24.00 |
| **1997** | 50 | 19.84 |
| **1998** | 65 | 26.42 |
| **1999** | 156 | 17.45 |
| **2000** | 395 | 38.73 |
| **2001** | 1,065 | 63.51 |
| **2002** | 1,596 | 74.03 |
| **2003** | 474 | 31.04 |
| **2004** | 638 | 26.03 |
| **2005** | 1,617 | 32.79 |
| **2006** | 3,333 | 50.44 |
| **2007** | 3,101 | 47.61 |
| **2008** | 2,486 | 44.14 |
| **2009** | 2,325 | 40.55 |
| **2010** | 1,009 | 34.10 |

- **Ruby**

## Statistical Data

| Year | # of Vulns | % of Total |
|------|-----------|-----------|
| 2004 | 1 | 0.04 |
| 2005 | 3 | 0.06 |
| 2006 | 8 | 0.12 |
| 2007 | 11 | 0.17 |
| 2008 | 19 | 0.34 |
| 2009 | 12 | 0.21 |
| 2010 | 6 | 0.20 |

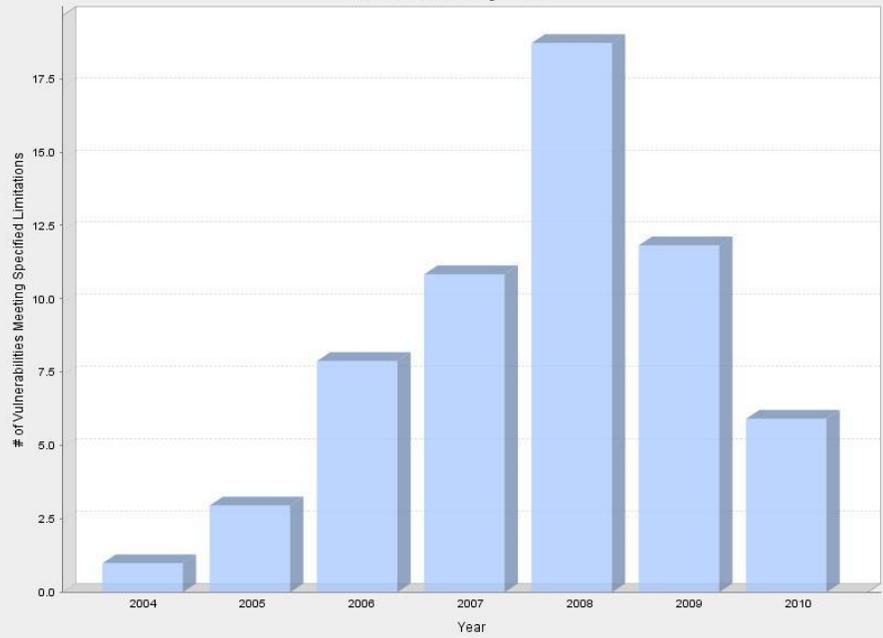

## [Appendix G] – Composition example in PHP

A. Gervasio. *Object Interaction in PHP: Introduction to Composition.* Available at
http://www.devshed.com/c/a/PHP/Object-Interaction-in-PHP-Introduction-to-Composition/.

```php
class Page {
    var $title;
    var $html;
    function Page($title='Default Page'){
        $this->title=$title;
        $this->html='';
    }
    function makeHeader($header){
        $this->html.='<html><head><title>'. $this-
>title.'</title></head><body>'.$header;
    }
    function makeBody($content=array()){
        return new Table($this,$content);
    }
    function makeFooter($footer){
        $this->html.=$footer.'</body></html>';
    }
    function display(){
        return $this->html;
    }
}

class Table {
    var $page;
    var $content;
    var $id;
    function Table(&$page,$content){
        $this->page=&$page;
        $this->content=$content;
        $this->id='defaultID';
    }
    function setId($id){
        $this->id=$id;
    }
    function build($colorA,$colorB){
        $this->page->html.='<table id="'.$this->id.'" width="100%">';
        $i=0;
        foreach($this->content as $row){
            $bgcolor=($i%2)?$colorA:$colorB;
                $this->page->html.='<tr bgcolor="'.$bgcolor.'">
             <td>'.$row.'</td></tr>';
            $i++;
        }
        $this->page->html.='</table>';
        return $this->page->html;
    }
}
```

## [Appendix H] – Inheritance in AspectJ

```
public abstract MultipleReaderSingleWriter perthis (readMethod() ||
writeMethod())
{
//declaring the lock
private ReadWriteLock rw = new WriterPreferenceReadWriteLock();
//two abstract named pointcuts that the extending aspect must redefine
protected abstract pointcut readMethod();
protected abstract pointcut writeMethod();

//advice before the execution of the readMethod that check if it is
possible to acquire the lock then it is possible to read
before() : readMethod(){
      try {rw.readLock().acquire();}
      catch (InterruptedException intEx)
      {
            throw new RuntimeException("Unable to obtain R Lock");
      }
}
// advice after reading
after() : readMethod() {
      rw.readLock().release();
}
// before writing the writer should acquire the write lock
before() : writeMethod() {
      try{
      rw.writeLock().acquire();
      }
      catch (InterruptedException intEx){
      throw new RuntimeException("Unable to obtain W lock");
      }
}
after() : writeMethod(){
      rw.writeLock.release();
}
declare warning: readMethod() && writeMethod() : "A method can not be
read and write method";
}
```

Parent aspect

```
public aspect MyResourceAccessProtection extends
MultipleReadersSingleWriter
{
Protected pointcut readMethod() : execution(* MyResource.*(..)) &&
!writeMethod();
Protected pointcut writeMethod() : execution(*
MyResource.update*(..));)
}
```

## [Appendix I] – Object Oriented COBOL example:

```
Identification Division.
*class name DemoCoo
Class-Id. DemoCOO.
Environment Division.
Configuration Section.
Repository.
Class BaseClasses.
Identification Division.
*definition of the class data
Object.
Data Division.
Working-Storage Section.
*string msg
01 msg pic x(30).
Procedure Division.
Identification Division.
*first method which simply display a message
Method-Id. MethodOne.
Procedure Division.
Display msg.
End Method MethodOne.
*end of definition of the first method
Identification Division.
*second method, the setter method
Method-Id. setMsg.
Data Division.
Linkage Section.
*string in-msg, this string is local, only the second method can use
it
01 in-msg pic x(30).
*the method signature
Procedure Division using in-msg.
Move in-msg to msg.
End Method setMsg.
*end of definition of second method
End Object.
End Class DemoCOO.
**###Main Program###
Identification Division.
      Program-Id. Client.
Environment Division.
Configuration Section.
Repository.
Class DemoCOO.
Data Division.
Working-Storage Section.
01 DC usage object reference DemoCOO.
Procedure Division.
Invoke DemoCOO "new" returning DC
Invoke DC "setMsg" using by content "This is a message when MethodOne
```

```
of DC object call"
Invoke DC "MethodOne"
      Exit Program.
End Program Client.
```

## [Appendix J] – Aspect Oriented Programming using AspectJ:

```
aspect PointBoundsEnforcement {

    private static final int MIN_X = 0;
    private static final int MAX_X = 15;
    private static final int MIN_Y = 0;
    private static final int MAX_Y = 15;
    //the around is executed whenever the method setX is called and
the new X parameter is passed to the clip method which verify if X is
between 0 and 15, if it is not so it is set to 0 or 15
        void around(int newX):
            call(void Point.setX(int)) && args(newX) {
          proceed( clip(newX, MIN_X, MAX_X) );
          System.out.println("Around Advice: point X checked, and it is
valid");
        }
    //the around is executed whenever the method setY is called and
the new Y parameter is passed to the clip method which verify if Y is
between 0 and 15, if it is not so it is set to 0 or 15

        void around(int newY):
            call(void Point.setY(int)) && args(newY) {
          proceed( clip(newY, MIN_Y, MAX_Y) );
          System.out.println("Around Advice: point Y checked, and it is
valid");
        }
// we can see in this example how the aspect could contain methods
used by the advices
        private int clip(int val, int min, int max) {
          return Math.max(min, Math.min(max, val));
        }
    }

package com.simple.aop.demo;

aspect PointBoundsPostCondition {
//named pointcut which associated with execution of any method that
starts with set within the object
    pointcut beforeAnySetCall(): call(void Point.set*(int)) &&
target(p) && args(newX);

// this is executed after the method setX has returned , it checks if
the new value is the that the user entered, it prints a message when
the method returns
  after(Point p, int newX) returning:
            call(void Point.setX(int)) && target(p) && args(newX) {
          assert p.getX() == newX;
          System.out.println("after Advice : Postcondition for X value
from the PointBoundsPostCondition aspect");
```

```
            }
// the same thing but for setY
        after(Point p, int newY) returning:
            call(void Point.setY(int)) && target(p) && args(newY) {
            assert p.getY() == newY;
            System.out.println("after Advice : Postcondition for Y value
from the PointBoundsPostCondition aspect");
        }
    }

package com.simple.aop.demo;

aspect PointBoundsPreCondition {

    private static final int MIN_X = 0;
    private static final int MAX_X = 15;
    private static final int MIN_Y = 0;
    private static final int MAX_Y = 15;

    before(int newX):
            call(void Point.setX(int)) && args(newX) {
        assert newX >= MIN_X;
        assert newX <= MAX_X;
        System.out.println("before Advice : precondition for X value
from the PointBoundsPreCondition aspect");
        }
        before(int newY):
            call(void Point.setY(int)) && args(newY) {
        assert newY >= MIN_Y;
        assert newY <= MAX_Y;
        System.out.println("before Advice : precondition for Y value
from the PointBoundsPreCondition aspect");}}
```

## [Appendix K] – Aspect Oriented Programming using COBOL

```
DECLARATIVES.
HANDLE-F0815-ASPECT SECTION.
USE AFTER ERROR ON FILE-F1.
HANDLE-F0815-ADVICE.
MOVE "F0815" TO PANIC-RESOURCE.
MOVE "FILE ERROR" TO PANIC-CATEGORY.
MOVE FILE-STATUS TO PANIC-CODE.
GO TO PANIC-STOP.
END DECLARATIVES.
```

## [Appendix L] – Reflection Using AspectJ

```
package com.demo;

public class Demo {

    static Demo d;

    void run(){
        d = new Demo();
        d.method1(1,"whatever string ");
        System.out.println(d.method2(new Integer(3)));
    }

    void method1(int i, String o){
        System.out.println("Demo.method1(" + i + ", " + o + ")\n");
    }

    String method2 (Integer j){
        System.out.println("Demo.method2(" + j + ")\n");
        return "Demo.method2(" + j  + ")";
    }

    public static void main(String[] args){
        new Demo().run();
    }
}

package com.demo;

import org.aspectj.lang.JoinPoint;
import org.aspectj.lang.reflect.CodeSignature;

aspect GetInfo {

        static final void println(String s){ System.out.println(s); }
// this is a to intercept any execution within the Demo object of
whatever object
        pointcut demoExecs(): within(Demo) && execution(* *(..));
// this is the advice during execution of a method
        Object around(): demoExecs() {
      // the case if the executing method is the main  method
        if
(!thisJoinPointStaticPart.getSignature().getName().equalsIgnoreCase("m
ain"))
              {
// this line return the name of the executing class
            println("Intercepted message: " +
                thisJoinPointStaticPart.getSignature().getName());
```

```
        println("in class: " +

thisJoinPointStaticPart.getSignature().getDeclaringType().getName());
// this call to the printPramas method which print the information
about the executing method. Information are the name of the method the
value of the parameters and their value.
        printParameters(thisJoinPoint);
        println("original method in running: \n" );
            }
        return proceed();
    }

    static private void printParameters(JoinPoint jp) {
        println("Arguments: " );
        Object[] args = jp.getArgs();
        String[] names =
((CodeSignature)jp.getSignature()).getParameterNames();
        Class[] types =
((CodeSignature)jp.getSignature()).getParameterTypes();
        for (int i = 0; i < args.length; i++) {
            println("   "  + i + ". " + names[i] +
                " : " +            types[i].getName() +
                " = " +            args[i]);
        }
    }
}
```

## [Appendix M] – Annotation Using AspectJ

```
@Aspect
Public class world{}
```

Is equivalent to:

```
Public aspect world {}
```

Another example

```
import project.pack1.Class1;

public aspect Foo {
        // any call of Class1 methods
        pointcut Class1Operation() : call(* class1.*(..));
        // any call of methods in classes under the pack1 package
        pointcut anyPack1Call() : call(* project.pack1..*(..));
}
```

Using the annotation style this is equivalent to :

```
@Aspect
public class Foo {
        @Pointcut("call(* project.pack1.Class1.*(..))")
        void Class1Operation() {}

        @Pointcut("call(* project.pack1..*(..))")
        void anyPack1Call() {}
}
```